%% file: r3.tex
\documentclass[a4paper,10pt]{iopart}
\usepackage[utf8x]{inputenc}
\usepackage{color}
\usepackage{graphicx}
\usepackage{amssymb}
\usepackage[bookmarks, bookmarksopen, bookmarksnumbered, colorlinks]{hyperref}

\newcommand{\eff}[1]{\ensuremath{{#1}_{\rm eff}}}
\newcommand{\beq}{\begin{equation}}
\newcommand{\eeq}{\end{equation}}
\newcommand{\bea}{\begin{eqnarray}}
\newcommand{\eea}{\end{eqnarray}}
\newcommand{\bit}{\begin{itemize}}
\newcommand{\eit}{\end{itemize}}
\newcommand{\bfi}{\begin{figure}}
\newcommand{\efi}{\end{figure}}
\newcommand{\bfic}{\begin{figure*}}
\newcommand{\efic}{\end{figure*}}
\newcommand{\bce}{\begin{center}}
\newcommand{\ece}{\end{center}}
\newcommand{\bt}{\begin{table}}
\newcommand{\et}{\end{table}}
\newcommand{\btb}{\begin{tabular}}
\newcommand{\etb}{\end{tabular}}

\newcommand{\qed}{\nobreak \ifvmode \relax \else
      \ifdim\lastskip<1.5em \hskip-\lastskip
      \hskip1.5em plus0em minus0.5em \fi \nobreak
      \vrule height0.75em width0.5em depth0.25em\fi}

\begin{document}

\title{Evolution of a family of expanding cubic black-hole lattices in numerical relativity}
\author{Eloisa Bentivegna}
\address{
Max-Planck-Institut f\"ur Gravitationsphysik\\
Albert-Einstein-Institut\footnote{\mbox{AEI preprint number: AEI-2013-218}} \\ 
Am M\"uhlenberg 1, D-14476 Golm \\
Germany}
\ead{eloisa.bentivegna@aei.mpg.de}

\author{Miko\l{}aj Korzy\'nski}
\address{
Center for Theoretical Physics \\
Polish Academy of Sciences \\
Al. Lotnik\'o{}w 32/46, 02-668 Warsaw \\
Poland}
\ead{korzynski@cft.edu.pl}

\begin{abstract}
We present the numerical evolution of a family of conformally-flat, infinite,
expanding cubic black-hole lattices. We solve for the initial data 
using an initial-data prescription presented recently, 
along with a new multigrid solver developed for this purpose.
We then apply the standard
tools of numerical relativity to calculate the time development of this
initial dataset and derive quantities of cosmological relevance, such
as the scaling of proper lengths. Similarly to the case of $S^3$ 
lattices, we find that the length scaling remains close to the analytical
solution for Friedmann-Lema\^itre-Robertson-Walker cosmologies throughout
our simulations, which span a window of about one order of magnitude
in the growth of the scale factor.
We highlight, however, a number of important departures from the 
Friedmann-Lema\^itre-Robertson-Walker class.
\end{abstract}


\section{Introduction}

Black-hole lattices have recently attracted some attention as simple
models of a universe in which inhomogeneities coexist with a large-scale
symmetry represented by an exact, discrete invariance under
rigid translations taking each black hole into one of its neighbours
~\cite{RevModPhys.29.432,springerlink:10.1007/BF01889418,Clifton:2009jw,
Clifton:2012qh,Yoo:2012jz,Bruneton:2012cg,Bentivegna:2012ei,Bruneton:2012ru}. They serve as relatively
simple test beds to study whether small scale inhomogeneities can influence the large scale behavior of the model in a
significant way and produce various backreaction effects \cite{Ellis:2011hk}. 
Despite their lack of Killing vectors, they retain a discrete group of isometries and thus remain homogeneous on large scales.

Initial
data for both contracting~\cite{springerlink:10.1007/BF01889418,Clifton:2012qh,
Bentivegna:2012ei} and expanding~\cite{Yoo:2012jz} lattices have appeared
in the literature, along with the analysis of their properties, such as 
the mass-to-expansion-rate proportionality and the associated backreaction
problem. In the case of lattices which are momentarily at rest, the full numerical-relativity
evolution of the 8-black-hole model has been carried out for roughly one
half of the collapsing time of the corresponding Friedmann-Lema\^itre-Robertson-Walker (FLRW) 
spacetime, exhibiting a length scaling that 
remained close to the FLRW counterpart at essentially all times~\cite{Bentivegna:2012ei}.

In this work, we build on~\cite{Bentivegna:2012ei} and apply the same
analysis to the evolution of initial data for an expanding cubic lattice, which 
we construct using the prescription from~\cite{Yoo:2012jz}. We then compare the result with its natural 
homogeneous counterpart, i.e. flat FLRW model with dust. 

The paper is organized as follows: in the next section we discuss the numerical construction of the initial data. In section~\ref{sec:results}
we discuss the results of the initial data construction and the numerical evolution. In section~\ref{sec:comp} 
we compare the results with the corresponding FLRW universe, and draw our conclusions in section~\ref{sec:concl}. 
A number of details, such as the convergence study and the gauge conditions necessary to evolve black holes
in an expanding lattice, are discussed in the appendices.

\section{Constructing initial data for a periodic, expanding black-hole lattice}
\label{sec:id}

In order to solve for initial data, we start with the standard 3+1 decomposition
of the metric tensor, where one works with the spatial metric $\gamma_{ij}$ and 
extrinsic curvature $K_{ij}$. In these terms, the Einstein constraints take
the form:
\bea
\label{eq:constraints}
R+K^2-K_{ij}K^{ij} = 0\\
D_j K^j_i-D_i K    = 0
\eea

In~\cite{Yoo:2012jz}, an expanding lattice solution is presented where the
trace of the extrinsic curvature $K_{ij}$ on the initial slice transitions 
from zero near the center to a constant, negative value next to the periodic
faces. One can thus expect to recover the Schwarzschild solution in the
familiar isotropic coordinates next to the black holes, whilst the volume expansion 
of the cell faces, represented (up to a sign) by the trace of the extrinsic 
curvature, is constant and positive, thus mimicking a homogeneous and 
isotropic spacetime, at least initially.

We generate similar initial data, following some of the prescriptions of~\cite{Yoo:2012jz},
in~\cite{Bentivegna:2013xna}.
The results are obtained by solving the constraints
equation according to the Lichnerowicz-York scheme:
\begin{eqnarray}
\label{eq:CTTconstraints}
 \tilde \Delta \psi - \frac{\tilde R}{8}\,\psi - \frac{K^2}{12}\,\psi^5 + \frac{1}{8} {\tilde A}_{ij} {\tilde A}^{ij} \psi^{-7} = 0 \\
 \tilde D_i \tilde A^{ij} - \frac{2}{3} \psi^6 \tilde \gamma^{ij} \tilde D_i K = 0
\end{eqnarray}
We choose a flat metric as the conformal ``background'' metric:
\bea
\tilde \gamma_{ij} = \delta_{ij}\\
\tilde R = 0
\eea
and the following form for the traceless part of the extrinsic curvature:
\beq
\tilde A_{ij} = \tilde D_i X_j + \tilde D_j X_i - \frac{2}{3} \tilde \gamma_{ij} \tilde D_k X^k
\eeq
$X_i$ being a new variable to solve for. The trace of the extrinsic curvature takes the form
\beq
K=K_c W
\eeq
where $W$ is a function that vanishes near the center and is equal to one on the faces,
with a smooth transition region in between:
\bea
W(r) &=& \left\{
  \begin{array}{ll}
  0 & \textrm{for } 0 \le r \le l \\
  \left(\frac{(r-l-\sigma)^6}{\sigma^6}-1\right)^6&\textrm{for } l \le r \le l + \sigma \\
  1 & \textrm{for } l + \sigma \le r
  \end{array} 
\right.
\eea
with $0 \le l \le l+\sigma \le L$ (see equation (13) in~\cite{Yoo:2012jz}).
The conformal factor is further decomposed in its singular part, with a simple $m/r$
behaviour, and a regular part $\psi_r$ which is solved for:
\beq
\psi=\psi_r+\frac{m}{2r}\left(1-W(r)\right)
\eeq
Notice that a non-constant $K$ implies that the full coupled system of 
Hamiltonian and momentum constraints~\cite{lichnerowicz:1944,York:1971hw} has to be 
solved concurrently. This is one of the main differences with the work in~\cite{Bentivegna:2012ei}.
The resulting system reads (dropping all the tildes as now all quantities refer
to the flat, conformal metric):
\begin{eqnarray}
 \Delta \psi_r - \Delta \left(\frac{m}{2r}W(r)\right)- \frac{K^2}{12}\,\psi^5 + \frac{1}{8} {A}_{ij} {A}^{ij} \psi^{-7} = 0  \label{eq:YooHam}\\
 \Delta X^i +\frac{1}{3} \partial^i \partial_j X^j - \frac{2}{3} \psi^6 \partial^i K = 0 \label{eq:YooMom}
\end{eqnarray}
In~\cite{Yoo:2012jz}, periodic boundary conditions are imposed on $\psi_r$. Notice that $X^i$ 
needs not be periodic in order to have a periodic $\tilde A_{ij}$. However, we have
found that imposing periodic boundary conditions on $X^i$ does not affect the generality 
of the solution; therefore, we impose periodic boundary conditions on all the variables of
our system.
Furthermore, in~\cite{Yoo:2012jz} the divergence of $X^i$ is introduced as an auxiliary 
variable named $Z=\partial_i X^i$. We have found no particular benefit in this choice;
as a matter of fact, we have found that solving the new system with relaxation
leads to rather large violations of the constraint $Z=\partial_i X^i$, which slow down convergence 
unless the constraint is enforced after every step.

All the quantities in the problem have to be such that the integrability condition
is satisfied:
\beq
\label{eq:integr}
K_c^2 \int W^2 \psi^5 = 2 \pi \left( m + \frac{1}{8}\int \tilde A_{ij} \tilde A^{ij} \psi^{-7}\right )  
\eeq
As explained in~\cite{Bentivegna:2013xna}, one only has to worry about the condition associated to the 
Hamiltonian constraint as the one for momentum constraint is satisfied identically. 
In~\cite{Yoo:2012jz}, this condition is satisfied by solving $K_c$. In~\cite{Bentivegna:2013xna}, 
however, we find that fixing $K_c$ at the start and tuning the zero-frequency part of
$\psi$ leads to faster convergence of the relaxation algorithm (this comes at the price of 
introducing a root-finding step at the end of each relaxation step, and also of giving up
control of the scale of $\psi$ at the origin, which is directly related to the ADM
mass of the black hole, as will be shown below). This is what we ultimately
use to generate the initial data evolved in this paper.
We also perform the resetting:
\beq
X^i \to X^i - X^i(O)
\eeq
to fix the zero mode of $X^i$, not determined by the periodic linear equation.

As opposed to~\cite{Bentivegna:2012ei}, we have three free parameters to specify
in this class of solutions: the mass parameter of each black hole $m$, the coordinate semilength $L$
of the lattice edge, and the value of $K_c$. Physically, however, they are not entirely independent because the equations
are invariant under the rescaling $m \to \lambda\,m$, $L\to \lambda\,L$ and $K_c \to \lambda^{-1} K_c$. Therefore we will parametrize our solution by a single scale--invariant number instead of the bare mass $m$. The parameter $\mu$, 
representing the degree of lumpiness of the solution, can e.g.~be defined as:
\beq
\mu = \frac{m_{\rm BH}(m,L,K_c)}{D_{\rm edge}(m,L,K_c)}
\eeq
where $m_{\rm BH}$ is the ADM mass of the black hole, and $D_{\rm edge}$ is the 
initial proper length of a cell's edge.

\section{Results}
\label{sec:results}

In order to solve for the initial data of a black hole in a cubic 
cell with periodic boundaries enforced at the faces, we apply the
multigrid elliptic solver described in~\cite{Bentivegna:2013xna}. This code is  
designed to be used with the infrastructure of the Einstein Toolkit~\cite{Loffler:2011ay}; in particular,
it is implemented as a module for the \texttt{Cactus} framework and relies on
the \texttt{Carpet} AMR package~\cite{Schnetter:2003rb}.

The multigrid implementation is rather standard: on each refinement level, 
the constraint system or the associated error equation is smoothed using a 
non-linear Gauss-Seidel solver, and information is passed between levels
through the \texttt{Carpet} restriction and prolongation operators, and following
the Full Approximation Scheme (FAS). The levels are traversed according to
Full Multigrid (FMG) prescription. We refer to~\cite{Bentivegna:2013xna} for 
the details.

Once the initial data are generated, we evolve them using the Einstein evolution 
code \texttt{McLachlan}.
The evolution and analysis tools used here are almost identical 
to those presented in~\cite{Bentivegna:2012ei}: we use the same BSSN formulation 
of the Einstein evolution system, but we slightly modify the usual moving-puncture gauge 
to prevent the exponential increase of the lapse in regions with negative $K$ 
(see~\ref{appendix:B}). In order to carry out the coordinate transformation from 
numerical to proper time, we use the same strategy and tools described 
in~\cite{Bentivegna:2012ei}.

The grid setup involves a base grid extending to $x=\pm 5$, $y=\pm 5$ and $z=\pm 5$,
with spacing $\Delta_0 = 1$ and with four 
additional grids covering the same domain, each at double the spatial resolution 
as the previous one, so that the finest resolution at the center is given by 
$\Delta_4 = 0.0625$. In the $m=1$ case, we also run two additional resolutions,
with $\Delta_0 = 5/6$ and $\Delta_0 = 5/8$, in order to quantify the numerical error.

Unlike regular lattices on $S^3$, in a regular lattice on a flat background 
we have the freedom to vary the density parameter continuously. We choose four 
values of $m={0.5, 1, 2, 5}$ and set $K_c=-0.21$, obtaining the 
ADM masses and initial proper cell-edge lengths in Table~\ref{table:IDpars}.
\begin{table*}
\bce
\caption{Initial data parameters for the four runs with $m={0.5, 1, 2, 5}$ and 
$K_c=-0.21$. $m_{\rm BH}$ is the ADM mass of the black hole at the center, 
and $|H|$ the $L_2$-norm of the Hamiltonian constraint.
For the $m=1$ configuration, three resolutions $\Delta_0={5/8, 5/6, 1}$ 
have been run to quantify the numerical error.\label{table:IDpars}}
\begin{tabular}{llllll}
\hline
Run & $m$ & $m_{\rm BH}$ & $D_{\rm edge}(t=0)$ & $|H|$                & $\Delta_0$     \\
\hline
L1  &   1 &  1.00797     &  12.2607            & $1.29 \cdot 10^{-4}$ & 1            \\
L2  &   1 &  1.00800     &  12.2607            & $7.03 \cdot 10^{-5}$ & 5/6          \\
L3  &   1 &  1.00801     &  12.2604            & $2.72 \cdot 10^{-5}$ & 5/8          \\
L4  & 0.5 &  0.46060     &  9.41388            & $1.61 \cdot 10^{-3}$ & 1            \\
L5  &   2 &  2.13032     &  15.9943            & $2.36 \cdot 10^{-5}$ & 1            \\
L6  &   5 &  6.34535     &  26.9543            & $8.99 \cdot 10^{-4}$ & 1            \\ 
\hline
\end{tabular}
\ece
\end{table*}


The evolution of the conformal factor $W = \psi^{-2}$ is rather featureless, with an overall 
change in scale as the most prominent effect. Profiles of $W$ at $t=0, 30, 60, 90, 120$ and $150$
are displayed in Figure~\ref{fig:psi}.
\bfi
\bce
\includegraphics[width=0.45\textwidth, trim=140 0 10 0, clip=true]{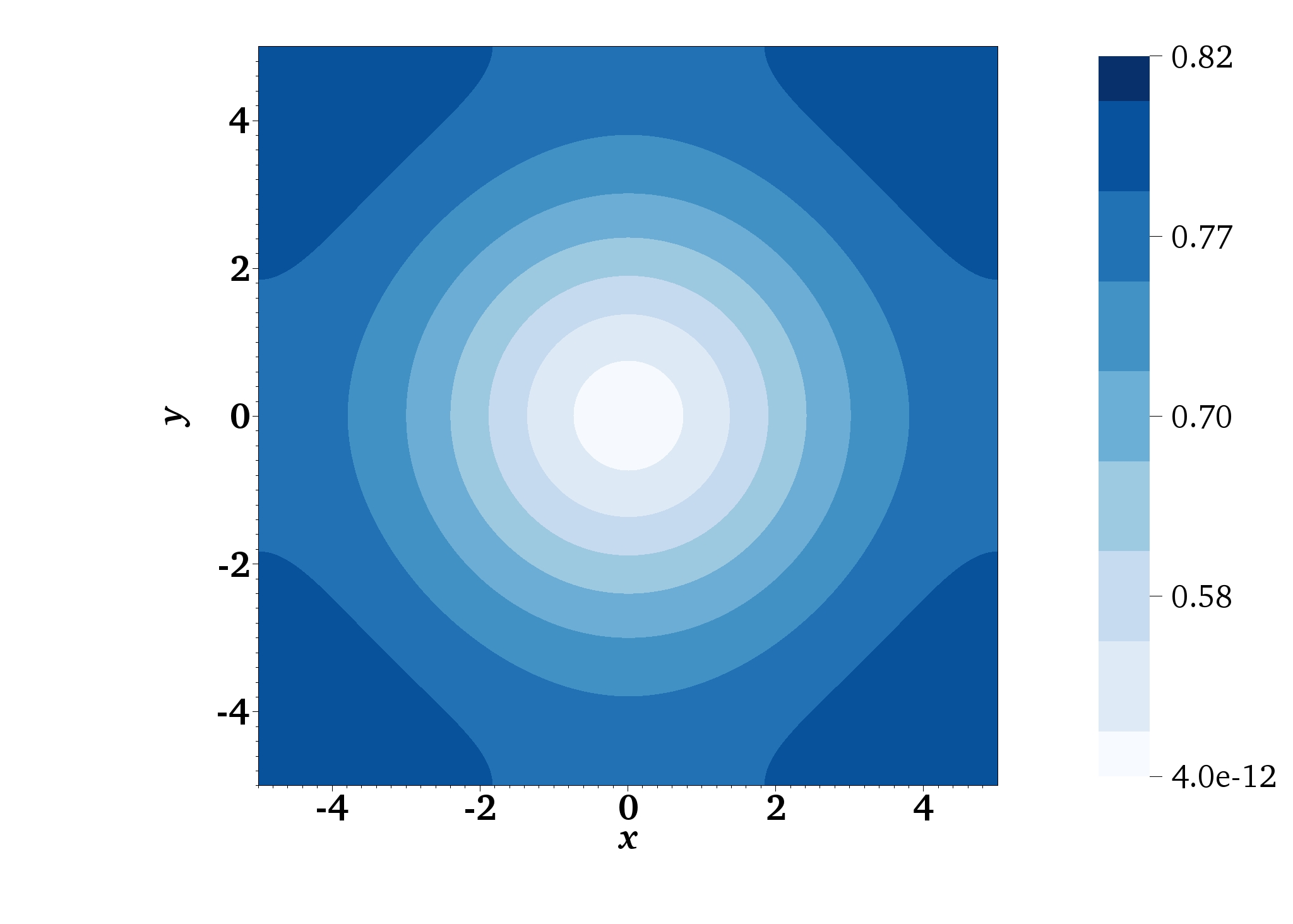}
\includegraphics[width=0.45\textwidth, trim=140 0 10 0, clip=true]{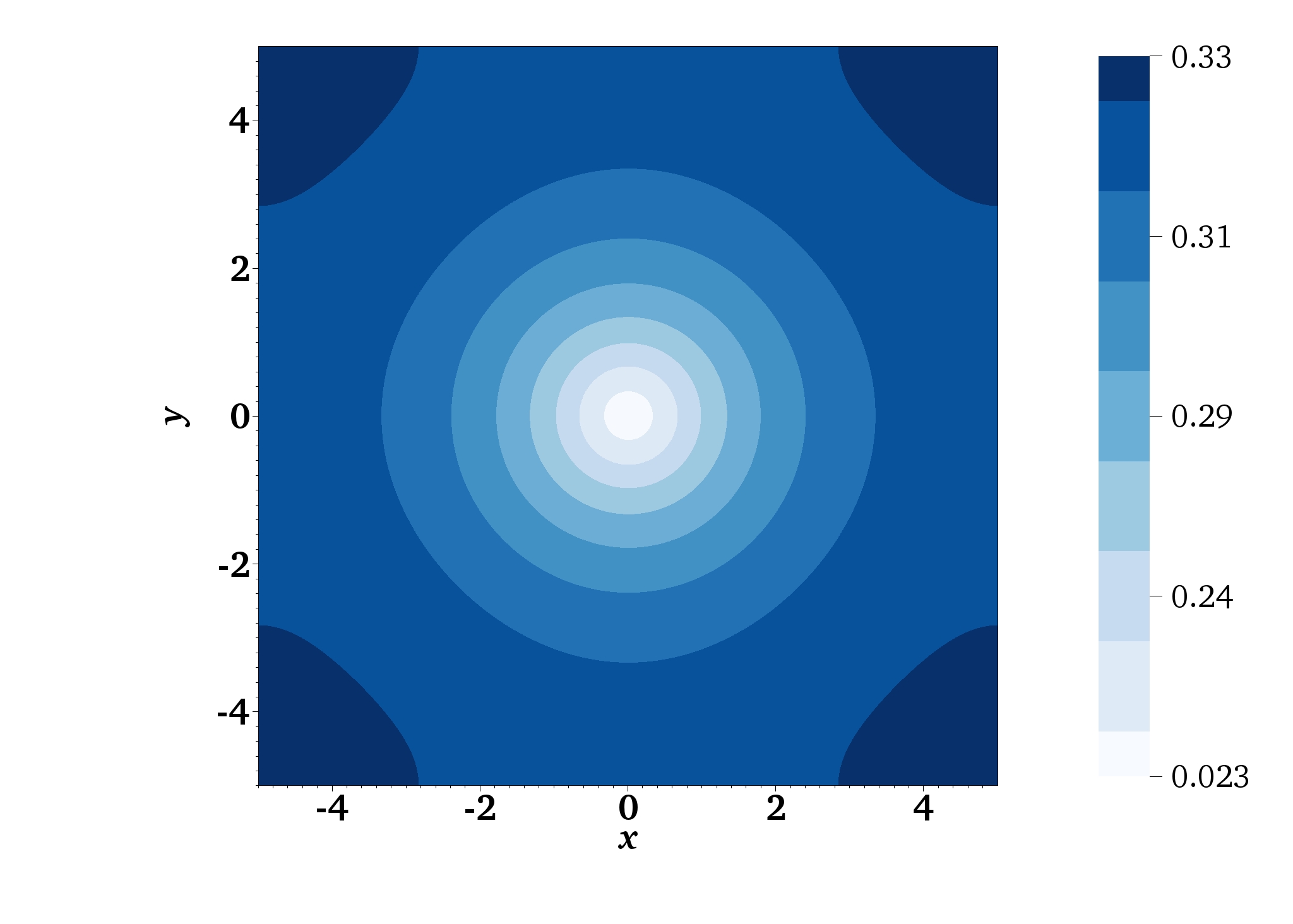}
\includegraphics[width=0.45\textwidth, trim=140 0 10 0, clip=true]{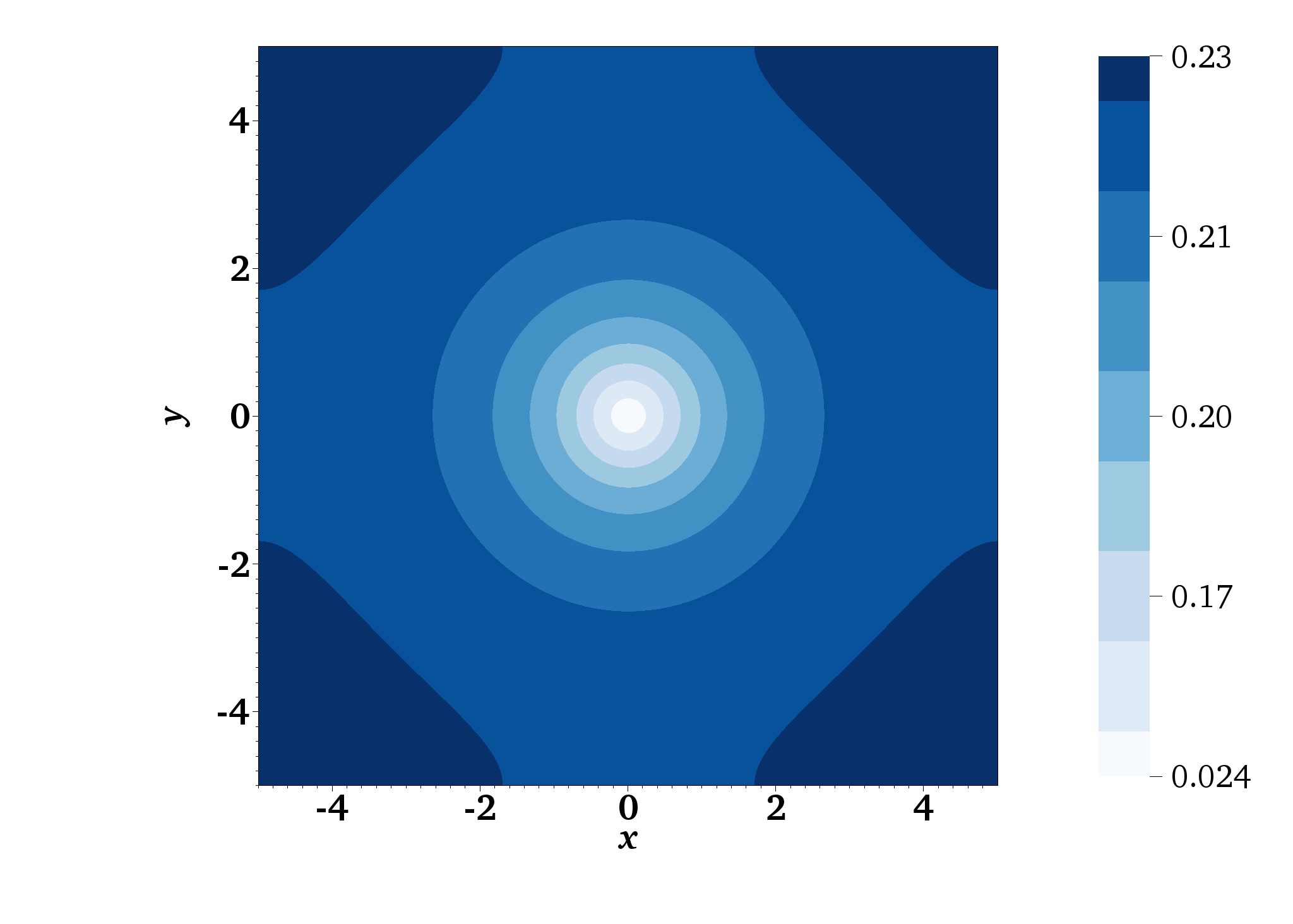}
\includegraphics[width=0.45\textwidth, trim=140 0 10 0, clip=true]{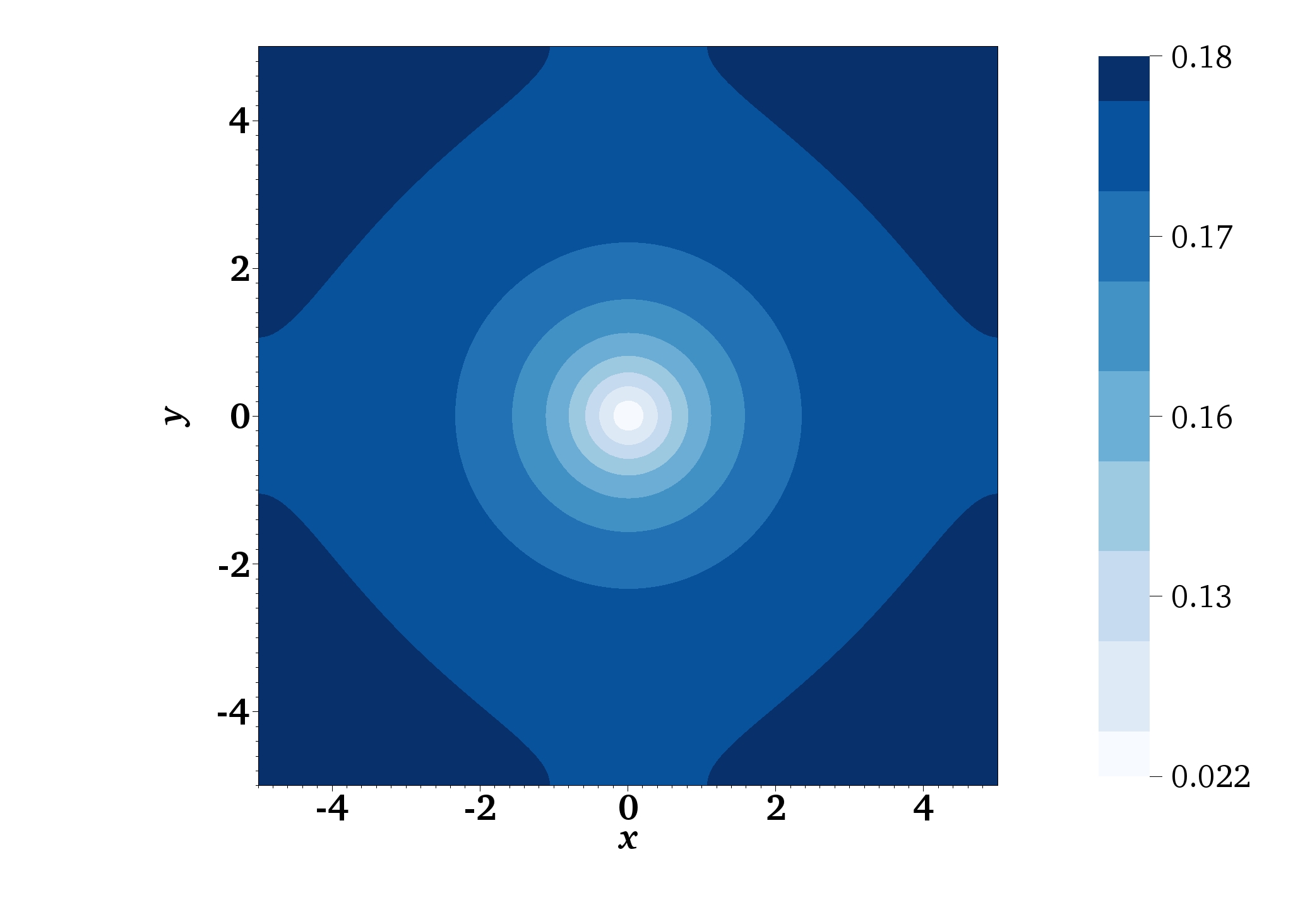}
\includegraphics[width=0.45\textwidth, trim=140 0 10 0, clip=true]{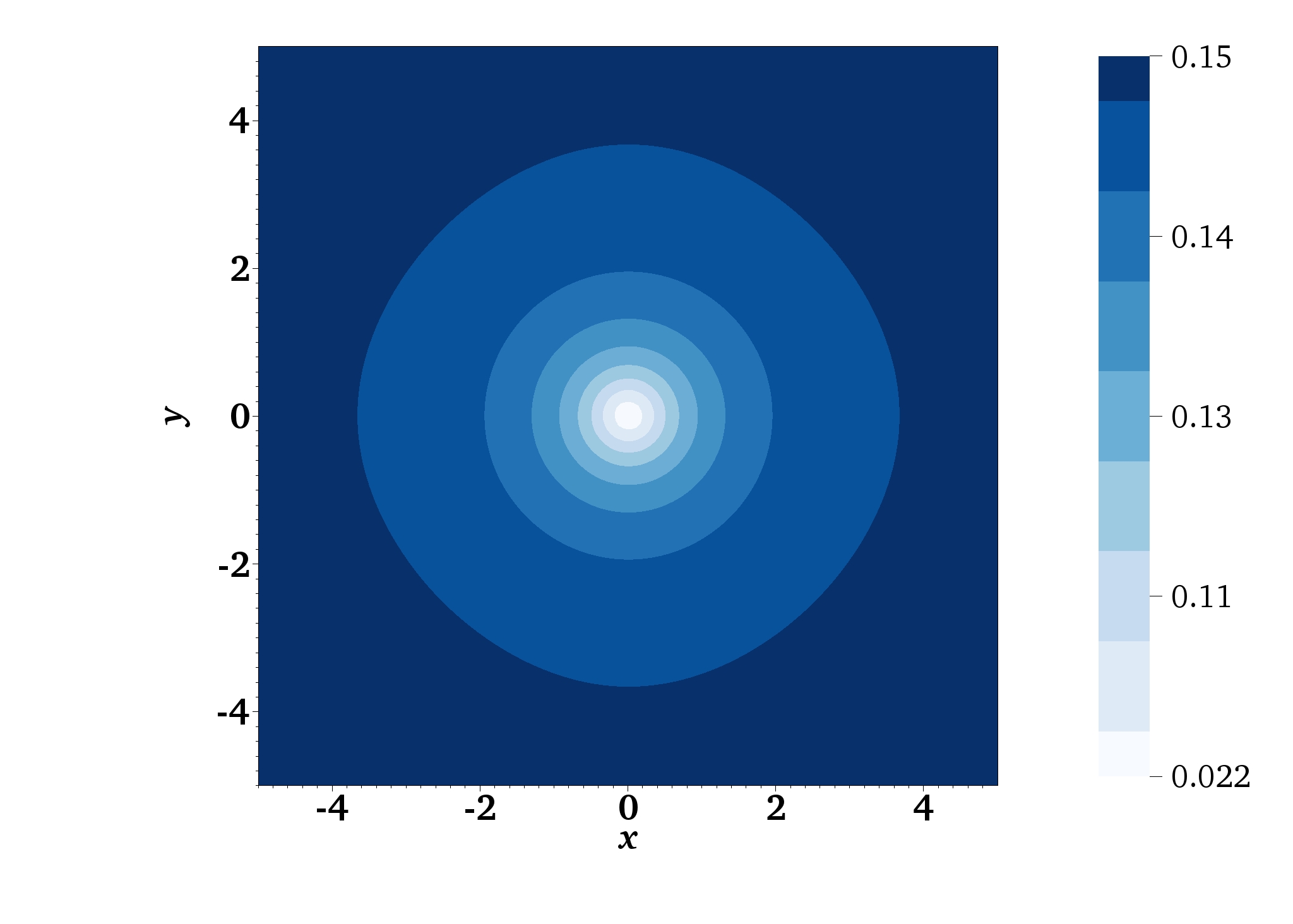}
\includegraphics[width=0.45\textwidth, trim=140 0 10 0, clip=true]{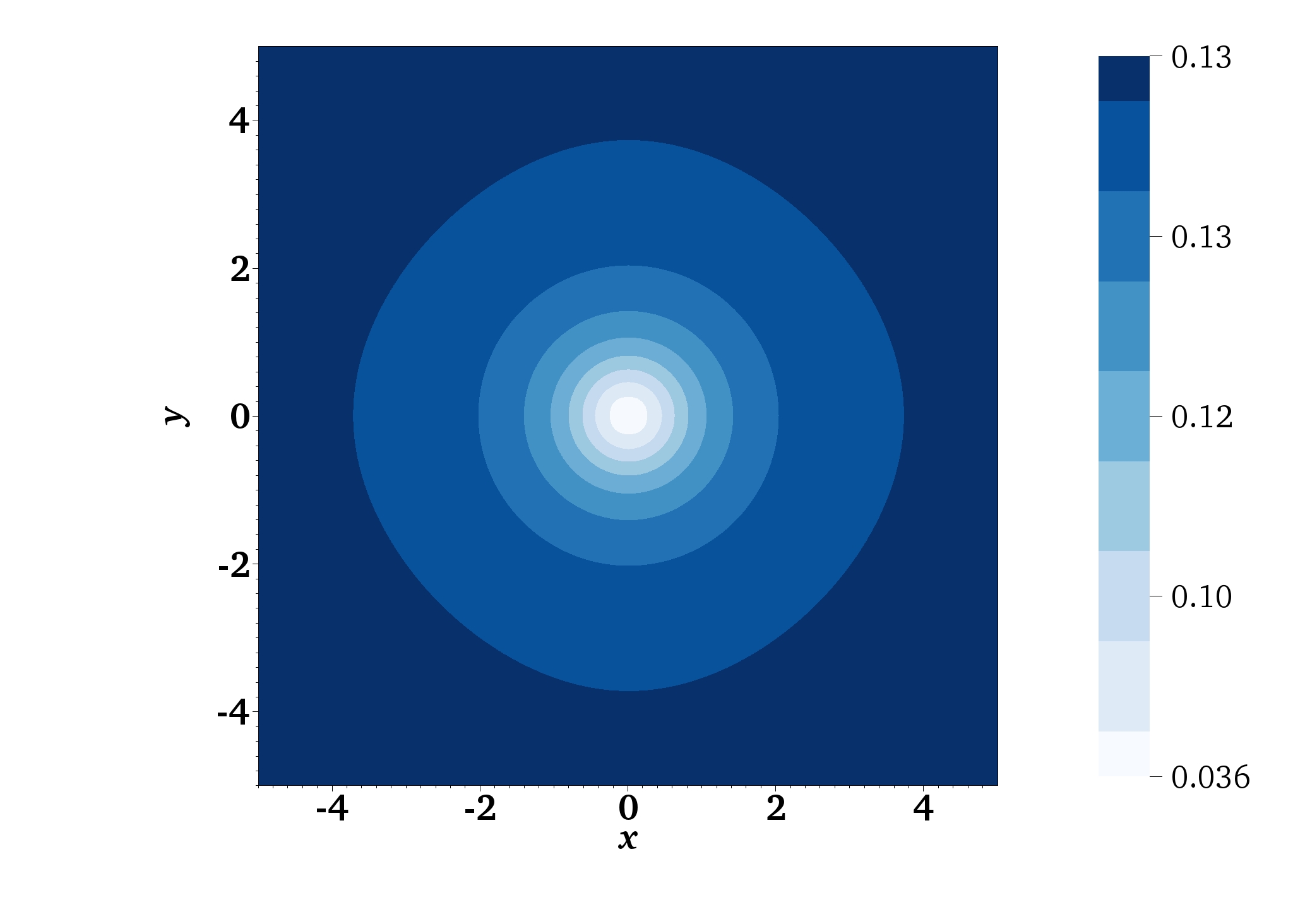}
\caption{Profiles of $W = \psi^{-2}$ on the equatorial plane at $t=0, 30, 60, 90, 120$ and $150$
for the $m=1$, $\Delta_0=1$ run.
\label{fig:psi}}
\ece
\efi
The trace of the extrinsic curvature $K$, on the other hand, transitions from a 
nearly-spherically-symmetric initial profile, with zero $K$ at the center, 
to a configuration where $K$ is negative everywhere, with a maximum in 
magnitude on the faces and a minimum at the center (see Figures~\ref{fig:K} 
and ~\ref{fig:K3D}). There is also an overall change in scale corresponding 
to that observed in $W$.

\bfi
\bce
\includegraphics[width=0.45\textwidth, trim=140 0 10 0, clip=true]{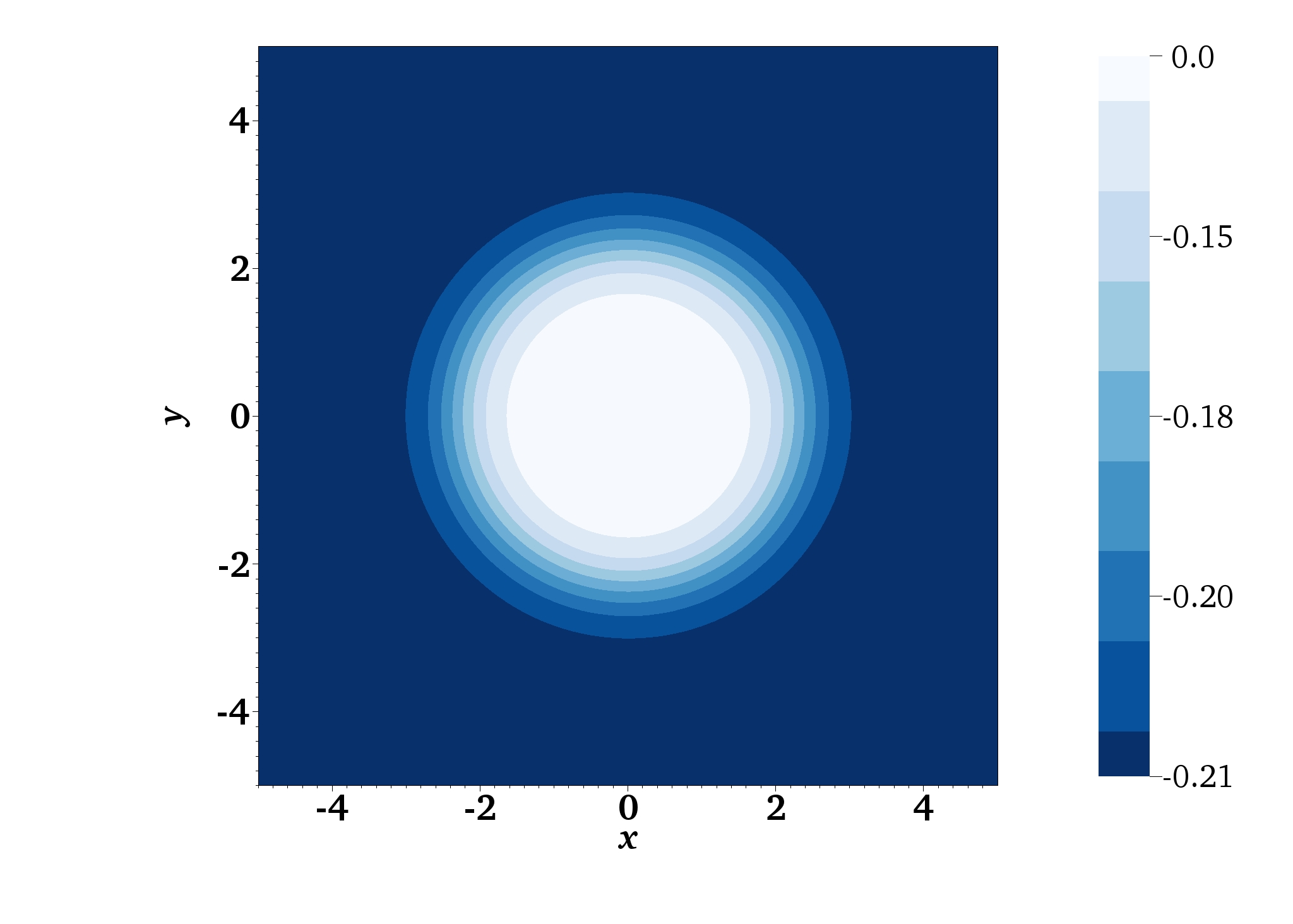}
\includegraphics[width=0.45\textwidth, trim=140 0 10 0, clip=true]{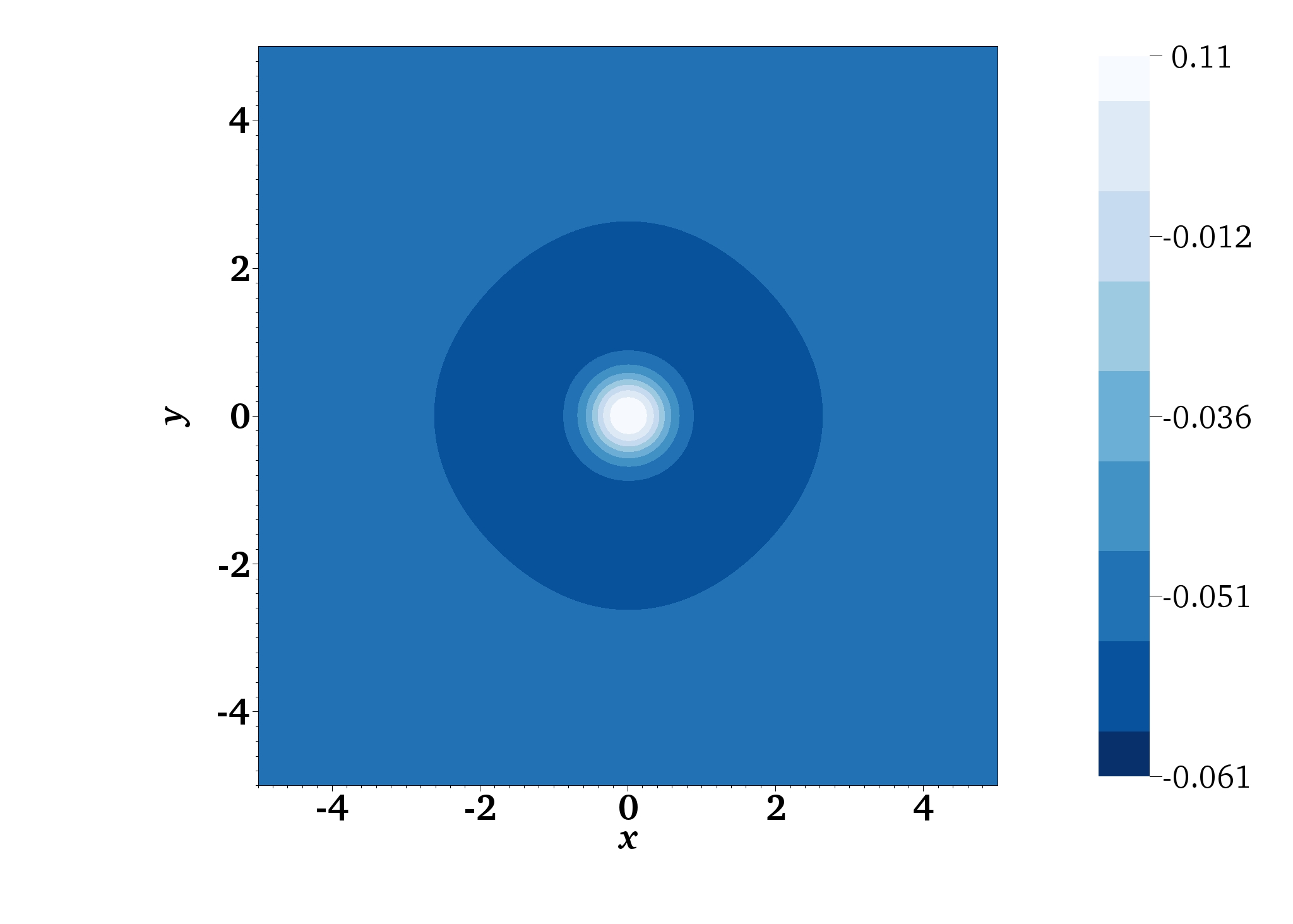}
\includegraphics[width=0.45\textwidth, trim=140 0 10 0, clip=true]{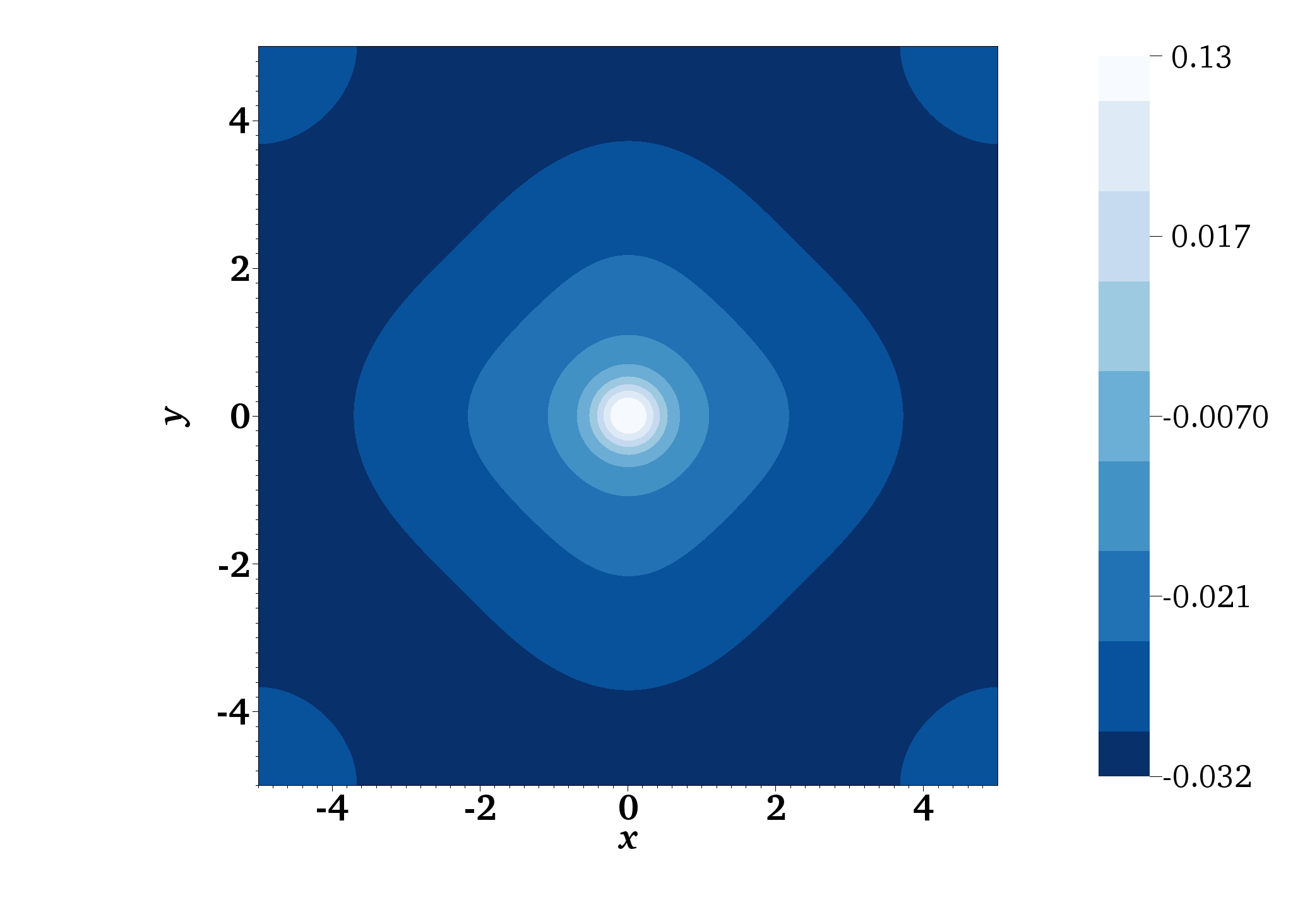}
\includegraphics[width=0.45\textwidth, trim=140 0 10 0, clip=true]{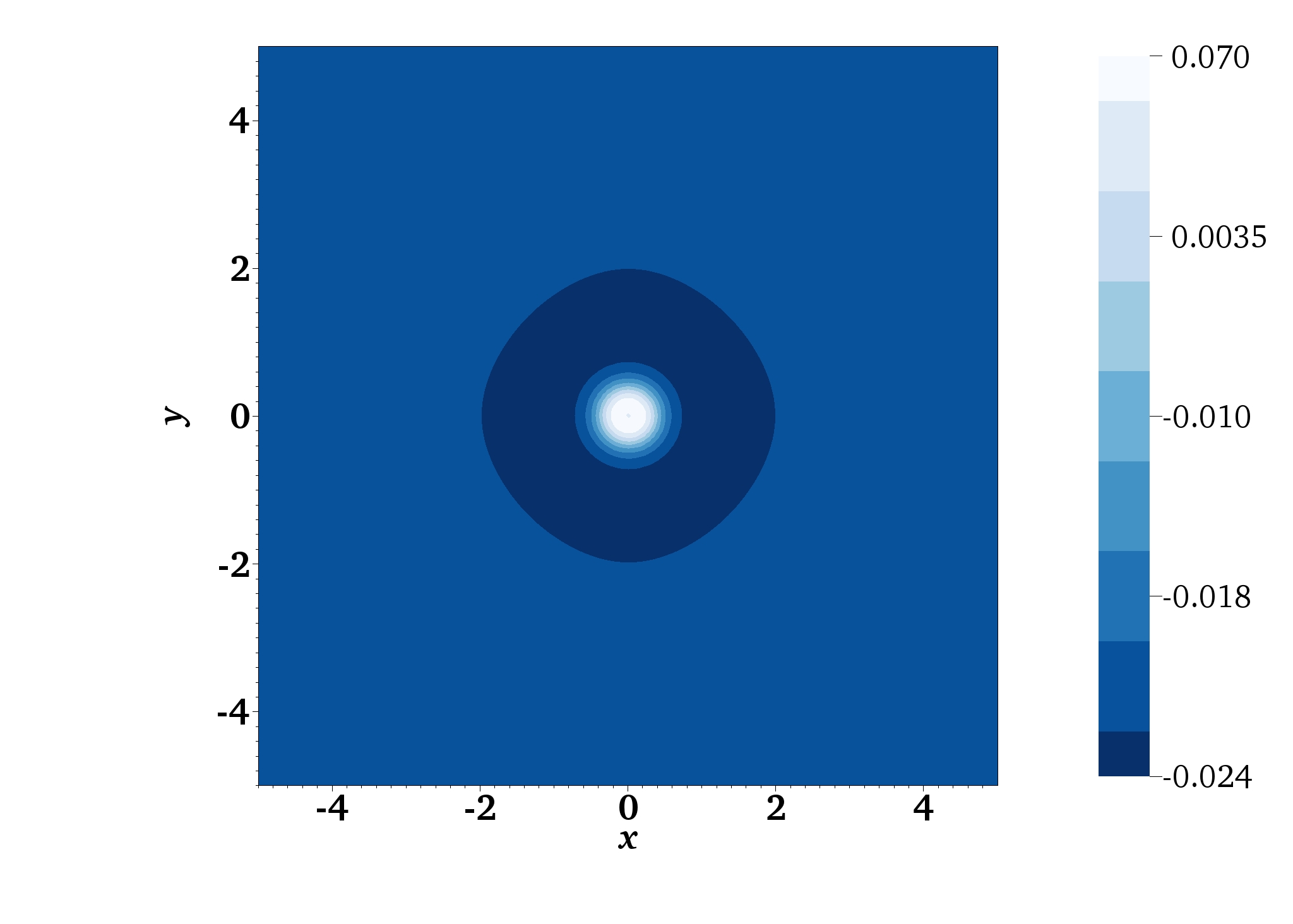}
\includegraphics[width=0.45\textwidth, trim=140 0 10 0, clip=true]{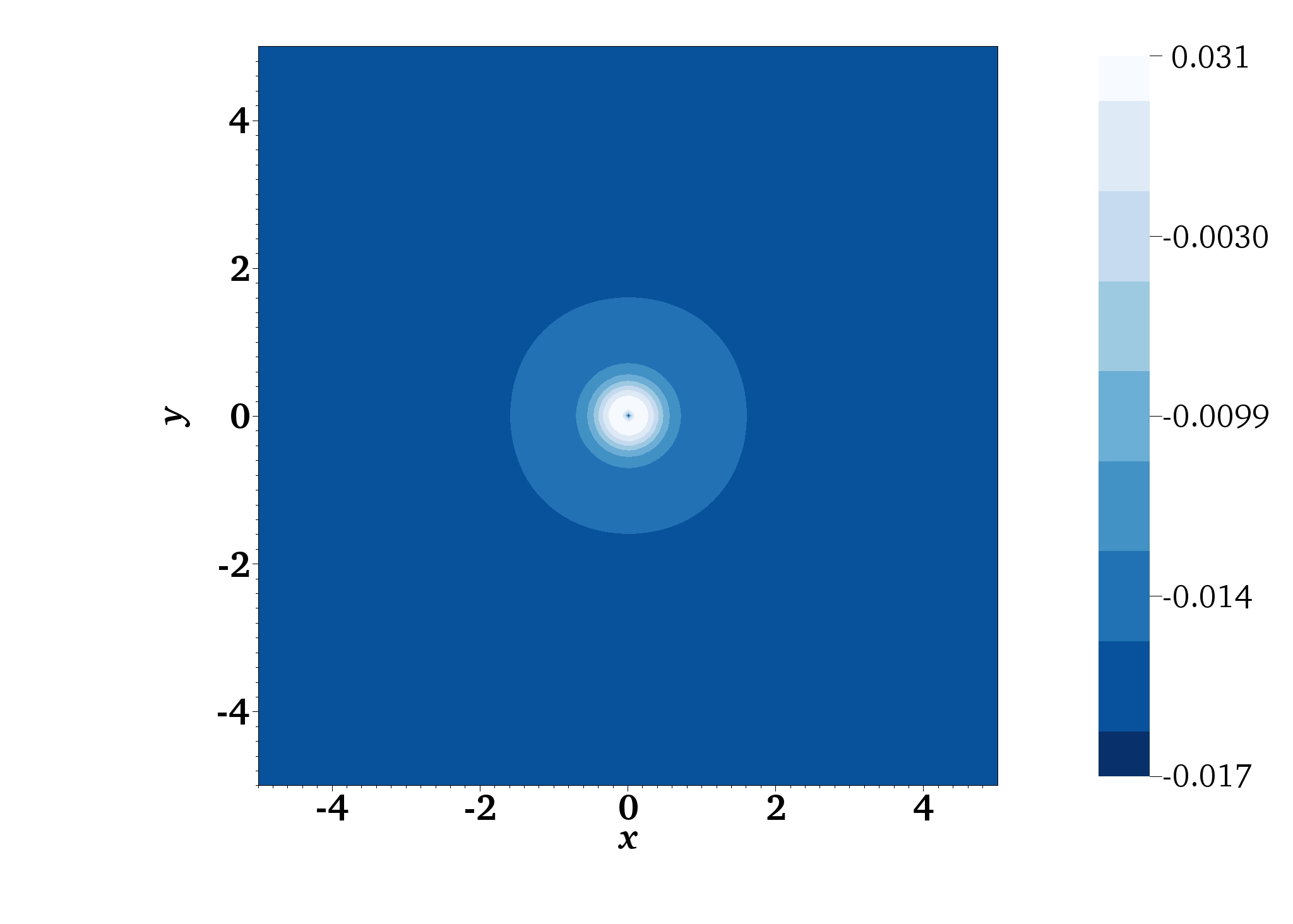}
\includegraphics[width=0.45\textwidth, trim=140 0 10 0, clip=true]{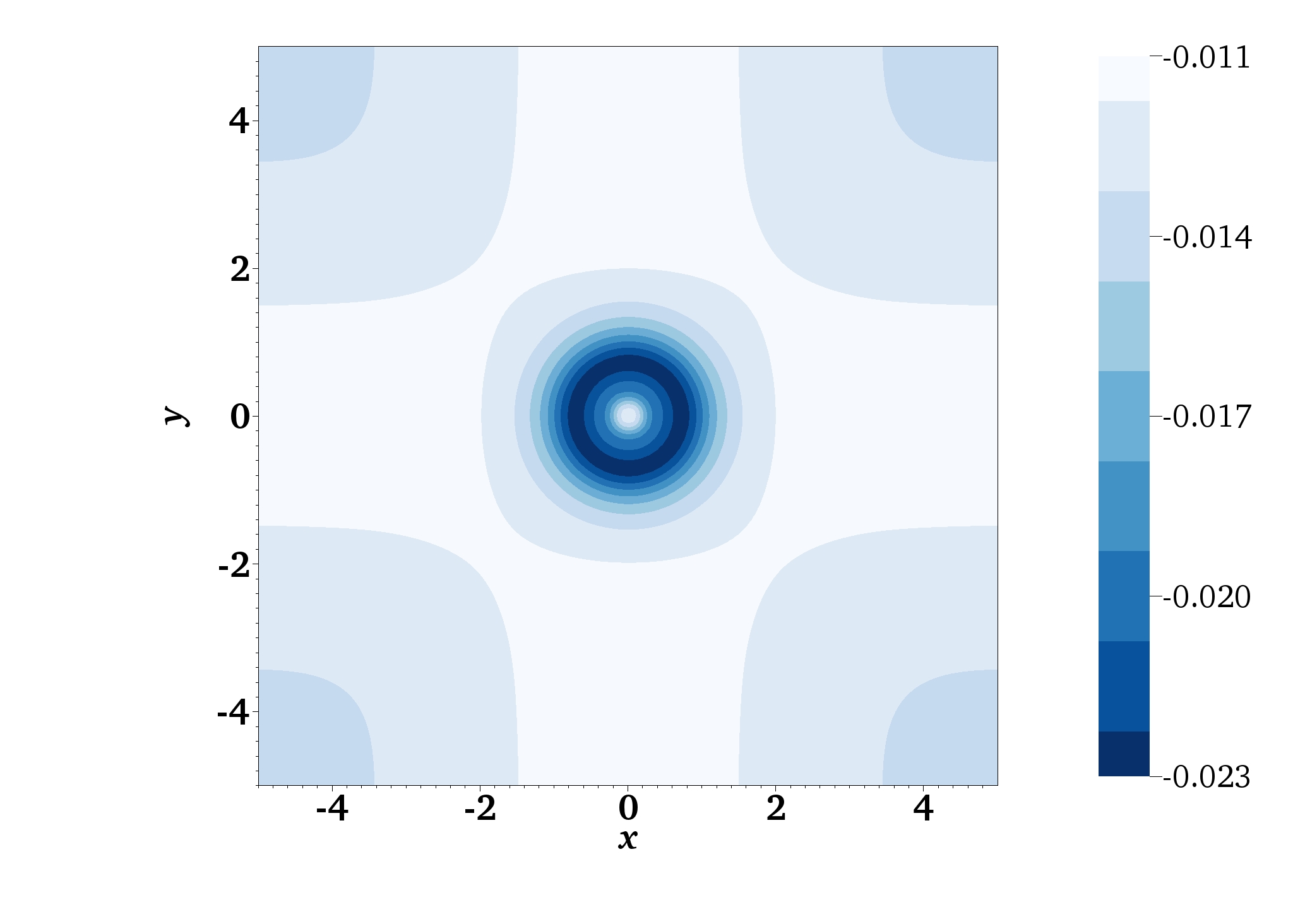}
\caption{Profiles of $K$ on the equatorial plane at $t=0, 30, 60, 90, 120$ and $150$
for the $m=1$, $\Delta_0=1$ run.
\label{fig:K}}
\ece
\efi

\bfi
\bce
\includegraphics[width=0.85\textwidth, trim=140 0 10 0, clip=true]{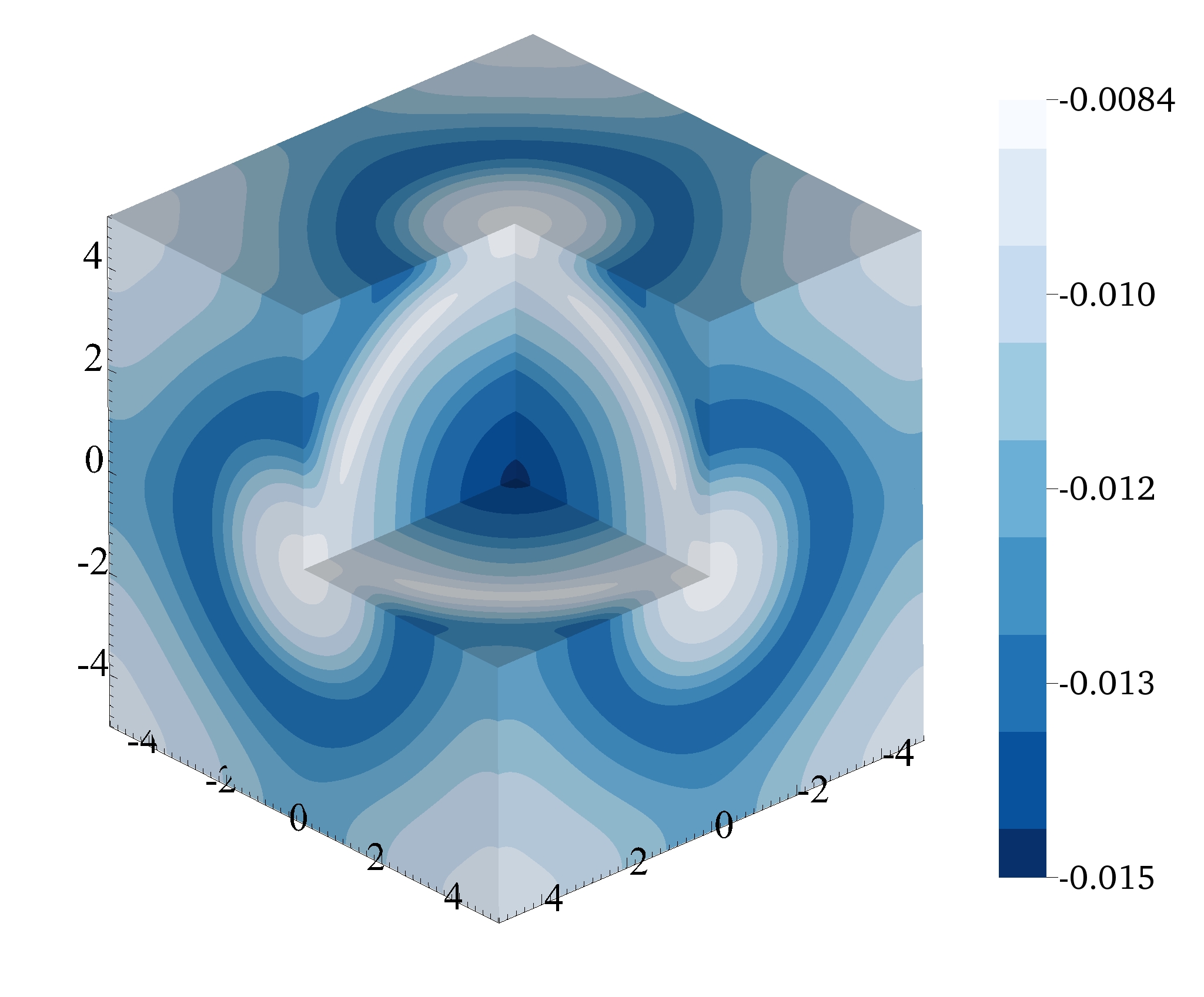}
\caption{The mean curvature $K$ at time $t=180$ for the $m=1$, $\Delta_0=1$ run. The $z$-axis runs vertically.
\label{fig:K3D}}
\ece
\efi

The $L_2$ norm of the Hamiltonian constraint and the $x$-component of the momentum
constraint are shown in Figure~\ref{fig:constr} for L1, L2 and L3, the three 
resolutions of the $m=1$ case.
Two-dimensional, equatorial sections of the Hamiltonian constraint for L1 are shown in 
Figure~\ref{fig:H2D}: this shows that the violations do not significantly
grow in time, remaining, outside the apparent horizon, always lower than $10^{-2}$. 
The constraint violations, however, only converge to zero for $t<70$; the Hamiltonian
constraint converges at fourth order first, and eventually at first order; the 
convergence order of the momentum constraint is much less clear.

\bfi
\bce
\include{figs/ham}
\include{figs/mom}
\caption{$L_2$ norms of the Hamiltonian and momentum constraint for runs L1, L2 and L3.
The constraint violations only converge to zero for $t<70$.
\label{fig:constr}}
\ece
\efi

\bfi
\bce
\includegraphics[width=0.45\textwidth, trim=140 0 10 0, clip=true]{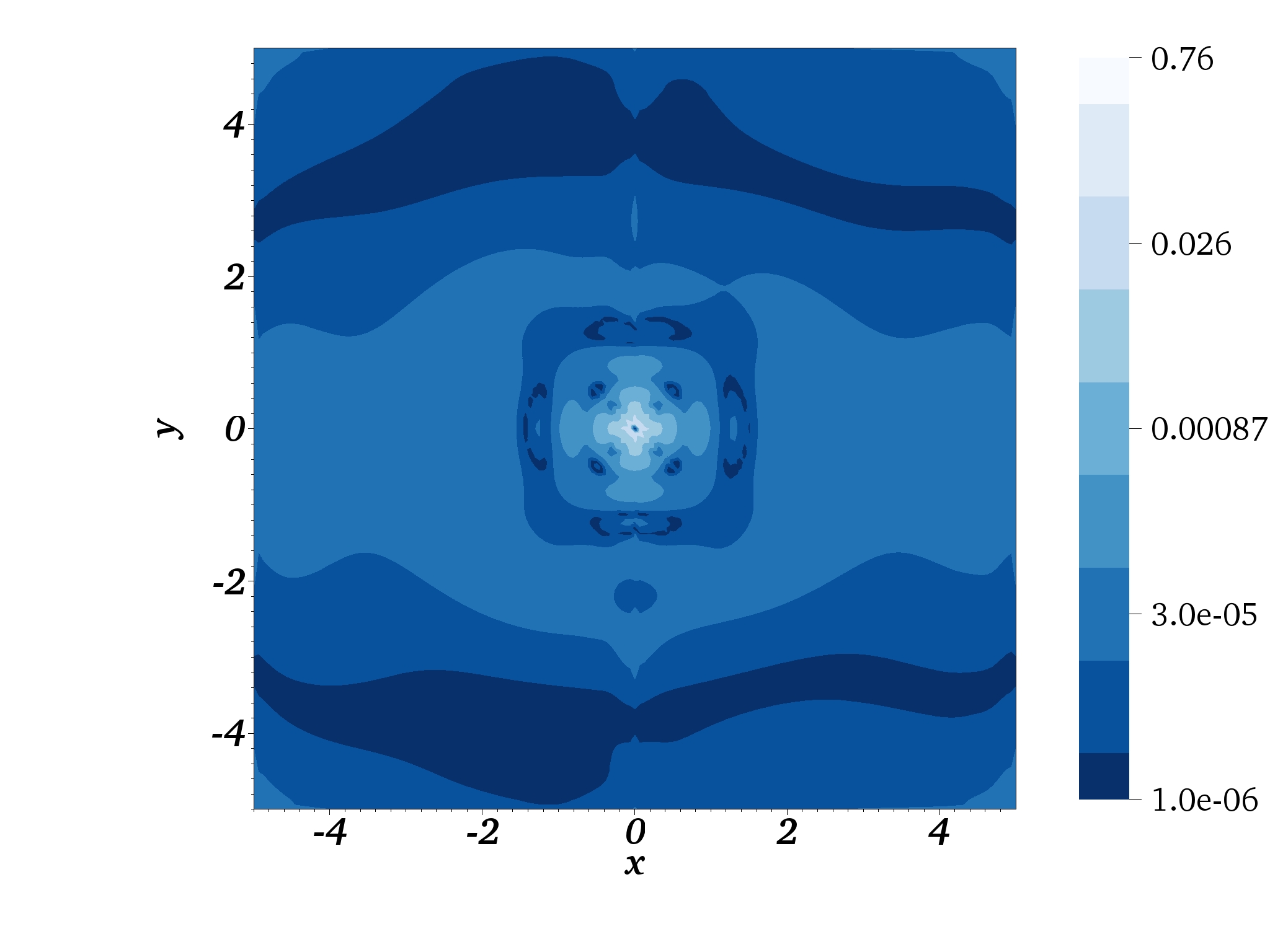}
\includegraphics[width=0.45\textwidth, trim=140 0 10 0, clip=true]{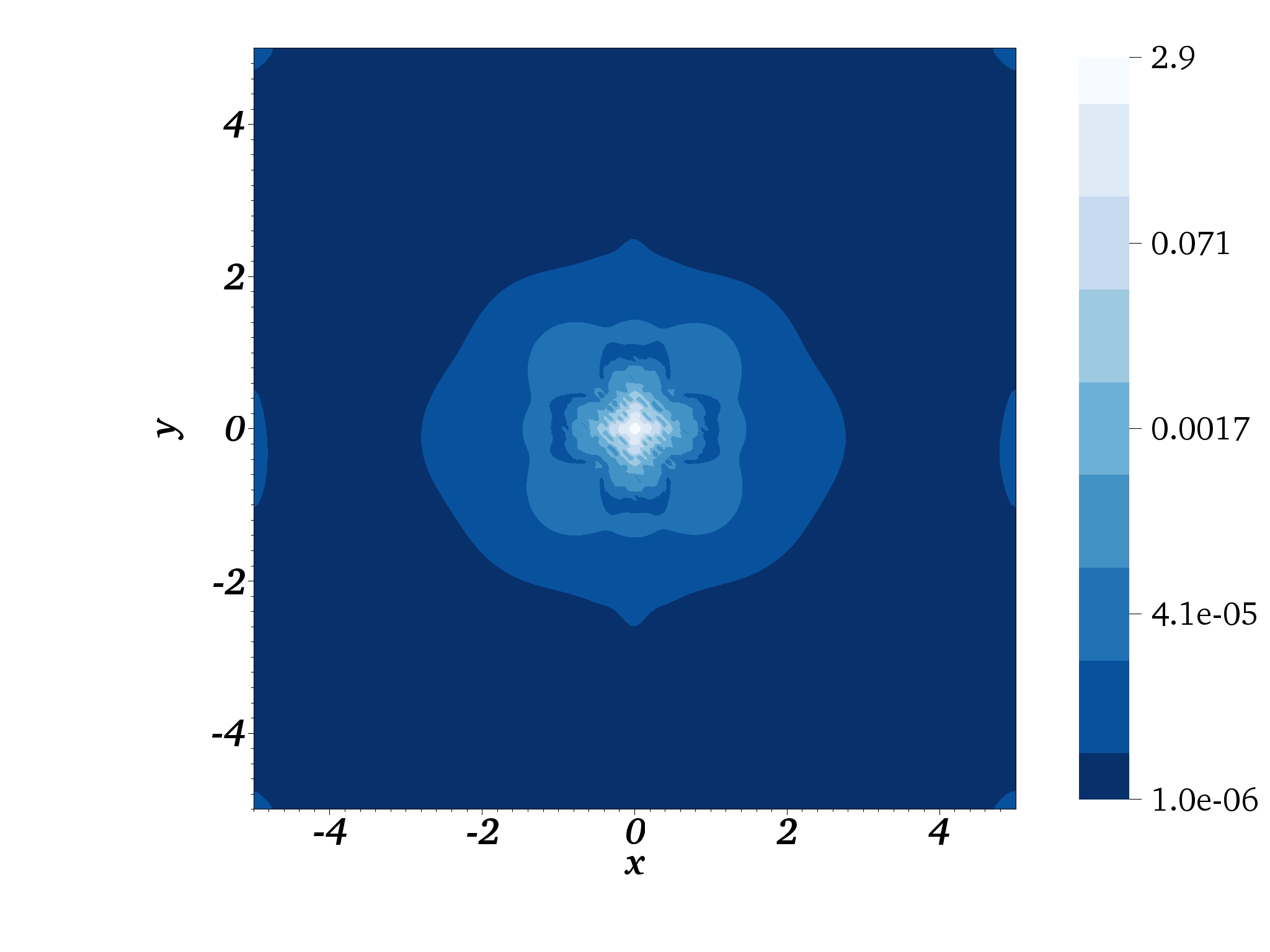}
\includegraphics[width=0.45\textwidth, trim=140 0 10 0, clip=true]{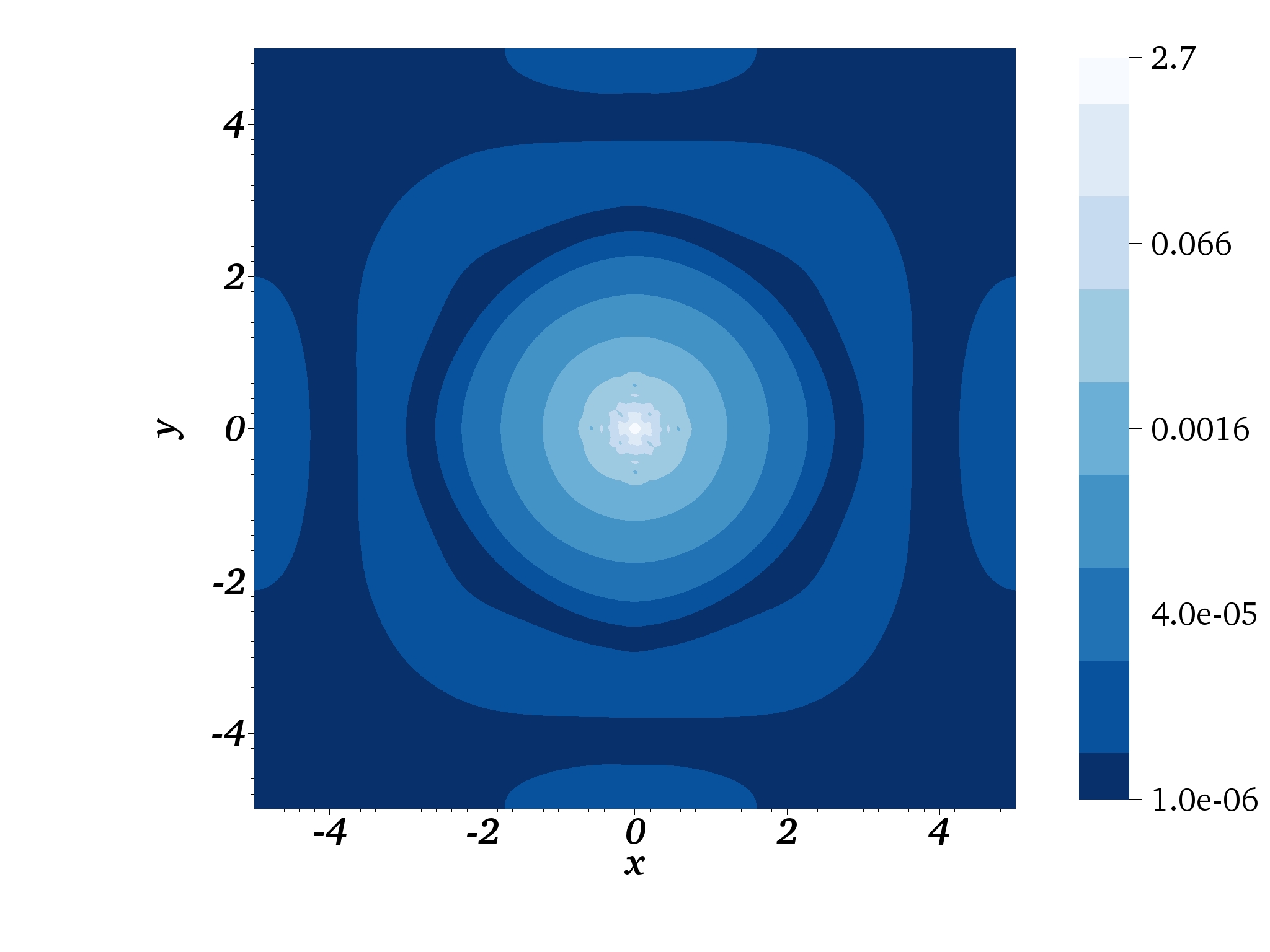}
\includegraphics[width=0.45\textwidth, trim=140 0 10 0, clip=true]{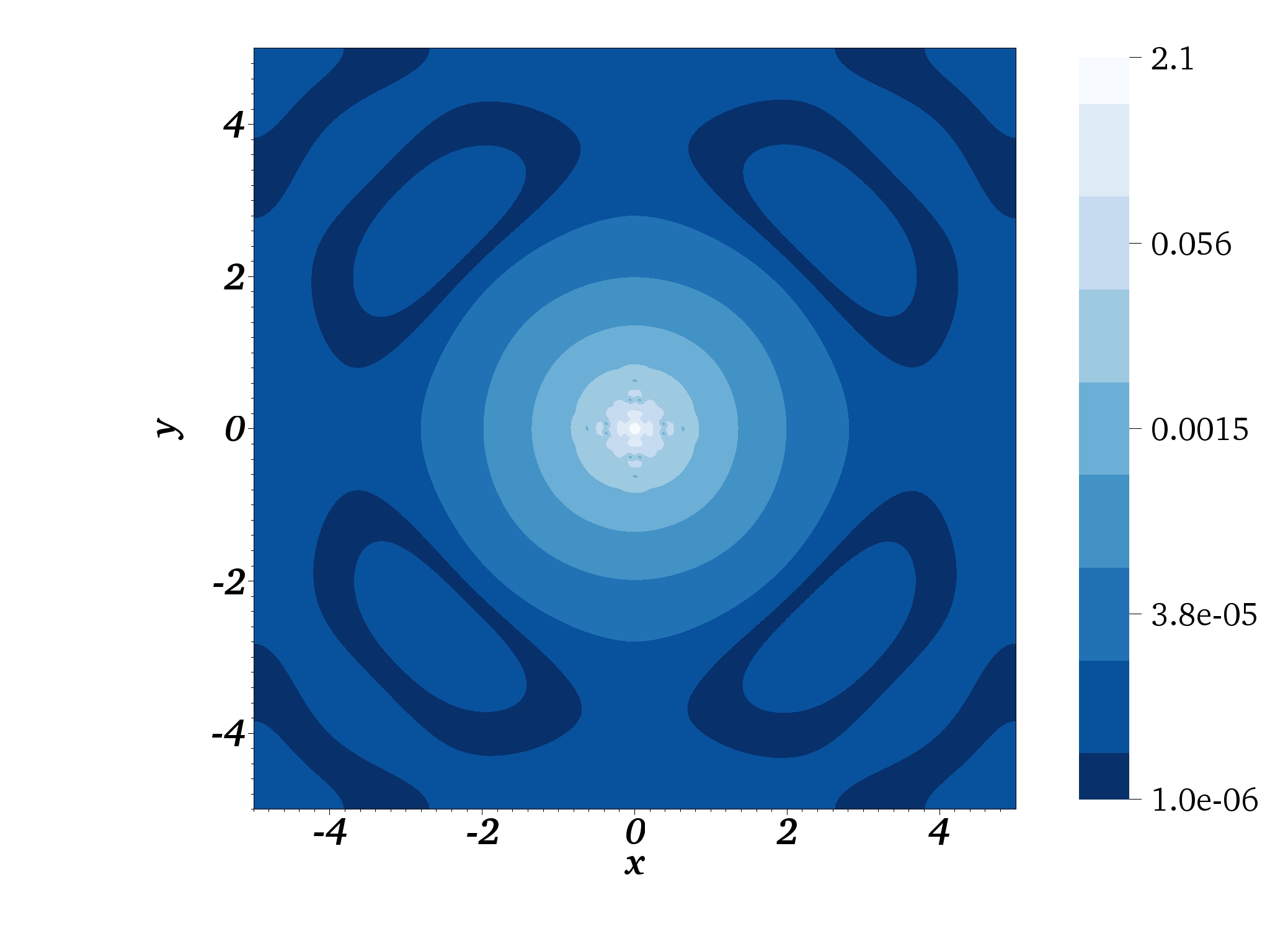}
\includegraphics[width=0.45\textwidth, trim=140 0 10 0, clip=true]{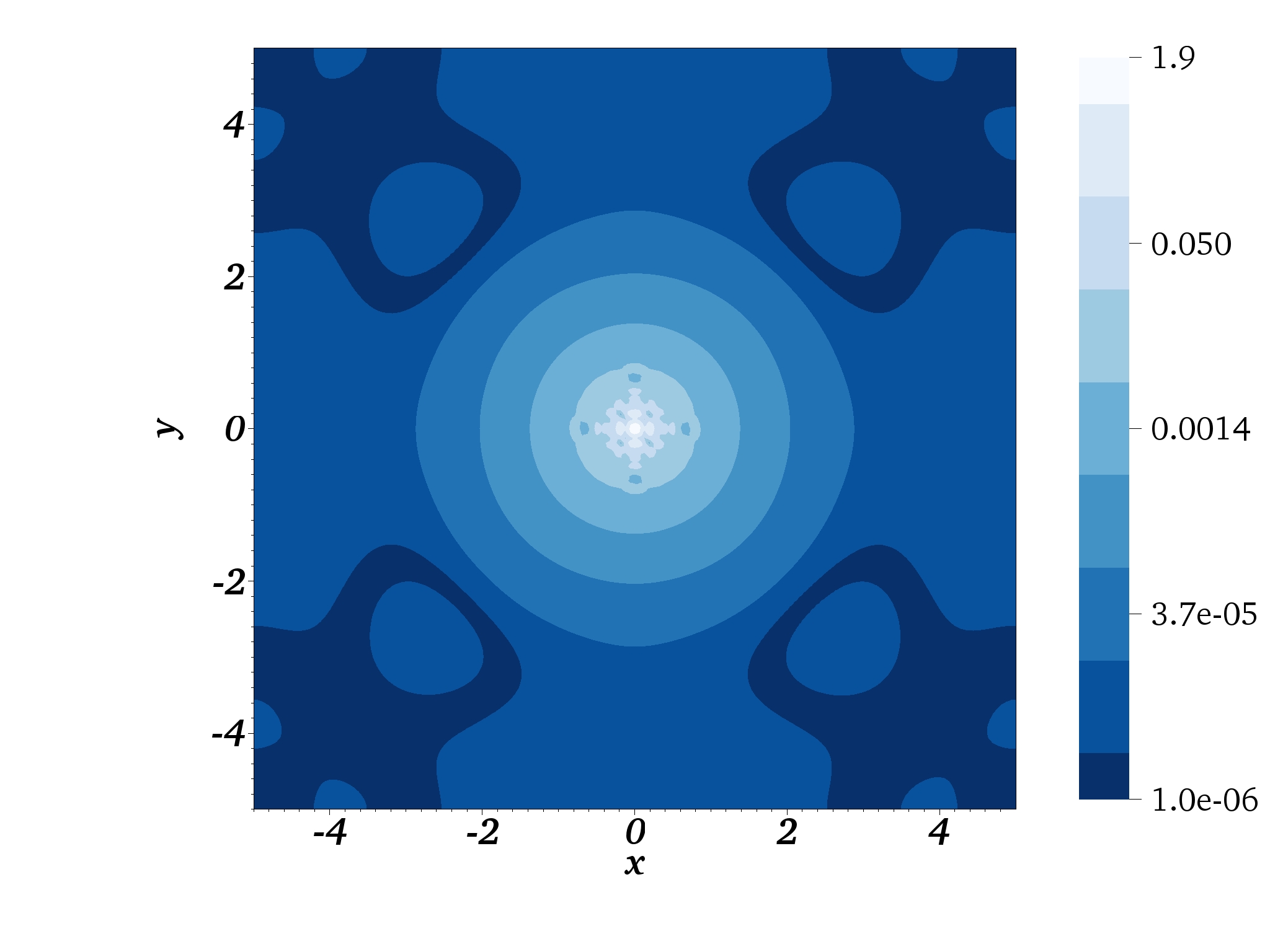}
\includegraphics[width=0.45\textwidth, trim=140 0 10 0, clip=true]{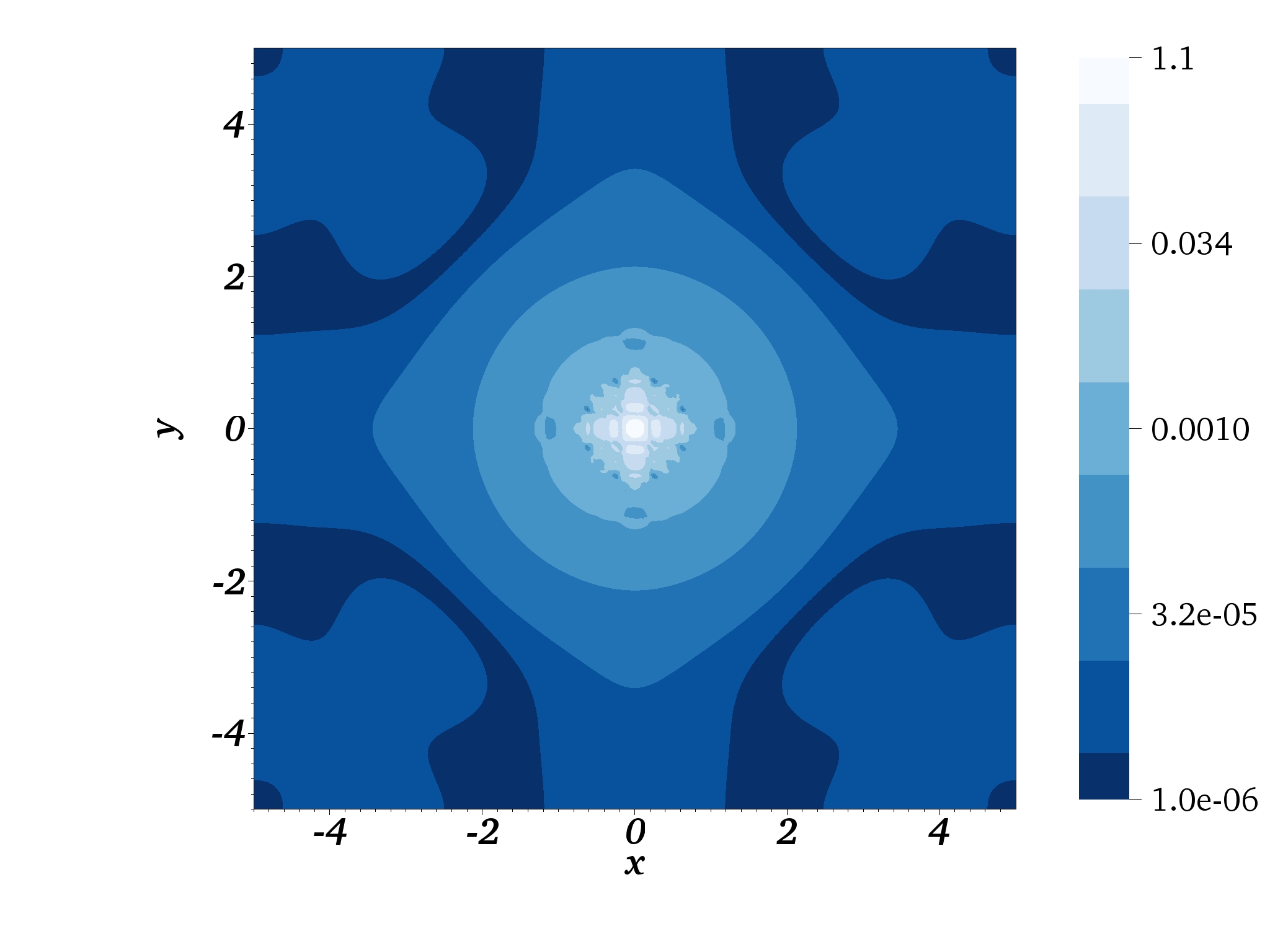}
\caption{Profiles of $|H|$ on the equatorial plane at $t=0, 30, 60, 90, 120$ and $150$
for the $m=1$, $\Delta_0=1$ run.
\label{fig:H2D}}
\ece
\efi

These quantities, like a number of others below, show considerable oscillations
throughout the evolution. This is not surprising as, due to the periodic
boundaries, travelling modes will not be able to disperse at infinity, but will
rather be trapped inside the domain and keep interfering with themselves. If
the result acquires a high-frequency component, due to the large truncation error the
constraints may increase accordingly. This effect would decrease if we increased
the resolution -- as do some of the features in Figure~\ref{fig:constr}.

It is interesting to ask where these modes could originate. An obvious culprit
is the assumption of conformal flatness that underlies our initial-data
construction; a similar connection exists in the conformally-flat initial data
for binary black-hole systems in asymptotically-flat spaces, which are known
to contain spurious gravitational radiation that travels out from the binary 
region right at the beginning of the evolution. Whilst defining gravitational waves in
our setting is not as straightforward, it is conceivable that a similar
mechanism is at play here, and that the assumption of conformal flatness 
sets up the lattice in a state which is mixed with oscillatory modes. 


\section{Comparison with the FLRW class}
\label{sec:comp}

In this section we will compare the results with the FLRW class of solutions 
with dust and evaluate all possible discrepancies, called in this context the 
cosmological backreaction. We focus on three aspects: the scaling of proper
lengths, and the cell's effective mass and pressure which can be defined via
the first and second Friedmann equations respectively.

We will examine these three quantities in the following subsections. For the purpose of this paper we will 
divide the backreaction effects into kinematical, i.e. those that can be read out directly 
from the initial data, and dynamical, which appear as we evolve the data in time. 
It turns out that the kinematical backreaction effects can be interpreted as a ``dressing'' of the bare black hole masses while
the dynamical ones can be thought of as an additional effective-pressure term in the Friedmann equations.

In addition to a comparison with homogeneous cosmologies, the analysis
highlights some of the technical issues arising in the simulation of lattices 
with too high, or too low, a ratio of mass to cell size. As this ratio grows, so does the ratio of 
the apparent horizon radius to the cell linear size $L$; this leaves less and less
room for the transition between the Schwarzschild solution at the center
and the constant-mean-curvature (CMC) region at the edges. Correspondingly, it is harder and harder to 
reduce the constraint violations in the initial data: as illustrated in Table~\ref{table:IDpars},
the $L_2$-norm of the Hamiltonian constraint violation for $m=5$ is almost an order of
magnitude larger than for $m=1$. At the opposite end of the spectrum, smaller values of 
the mass-to-size ratio require correspondingly finer grids in order to resolve the central black hole;
decreasing this ratio at constant resolution leads to a length scaling that tends to
a linear law -- the length scaling of empty space in a CMC foliation.

\subsection{Proper length of a cell edge}

A comparison with the homogeneous Friedmann-Lema\^itre-Robertson-Walker class with dust requires 
solving Ellis's fitting problem, i.e. finding the FLRW model which resembles each black-hole lattice most closely. 

In this paper we will follow the approach from
\cite{Bentivegna:2012ei}, also similar in spirit to  \cite{Clifton:2012qh}.
Consider the edges of lattice cells at $t=0$: obviously they constitute identical geodesic segments, each meeting five others 
at right angles at each vertex. While the lattice faces are not isometric to flat squares, the edges still constitute a regular lattice which is entirely isometric to a cubic lattice of edge length $D_{\rm edge}(0)$ in flat space. Due to the discrete symmetry of the initial data this property carries over to all times provided that the shift function preserves all the symmetries. We will assume for the purpose of the FLRW fitting that the gauge is such that $\alpha=1$ and $\beta^i = 0$. This requires of course a numerical refoliation of at least a part of the spacetime.

Clearly, an arrangement of cells with edges isometric to a cubic lattice can only be found in a flat FLRW model. Thus we will take a flat FLRW as our reference model: we fix a flat, three--dimensional space with metric $\delta_{ij}$ containing a cubic lattice of length $D_{\rm edge}(0)$.  In order to single out exactly one flat dust FLRW model, we also need 
to set the Hubble parameter at $t=0$, which we fix to $\dot D_{\rm edge}(0)/D_{\rm edge}(0)$. We label these models with FLRW($m$).

As in~\cite{Bentivegna:2012ei}, given the spatial metric, the lapse and the shift on an edge, we can
calculate the expansion of its proper length as a function of proper time. 
The result is shown in Figure~\ref{fig:pd} for the
four models considered here, for the edge at $x=5$ (the others differ for less than $10^{-3}$ for $m=1$)
and with error bars deriving from the error analysis (see~\ref{appendix:A}).
In the same plot, we show the length scaling of the corresponding FLRW universes defined above, through the fitting
of the initial edge length and its first derivative.
The length of the black-hole lattice edges remains close to that of the FLRW counterpart
at all times, except for the $m=5$ case, where the value of $\dot D_{\rm edge}(0)$ does
not seem to single out a well-fitting FLRW model; we attribute this to the fact that,
for $m=5$, $D_{\rm edge}(t)$ presents high-amplitude oscillations on top of the leading-order length expansion, 
so that the value of its tangent at a single point may be rather different from the 
``average'' expansion rate. Choosing a different FLRW counterpart, such as the one with initial density equal
to $m_{\rm BH}/D^3_{\rm edge}(0)$ (labelled FLRWb(5)), improves the fit. 

\bfi
\bce
\include{figs/pd}
\caption{Scaling of the edge proper length as a function of proper time, for $m={0.5, 1, 2, 5}$.
\label{fig:pd}}
\ece
\efi




As we were preparing this manuscript, a paper appeared~\cite{Yoo:2013yea} in which the authors address the
same issue of proper-length scaling in an expanding black-hole lattice. There, however, the FLRW
counterpart is chosen by a parameter fit of the evolution data. 
Given the evidence in this section, it turns out that this is unnecessary: choosing the mapping based on the kinematical properties of
the initial data alone automatically leads to a well-fitting expansion at subsequent times.

\subsection{Effective mass}
\label{sec:Meff}
We begin by introducing a dimensionless inhomogeneity parameter:
\beq
\mu = \frac{m_{\rm BH}}{D_{\rm edge}(0)}
\eeq
measuring the lumpiness of a given mass distribution. Here $m_{\rm BH}$ denotes the ADM mass of the black hole evaluated at the asymptoticaly flat region near $r=0$. This mass is related to the bare mass parameter by $m_{\rm BH} = m\, \psi_r(0)$. This is easy to see if we transform to spherical coordinates $(r,\theta,\varphi)$ and then change the radial coordinate to $\hat r = \frac{m^2}{4r}$. In the new coordinates the asymptoticaly flat region of the black hole corresponds to $\hat r \to \infty$ and the metric takes again a conformally-flat form with the conformal factor $\hat \psi = 1 + \frac{m\psi_r}{2\hat r}$. The ADM mass can now be read out from the asymptotic behavior of $\hat \psi$ at infinity. 

 The physical interpretation of $\mu$ as the inhomogeneity parameter is easy to understand if we note that, for a fixed energy density $\rho=m_{\rm BH}/{D^3_{\rm edge}}$, we have 
$\mu = \rho D_{\rm edge}^2$ and the parameter decreases as the size of the cell goes to zero. 
We expect that for a fixed $\rho$ and $\mu$ going to zero the properties of the lumpy model will match those of the homogeneous model. In Figure \ref{fig:mu} we plot  $\mu$ as a function of the ADM mass of the black hole for four values of the bare mass $m=0.5, 1, 2, 5$, and with $K_c = -0.21$. The dependence is not exactly linear because the physical size of the cell $D_{\rm edge}$ depends  on the conformal factor $\psi$, which in turn depends on the ADM mass in a non-trivial way.
\bfi
\bce
\include{figs/mu}
\caption{$\mu$ parameter as a function of the ADM mass of the black hole, for $m={0.5, 1, 2, 5}$.
\label{fig:mu}}
\ece
\efi

Note that the procedure outlined above for fitting an FLRW model to the initial data can be repeated at any other time $t$ by an appropriate rescaling of the metric of the FLRW counterpart,
such that the proper length of the edge in the FLRW model becomes equal to $D_{\rm edge}(t)$. In this way we obtain a family of ``tangent'' flat FLRW models parametrized by $t$.
Accordingly, we can define a time--dependent effective Hubble parameter to be $\eff{H}(t)=\dot D_{\rm edge}(t)/D_{\rm edge}(t)$ at any time $t$. We take the effective
3--dimensional Ricci scalar $\eff{R}$ of our solution to be equal to the Ricci scalar of the corresponding FLRW, which is of course zero at all times. 
The first Friedmann equation 
\beq
\eff{H}^2 + \frac{\eff{R}}{6} = \frac{8\pi}{3} \,\eff{\rho}  \label{eq:efffried1}
\eeq
will now serve as the definition of the effective energy density $\eff{\rho}$. In our case it simply reads
\beq
\eff{\rho}(t) = \frac{3\dot D_{\rm edge}^2(t)}{8\pi \,D_{\rm edge}^2(t)}.
\eeq 
We may now compare the effective mass enclosed in a single cell, given by $\eff{M} = \eff{\rho}\,D_{\rm edge}(t)^3$, with the mass of the black hole as measured is its asymptotically flat end at time $t=0$. 
This gives the overall mass dressing effect due to the self--interaction of the strong gravitational field. It manifests itself
as a difference between the ADM mass, intrinsic to the black hole, and the effective mass contained in a  cell as it is ``felt'' by the Friedmann equations.

\bfi
\bce
\include{figs/x}
\caption{Relative mass deficit $x = \frac{M_{\rm eff} - m_{\rm BH}}{M_{\rm eff}}$  as a function of $\mu$, for $m={0.5, 1, 2, 5}$.
\label{fig:x}}
\ece
\efi

\bfi
\bce
\include{figs/meff}
\caption{Effective mass $M_{\rm eff}$ as a function of proper time, for $m={0.5, 1, 2, 5}$.
\label{fig:meff}}
\ece
\efi

In Figure \ref{fig:x}, we have plotted the relative mass deficiency parameter $x$, defined as
\bea 
x = \frac{M_{\rm eff} - m_{\rm BH}}{M_{\rm eff}}, \label{eq:defx}
\eea
against the values of $\mu$ for the four cases we consider.
We have found out that the effective mass  is smaller than the ADM mass, unlike the spherical case we investigated in \cite{Bentivegna:2012ei}. Rather than 
the "gravitating gravity" effect, in which the strong gravitational field contributes positively to the effective mass, we see an effective shielding of the black-hole mass whose magnitude grows nonlinearly with $\mu$.

During the evolution, $M_{\rm eff}$ oscillates, sometimes reaching values larger than
the ADM mass, as shown in Figure~\ref{fig:meff}. This effect is, for the proper time $\tau<70$, largely
independent from the resolution, as indicated by the error bars in Figure~\ref{fig:Mcont}.

Identifying the origin of this oscillations is not straightforward. Here we will only remark
that the initial part is dominated by a single frequency in the range $(0.3-0.5) m_{\rm BH}^{-1}$,
i.e.~the range of frequencies of the lowest quasi--normal modes 
of a Schwarzschild black hole.


\subsection{Effective pressure}
\label{sec:peff}
At any  time $t$,
 the reference FLRW spatial slice will have a rescaled metric $D_{\rm edge}(t)^2/D_{\rm edge}(0)^2\,\delta_{ij}$, so that the  size of the lattice edge matches exactly the size of the lattice cell of the inhomogeneous model at $t$. 

The second Friedmann equation for the effective parameters, i.e. the evolution equation
\beq
\frac{\ddot D_{\rm edge}}{D_{\rm edge}} = - \frac{4\pi}{3}\left(\eff{\rho} + 3\eff{p}\right) \label{eq:efffried2}
\eeq
can be used to define an effective pressure $p_{\rm eff}$. After the substitution of (\ref{eq:efffried1}) and setting $\eff{R}$ to zero, it reads
\beq
\eff{p} = -\frac{1}{4\pi}\frac{\ddot D_{\rm edge}(t)}{D_{\rm edge}(t)} -\frac{1}{8\pi}\frac{\dot D_{\rm edge}(t)^2}{D_{\rm edge}(t)^2}.
\eeq
If no backreaction effects were present, the pressure would naturally be zero at all times since the black holes do not interact with each other
by any means other than gravitational attraction. In Figure \ref{fig:peffcmf} we see the effective pressure plotted for $m=1$ and
three resolutions in runs L1, L2 and L3. The pressure starts out at zero and oscillates with relatively small amplitude (around $10^{-5}$),
and eventually gets damped out to zero. The behavior of $p_{\rm eff}$ is similar for $m=2$, whilst for $m=5$ one observes an initial
non-zero effective pressure, again probably due to the same effect discussed in Figure~\ref{fig:pd}.
For the largest mass $m=5$, the initial amplitude of the oscillations is around $10^{-4}$.

Since both effective pressure and density appear in equation (\ref{eq:efffried2}), 
it is reasonable to compare their influence on the dynamics of the model. It turns out that the
amplitude of the oscillations of $p_{\rm eff}$ is somewhat smaller in magnitude than 
the value of $\rho_{\rm eff}$ for  $m=1$ and $m=2$ (the ratio is around 
0.3 and 0.15 respectively), although their ratio is close to one in the $m=5$ case. 
Overall, the evolution of $D_{\rm edge}$, given by (\ref{eq:efffried2}),
is dominated  by $\rho_{\rm eff}$ in all cases, because even though the 
pressure term is comparable with the density term, it is at the same time 
oscillating rapidly around zero and its influence averages out to zero.

\bfi
\bce
\include{figs/peffcmf}
\caption{Top: effective pressure for runs L1, L2 and L3 ($m=1$, $\Delta_0=1, 5/6$ and $5/8$ respectively).
Bottom: effective pressure for runs L1, L5 and L6 ($\Delta_0=1$, $m=1, 2$, and $5$ respectively). 
\label{fig:peffcmf}}
\ece
\efi

\section{Conclusions}
\label{sec:concl}
We have presented the evolution of a family of cubic black-hole lattices, starting from 
a conformal transverse-traceless set of initial data proposed by~\cite{Yoo:2012jz}
and reproduced by us via a multigrid elliptic solver.

After a description of some of the 3+1 quantities during the evolution, we constructed
three observables to use as a ground of comparison between these solutions and some
suitably-chosen FLRW counterpart: the proper length of the periodic cell's edge and its
scaling according to geodesic observers, and an effective mass and pressure defined
through the first and second Friedmann equations.

We find that, while the scaling remains close to the dust-FLRW law
for the evolutions considered here, a number of important differences emerges:
first, the study of the $m=1$ case shows that the proper length, extrapolated to
infinite numerical resolution, differs from the FLRW law by up to $1.5\%$ for $t<70$;
this is over an order of magnitude higher than the numerical error bars. 
After this time, the numerical convergence becomes less clear and it is impossible
to evaluate the agreement.

Second, taking time derivatives of this observable brings
out the presence of oscillatory modes superimposed on the zero-frequency behavior.
Using an analogy to the similar occurrence in conformal-transverse-traceless 
initial data for multiple black holes in asymptotically-flat spaces, we conjecture 
that these vibrations may be due to the somewhat simplified initial data 
construction. Unlike the asymptotically-flat case, however, these modes cannot
diffuse to infinity, and remain present at all times.

The main backreaction effect we observe is the mass dressing, i.e. the difference between the
``bare'' ADM mass of the central black hole and the effective mass of the individual cell. In the initial data, the effective mass
is smaller than the ADM mass in all cases, the effect growing with the inhomogeneity parameter in a nonlinear manner. 
Furthermore, the effective mass turns out to be time--dependent; its value  oscillates during the evolution around values somewhat smaller than the ADM mass.
While most of the time it remains smaller than the ADM mass, it
can actually get larger for a short time. 
Analyzing its power spectrum, we find peaks in 
the band of the quasinormal frequencies of a Schwarzschild black hole of 
comparable mass.
Finally, the effective pressure, whose non--vanishing values are the main dynamical backreaction effect in these models, undergoes similar 
damped oscillations around the zero value with relatively small amplitude.

We conclude that this set of initial data has a leading-order behavior that 
remains close to the FLRW model of matching initial energy density, but
presents noticeable higher-order oscillations which affect the evolution.
Whether it would be feasible to prepare an expanding black-hole lattice in a 
``purer'' initial state remains a question for future studies.

\section*{Acknowledgements}
The authors would like to thank Lars Andersson and Jerzy Kijowski for valuable discussions. M.K.~is supported by the project \emph{
``The role of small-scale inhomogeneities in general relativity and cosmology''} (HOMING PLUS/2012-5/4), realized within the Homing Plus programme of Foundation for Polish Science, cofinanced by the European Union from the Regional Development Fund. E.B.~is supported by a Marie Curie International 
Reintegration Grant (PIRG05-GA-2009-249290). Computations were carried out on the MPI-GP Damiana and Datura 
clusters, as well as on SuperMUC at the Leibniz-Rechenzentrum in Munich.

\appendix
\section{Convergence and numerical errors}
\label{appendix:A}

We performed a three-point convergence test using the proper edge length
of run L1, L2 and L3. The convergence factor
\beq
c \equiv \frac{|D^{(\\\rm c)}-D^{(\\\rm m)|}}{|D^{(\\\rm m)}-D^{(\\\rm f)}|}
\eeq
remains close to the first-order value until $t\approx 50$, as illustrated
in Figure~\ref{fig:cfact}.
\bfi
\bce
\include{figs/cfact}
\caption{Convergence factor from the three-point convergence test.
\label{fig:cfact}}
\ece
\efi
Using this, one can extrapolate to the continuum limit, $\Delta\to 0$,
obtaining the curve in Figure~\ref{fig:cont}, with error bars given
by the difference between the continuum limit and the coarsest 
resolution. 
In Figure~\ref{fig:dvsFLRW}, the difference between
the edge length in L1, L2, L3, and the continuum solution and the
scale factor of a FLRW universe with the same initial expansion rate is
compared. 

Similarly, in Figures~\ref{fig:Mcont} and~\ref{fig:pcont}, the continuum limit of the effective
mass and effective pressure for $m=1$ is plotted, along with error bars given by the difference
between the continuum value and L1.
\bfi
\bce
\include{figs/cont}
\caption{Richardson extrapolation of the edge length in the limit 
of zero grid spacing, versus the FLRW scale factor for the same
initial density. The error bars are given by the difference between the
the Richardson-extrapolated value and L1.
\label{fig:cont}}
\ece
\efi
\bfi
\bce
\include{figs/dvsFLRW}
\caption{Difference between the proper length of an edge (in low, medium and high resolution, and in the
Richardson-extrapolated value) and the FLRW scale factor.
\label{fig:dvsFLRW}}
\ece
\efi
\bfi
\bce
\include{figs/Mcont}
\caption{Richardson extrapolation of the effective mass in the limit
of zero grid spacing, for $m=1$.
The error bars are given by the difference between the
the Richardson-extrapolated value and L1.
\label{fig:Mcont}}
\ece
\efi
\bfi
\bce
\include{figs/pcont}
\caption{Richardson extrapolation of the effective pressure in the limit
of zero grid spacing, for $m=1$.
The error bars are given by the difference between the
the Richardson-extrapolated value and L1.
\label{fig:pcont}}
\ece
\efi

Other values of $m$ lead to qualitatively similar results, with lower
masses losing convergence earlier than higher ones (this is reflected 
in the intervals chosen for Figure~\ref{fig:meff}).

\section{Gauge conditions}
\label{appendix:B}
In the numerical evolution of black holes without excision, the moving-puncture gauge,
based on a second-order $\beta$-driver for the shift and the ``1+log'' evolution equation 
for the lapse~\cite{Alcubierre:2002kk}, is the standard.
The lapse prescription, however, does not work when the mean curvature $K$ can assume
negative values, as the lapse would grow exponentially in these regions:
\bea
 \partial_t \alpha - \beta^i \partial_i \alpha = -2K\alpha. \label{eq:gaugealpha1}
\eea
We searched for a modification of (\ref{eq:gaugealpha1}) that avoids the exponential runaway. 
In principle, one could cap the value of $\alpha$ at, say, one, resetting it to this value 
whenever it becomes larger. Unfortunately this modification quickly produces a sharp discontinuity
in $K$, which leads to a shell of high Hamiltonian-constraint violation. This propagates then outwards,
eventually affecting the entire domain.

Another potential strategy is a polynomial modification of the type:
\bea
\partial_t \alpha - \beta^i \partial_i \alpha = -2K\alpha\left(1 - \left(\frac{\alpha}{\alpha_0}\right)^n\right). \label{eq:gaugealpha2}
\eea
In principle  this could provide a smoother mechanism to curb the growth of $\alpha$ in the outer regions, 
since the modified right-hand-side has two attractive fixed points at $\alpha_0$ and $-\alpha_0$ 
as long as $K$ is negative (at the same time, if $n$ is large enough, the modification does not affect
the behavior of the gauge near the puncture where $\alpha$ quickly becomes small). Unfortunately 
this gauge leads to development of a different kind of runaway behavior near the 
vertices of the cell, where at late times $K$ changes sign unexpectedly.

It turned out that the most successful strategy is simply to locate the maximum of $\alpha$ at each time step and normalize the lapse according to
\bea
\alpha \to \frac{\alpha}{\alpha_{\rm max}}.
\eea
This gauge leads to a well-behaved evolution,
its only drawback being the computational cost of the additional 
reduction operation necessary to find the maximum value of the lapse 
at each iteration.
As initial data, we set $\alpha = \psi^{-2}$ and $\beta^i=0$.

\section*{References}
\bibliographystyle{iopart-num}
\bibliography{references}

\end{document}

%% file: figs/ham.tex
\begingroup
  \makeatletter
  \providecommand\color[2][]{%
    \GenericError{(gnuplot) \space\space\space\@spaces}{%
      Package color not loaded in conjunction with
      terminal option `colourtext'%
    }{See the gnuplot documentation for explanation.%
    }{Either use 'blacktext' in gnuplot or load the package
      color.sty in LaTeX.}%
    \renewcommand\color[2][]{}%
  }%
  \providecommand\includegraphics[2][]{%
    \GenericError{(gnuplot) \space\space\space\@spaces}{%
      Package graphicx or graphics not loaded%
    }{See the gnuplot documentation for explanation.%
    }{The gnuplot epslatex terminal needs graphicx.sty or graphics.sty.}%
    \renewcommand\includegraphics[2][]{}%
  }%
  \providecommand\rotatebox[2]{#2}%
  \@ifundefined{ifGPcolor}{%
    \newif\ifGPcolor
    \GPcolortrue
  }{}%
  \@ifundefined{ifGPblacktext}{%
    \newif\ifGPblacktext
    \GPblacktexttrue
  }{}%
  \let\gplgaddtomacro\g@addto@macro
  \gdef\gplbacktext{}%
  \gdef\gplfronttext{}%
  \makeatother
  \ifGPblacktext
    \def\colorrgb#1{}%
    \def\colorgray#1{}%
  \else
    \ifGPcolor
      \def\colorrgb#1{\color[rgb]{#1}}%
      \def\colorgray#1{\color[gray]{#1}}%
      \expandafter\def\csname LTw\endcsname{\color{white}}%
      \expandafter\def\csname LTb\endcsname{\color{black}}%
      \expandafter\def\csname LTa\endcsname{\color{black}}%
      \expandafter\def\csname LT0\endcsname{\color[rgb]{1,0,0}}%
      \expandafter\def\csname LT1\endcsname{\color[rgb]{0,1,0}}%
      \expandafter\def\csname LT2\endcsname{\color[rgb]{0,0,1}}%
      \expandafter\def\csname LT3\endcsname{\color[rgb]{1,0,1}}%
      \expandafter\def\csname LT4\endcsname{\color[rgb]{0,1,1}}%
      \expandafter\def\csname LT5\endcsname{\color[rgb]{1,1,0}}%
      \expandafter\def\csname LT6\endcsname{\color[rgb]{0,0,0}}%
      \expandafter\def\csname LT7\endcsname{\color[rgb]{1,0.3,0}}%
      \expandafter\def\csname LT8\endcsname{\color[rgb]{0.5,0.5,0.5}}%
    \else
      \def\colorrgb#1{\color{black}}%
      \def\colorgray#1{\color[gray]{#1}}%
      \expandafter\def\csname LTw\endcsname{\color{white}}%
      \expandafter\def\csname LTb\endcsname{\color{black}}%
      \expandafter\def\csname LTa\endcsname{\color{black}}%
      \expandafter\def\csname LT0\endcsname{\color{black}}%
      \expandafter\def\csname LT1\endcsname{\color{black}}%
      \expandafter\def\csname LT2\endcsname{\color{black}}%
      \expandafter\def\csname LT3\endcsname{\color{black}}%
      \expandafter\def\csname LT4\endcsname{\color{black}}%
      \expandafter\def\csname LT5\endcsname{\color{black}}%
      \expandafter\def\csname LT6\endcsname{\color{black}}%
      \expandafter\def\csname LT7\endcsname{\color{black}}%
      \expandafter\def\csname LT8\endcsname{\color{black}}%
    \fi
  \fi
  \setlength{\unitlength}{0.0500bp}%
  \begin{picture}(7200.00,3528.00)%
    \gplgaddtomacro\gplbacktext{%
      \csname LTb\endcsname%
      \put(1342,704){\makebox(0,0)[r]{\strut{} 0}}%
      \put(1342,1070){\makebox(0,0)[r]{\strut{} 0.0005}}%
      \put(1342,1435){\makebox(0,0)[r]{\strut{} 0.001}}%
      \put(1342,1801){\makebox(0,0)[r]{\strut{} 0.0015}}%
      \put(1342,2166){\makebox(0,0)[r]{\strut{} 0.002}}%
      \put(1342,2532){\makebox(0,0)[r]{\strut{} 0.0025}}%
      \put(1342,2897){\makebox(0,0)[r]{\strut{} 0.003}}%
      \put(1342,3263){\makebox(0,0)[r]{\strut{} 0.0035}}%
      \put(1474,484){\makebox(0,0){\strut{} 0}}%
      \put(2806,484){\makebox(0,0){\strut{} 50}}%
      \put(4139,484){\makebox(0,0){\strut{} 100}}%
      \put(5471,484){\makebox(0,0){\strut{} 150}}%
      \put(6803,484){\makebox(0,0){\strut{} 200}}%
      \put(176,1983){\rotatebox{-270}{\makebox(0,0){\strut{}$|H|$}}}%
      \put(4138,154){\makebox(0,0){\strut{}$t$}}%
    }%
    \gplgaddtomacro\gplfronttext{%
      \csname LTb\endcsname%
      \put(5816,3090){\makebox(0,0)[r]{\strut{}L1}}%
      \csname LTb\endcsname%
      \put(5816,2870){\makebox(0,0)[r]{\strut{}L2}}%
      \csname LTb\endcsname%
      \put(5816,2650){\makebox(0,0)[r]{\strut{}L3}}%
    }%
    \gplbacktext
    \put(0,0){\includegraphics{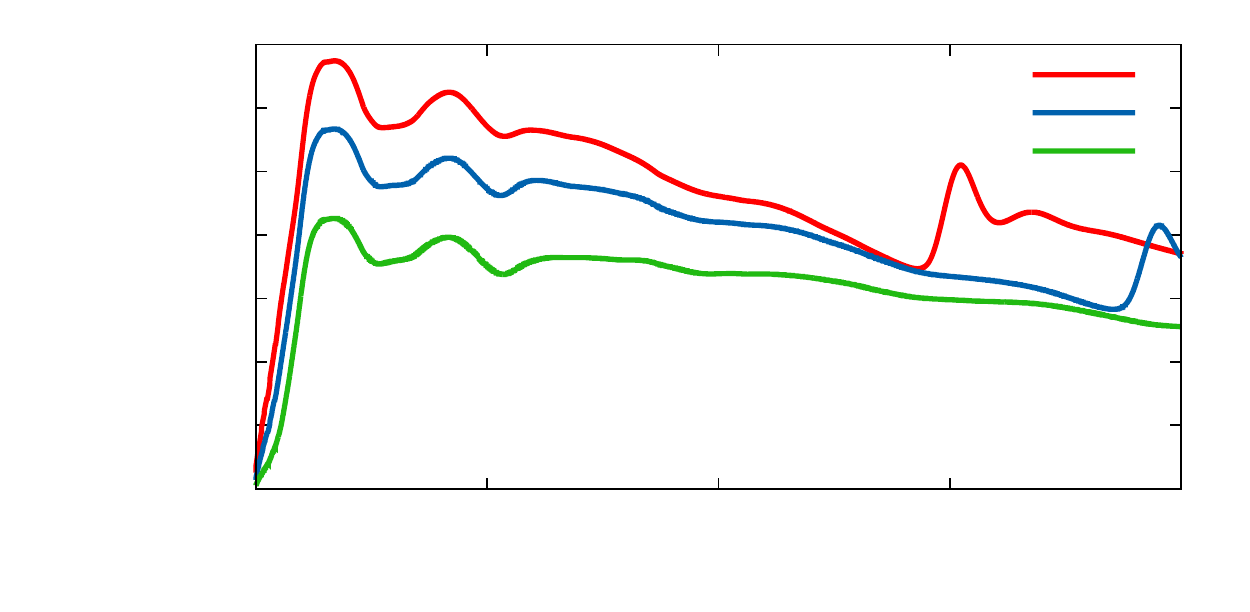}}%
    \gplfronttext
  \end{picture}%
\endgroup

%% file: figs/mom.tex
\begingroup
  \makeatletter
  \providecommand\color[2][]{%
    \GenericError{(gnuplot) \space\space\space\@spaces}{%
      Package color not loaded in conjunction with
      terminal option `colourtext'%
    }{See the gnuplot documentation for explanation.%
    }{Either use 'blacktext' in gnuplot or load the package
      color.sty in LaTeX.}%
    \renewcommand\color[2][]{}%
  }%
  \providecommand\includegraphics[2][]{%
    \GenericError{(gnuplot) \space\space\space\@spaces}{%
      Package graphicx or graphics not loaded%
    }{See the gnuplot documentation for explanation.%
    }{The gnuplot epslatex terminal needs graphicx.sty or graphics.sty.}%
    \renewcommand\includegraphics[2][]{}%
  }%
  \providecommand\rotatebox[2]{#2}%
  \@ifundefined{ifGPcolor}{%
    \newif\ifGPcolor
    \GPcolortrue
  }{}%
  \@ifundefined{ifGPblacktext}{%
    \newif\ifGPblacktext
    \GPblacktexttrue
  }{}%
  \let\gplgaddtomacro\g@addto@macro
  \gdef\gplbacktext{}%
  \gdef\gplfronttext{}%
  \makeatother
  \ifGPblacktext
    \def\colorrgb#1{}%
    \def\colorgray#1{}%
  \else
    \ifGPcolor
      \def\colorrgb#1{\color[rgb]{#1}}%
      \def\colorgray#1{\color[gray]{#1}}%
      \expandafter\def\csname LTw\endcsname{\color{white}}%
      \expandafter\def\csname LTb\endcsname{\color{black}}%
      \expandafter\def\csname LTa\endcsname{\color{black}}%
      \expandafter\def\csname LT0\endcsname{\color[rgb]{1,0,0}}%
      \expandafter\def\csname LT1\endcsname{\color[rgb]{0,1,0}}%
      \expandafter\def\csname LT2\endcsname{\color[rgb]{0,0,1}}%
      \expandafter\def\csname LT3\endcsname{\color[rgb]{1,0,1}}%
      \expandafter\def\csname LT4\endcsname{\color[rgb]{0,1,1}}%
      \expandafter\def\csname LT5\endcsname{\color[rgb]{1,1,0}}%
      \expandafter\def\csname LT6\endcsname{\color[rgb]{0,0,0}}%
      \expandafter\def\csname LT7\endcsname{\color[rgb]{1,0.3,0}}%
      \expandafter\def\csname LT8\endcsname{\color[rgb]{0.5,0.5,0.5}}%
    \else
      \def\colorrgb#1{\color{black}}%
      \def\colorgray#1{\color[gray]{#1}}%
      \expandafter\def\csname LTw\endcsname{\color{white}}%
      \expandafter\def\csname LTb\endcsname{\color{black}}%
      \expandafter\def\csname LTa\endcsname{\color{black}}%
      \expandafter\def\csname LT0\endcsname{\color{black}}%
      \expandafter\def\csname LT1\endcsname{\color{black}}%
      \expandafter\def\csname LT2\endcsname{\color{black}}%
      \expandafter\def\csname LT3\endcsname{\color{black}}%
      \expandafter\def\csname LT4\endcsname{\color{black}}%
      \expandafter\def\csname LT5\endcsname{\color{black}}%
      \expandafter\def\csname LT6\endcsname{\color{black}}%
      \expandafter\def\csname LT7\endcsname{\color{black}}%
      \expandafter\def\csname LT8\endcsname{\color{black}}%
    \fi
  \fi
  \setlength{\unitlength}{0.0500bp}%
  \begin{picture}(7200.00,3528.00)%
    \gplgaddtomacro\gplbacktext{%
      \csname LTb\endcsname%
      \put(1210,704){\makebox(0,0)[r]{\strut{} 0.001}}%
      \put(1210,960){\makebox(0,0)[r]{\strut{} 0.002}}%
      \put(1210,1216){\makebox(0,0)[r]{\strut{} 0.003}}%
      \put(1210,1472){\makebox(0,0)[r]{\strut{} 0.004}}%
      \put(1210,1728){\makebox(0,0)[r]{\strut{} 0.005}}%
      \put(1210,1984){\makebox(0,0)[r]{\strut{} 0.006}}%
      \put(1210,2239){\makebox(0,0)[r]{\strut{} 0.007}}%
      \put(1210,2495){\makebox(0,0)[r]{\strut{} 0.008}}%
      \put(1210,2751){\makebox(0,0)[r]{\strut{} 0.009}}%
      \put(1210,3007){\makebox(0,0)[r]{\strut{} 0.01}}%
      \put(1210,3263){\makebox(0,0)[r]{\strut{} 0.011}}%
      \put(1342,484){\makebox(0,0){\strut{} 0}}%
      \put(2707,484){\makebox(0,0){\strut{} 50}}%
      \put(4073,484){\makebox(0,0){\strut{} 100}}%
      \put(5438,484){\makebox(0,0){\strut{} 150}}%
      \put(6803,484){\makebox(0,0){\strut{} 200}}%
      \put(176,1983){\rotatebox{-270}{\makebox(0,0){\strut{}$|P^x|$}}}%
      \put(4072,154){\makebox(0,0){\strut{}$t$}}%
    }%
    \gplgaddtomacro\gplfronttext{%
      \csname LTb\endcsname%
      \put(5816,3090){\makebox(0,0)[r]{\strut{}L1}}%
      \csname LTb\endcsname%
      \put(5816,2870){\makebox(0,0)[r]{\strut{}L2}}%
      \csname LTb\endcsname%
      \put(5816,2650){\makebox(0,0)[r]{\strut{}L3}}%
    }%
    \gplbacktext
    \put(0,0){\includegraphics{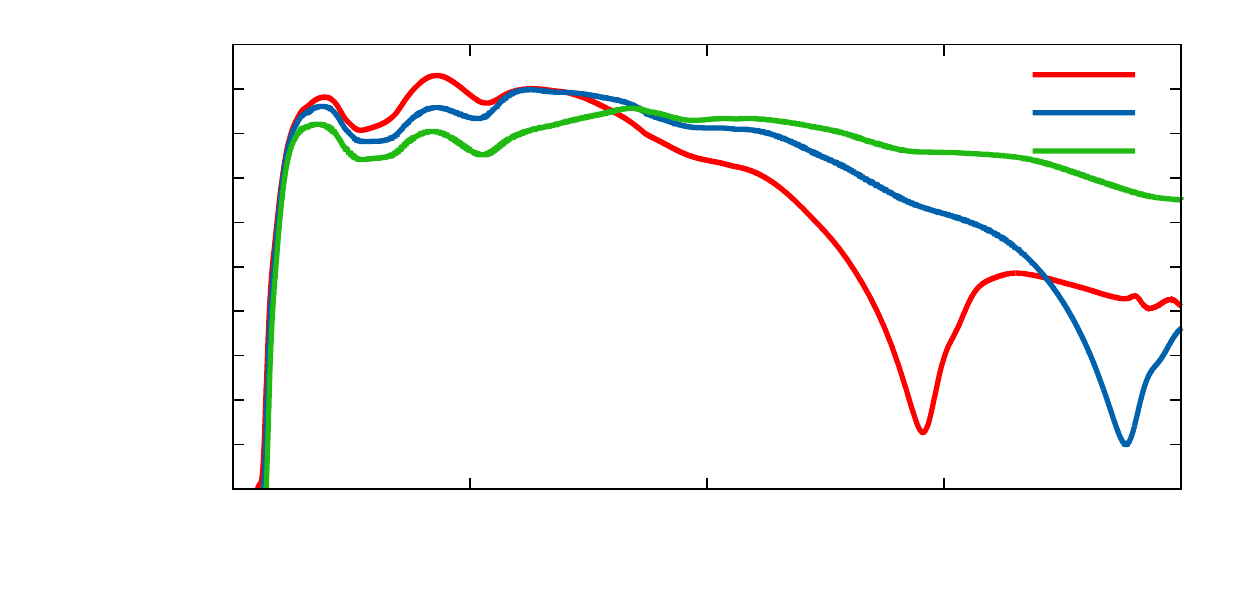}}%
    \gplfronttext
  \end{picture}%
\endgroup

%% file: figs/pd.tex
\begingroup
  \makeatletter
  \providecommand\color[2][]{%
    \GenericError{(gnuplot) \space\space\space\@spaces}{%
      Package color not loaded in conjunction with
      terminal option `colourtext'%
    }{See the gnuplot documentation for explanation.%
    }{Either use 'blacktext' in gnuplot or load the package
      color.sty in LaTeX.}%
    \renewcommand\color[2][]{}%
  }%
  \providecommand\includegraphics[2][]{%
    \GenericError{(gnuplot) \space\space\space\@spaces}{%
      Package graphicx or graphics not loaded%
    }{See the gnuplot documentation for explanation.%
    }{The gnuplot epslatex terminal needs graphicx.sty or graphics.sty.}%
    \renewcommand\includegraphics[2][]{}%
  }%
  \providecommand\rotatebox[2]{#2}%
  \@ifundefined{ifGPcolor}{%
    \newif\ifGPcolor
    \GPcolortrue
  }{}%
  \@ifundefined{ifGPblacktext}{%
    \newif\ifGPblacktext
    \GPblacktexttrue
  }{}%
  \let\gplgaddtomacro\g@addto@macro
  \gdef\gplbacktext{}%
  \gdef\gplfronttext{}%
  \makeatother
  \ifGPblacktext
    \def\colorrgb#1{}%
    \def\colorgray#1{}%
  \else
    \ifGPcolor
      \def\colorrgb#1{\color[rgb]{#1}}%
      \def\colorgray#1{\color[gray]{#1}}%
      \expandafter\def\csname LTw\endcsname{\color{white}}%
      \expandafter\def\csname LTb\endcsname{\color{black}}%
      \expandafter\def\csname LTa\endcsname{\color{black}}%
      \expandafter\def\csname LT0\endcsname{\color[rgb]{1,0,0}}%
      \expandafter\def\csname LT1\endcsname{\color[rgb]{0,1,0}}%
      \expandafter\def\csname LT2\endcsname{\color[rgb]{0,0,1}}%
      \expandafter\def\csname LT3\endcsname{\color[rgb]{1,0,1}}%
      \expandafter\def\csname LT4\endcsname{\color[rgb]{0,1,1}}%
      \expandafter\def\csname LT5\endcsname{\color[rgb]{1,1,0}}%
      \expandafter\def\csname LT6\endcsname{\color[rgb]{0,0,0}}%
      \expandafter\def\csname LT7\endcsname{\color[rgb]{1,0.3,0}}%
      \expandafter\def\csname LT8\endcsname{\color[rgb]{0.5,0.5,0.5}}%
    \else
      \def\colorrgb#1{\color{black}}%
      \def\colorgray#1{\color[gray]{#1}}%
      \expandafter\def\csname LTw\endcsname{\color{white}}%
      \expandafter\def\csname LTb\endcsname{\color{black}}%
      \expandafter\def\csname LTa\endcsname{\color{black}}%
      \expandafter\def\csname LT0\endcsname{\color{black}}%
      \expandafter\def\csname LT1\endcsname{\color{black}}%
      \expandafter\def\csname LT2\endcsname{\color{black}}%
      \expandafter\def\csname LT3\endcsname{\color{black}}%
      \expandafter\def\csname LT4\endcsname{\color{black}}%
      \expandafter\def\csname LT5\endcsname{\color{black}}%
      \expandafter\def\csname LT6\endcsname{\color{black}}%
      \expandafter\def\csname LT7\endcsname{\color{black}}%
      \expandafter\def\csname LT8\endcsname{\color{black}}%
    \fi
  \fi
  \setlength{\unitlength}{0.0500bp}%
  \begin{picture}(8640.00,3528.00)%
    \gplgaddtomacro\gplbacktext{%
      \csname LTb\endcsname%
      \put(946,704){\makebox(0,0)[r]{\strut{} 0}}%
      \put(946,1216){\makebox(0,0)[r]{\strut{} 50}}%
      \put(946,1728){\makebox(0,0)[r]{\strut{} 100}}%
      \put(946,2239){\makebox(0,0)[r]{\strut{} 150}}%
      \put(946,2751){\makebox(0,0)[r]{\strut{} 200}}%
      \put(946,3263){\makebox(0,0)[r]{\strut{} 250}}%
      \put(1078,484){\makebox(0,0){\strut{} 0}}%
      \put(1910,484){\makebox(0,0){\strut{} 50}}%
      \put(2741,484){\makebox(0,0){\strut{} 100}}%
      \put(3573,484){\makebox(0,0){\strut{} 150}}%
      \put(4404,484){\makebox(0,0){\strut{} 200}}%
      \put(5236,484){\makebox(0,0){\strut{} 250}}%
      \put(6067,484){\makebox(0,0){\strut{} 300}}%
      \put(176,1983){\rotatebox{-270}{\makebox(0,0){\strut{}$D_{\rm edge}(\tau)$}}}%
      \put(3572,154){\makebox(0,0){\strut{}$\tau$}}%
    }%
    \gplgaddtomacro\gplfronttext{%
      \csname LTb\endcsname%
      \put(7651,2863){\makebox(0,0)[r]{\strut{}$m$=0.5 lattice}}%
      \csname LTb\endcsname%
      \put(7651,2643){\makebox(0,0)[r]{\strut{}FLRW(0.5)}}%
      \csname LTb\endcsname%
      \put(7651,2423){\makebox(0,0)[r]{\strut{}$m$=1 lattice}}%
      \csname LTb\endcsname%
      \put(7651,2203){\makebox(0,0)[r]{\strut{}FLRW(1)}}%
      \csname LTb\endcsname%
      \put(7651,1983){\makebox(0,0)[r]{\strut{}$m$=2 lattice}}%
      \csname LTb\endcsname%
      \put(7651,1763){\makebox(0,0)[r]{\strut{}FLRW(2)}}%
      \csname LTb\endcsname%
      \put(7651,1543){\makebox(0,0)[r]{\strut{}$m$=5 lattice}}%
      \csname LTb\endcsname%
      \put(7651,1323){\makebox(0,0)[r]{\strut{}FLRW(5)}}%
      \csname LTb\endcsname%
      \put(7651,1103){\makebox(0,0)[r]{\strut{}FLRWb(5)}}%
    }%
    \gplbacktext
    \put(0,0){\includegraphics{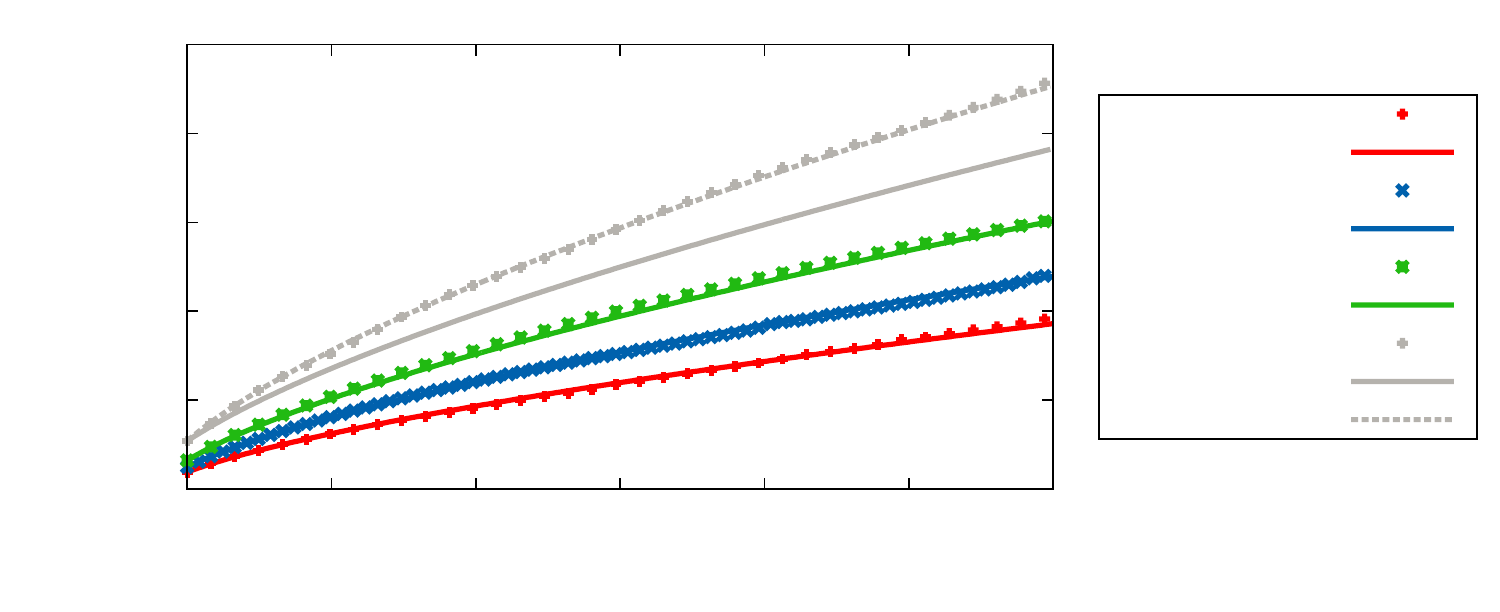}}%
    \gplfronttext
  \end{picture}%
\endgroup

%% file: figs/mu.tex
\begingroup
  \makeatletter
  \providecommand\color[2][]{%
    \GenericError{(gnuplot) \space\space\space\@spaces}{%
      Package color not loaded in conjunction with
      terminal option `colourtext'%
    }{See the gnuplot documentation for explanation.%
    }{Either use 'blacktext' in gnuplot or load the package
      color.sty in LaTeX.}%
    \renewcommand\color[2][]{}%
  }%
  \providecommand\includegraphics[2][]{%
    \GenericError{(gnuplot) \space\space\space\@spaces}{%
      Package graphicx or graphics not loaded%
    }{See the gnuplot documentation for explanation.%
    }{The gnuplot epslatex terminal needs graphicx.sty or graphics.sty.}%
    \renewcommand\includegraphics[2][]{}%
  }%
  \providecommand\rotatebox[2]{#2}%
  \@ifundefined{ifGPcolor}{%
    \newif\ifGPcolor
    \GPcolortrue
  }{}%
  \@ifundefined{ifGPblacktext}{%
    \newif\ifGPblacktext
    \GPblacktexttrue
  }{}%
  \let\gplgaddtomacro\g@addto@macro
  \gdef\gplbacktext{}%
  \gdef\gplfronttext{}%
  \makeatother
  \ifGPblacktext
    \def\colorrgb#1{}%
    \def\colorgray#1{}%
  \else
    \ifGPcolor
      \def\colorrgb#1{\color[rgb]{#1}}%
      \def\colorgray#1{\color[gray]{#1}}%
      \expandafter\def\csname LTw\endcsname{\color{white}}%
      \expandafter\def\csname LTb\endcsname{\color{black}}%
      \expandafter\def\csname LTa\endcsname{\color{black}}%
      \expandafter\def\csname LT0\endcsname{\color[rgb]{1,0,0}}%
      \expandafter\def\csname LT1\endcsname{\color[rgb]{0,1,0}}%
      \expandafter\def\csname LT2\endcsname{\color[rgb]{0,0,1}}%
      \expandafter\def\csname LT3\endcsname{\color[rgb]{1,0,1}}%
      \expandafter\def\csname LT4\endcsname{\color[rgb]{0,1,1}}%
      \expandafter\def\csname LT5\endcsname{\color[rgb]{1,1,0}}%
      \expandafter\def\csname LT6\endcsname{\color[rgb]{0,0,0}}%
      \expandafter\def\csname LT7\endcsname{\color[rgb]{1,0.3,0}}%
      \expandafter\def\csname LT8\endcsname{\color[rgb]{0.5,0.5,0.5}}%
    \else
      \def\colorrgb#1{\color{black}}%
      \def\colorgray#1{\color[gray]{#1}}%
      \expandafter\def\csname LTw\endcsname{\color{white}}%
      \expandafter\def\csname LTb\endcsname{\color{black}}%
      \expandafter\def\csname LTa\endcsname{\color{black}}%
      \expandafter\def\csname LT0\endcsname{\color{black}}%
      \expandafter\def\csname LT1\endcsname{\color{black}}%
      \expandafter\def\csname LT2\endcsname{\color{black}}%
      \expandafter\def\csname LT3\endcsname{\color{black}}%
      \expandafter\def\csname LT4\endcsname{\color{black}}%
      \expandafter\def\csname LT5\endcsname{\color{black}}%
      \expandafter\def\csname LT6\endcsname{\color{black}}%
      \expandafter\def\csname LT7\endcsname{\color{black}}%
      \expandafter\def\csname LT8\endcsname{\color{black}}%
    \fi
  \fi
  \setlength{\unitlength}{0.0500bp}%
  \begin{picture}(7200.00,3528.00)%
    \gplgaddtomacro\gplbacktext{%
      \csname LTb\endcsname%
      \put(1078,704){\makebox(0,0)[r]{\strut{} 0.04}}%
      \put(1078,960){\makebox(0,0)[r]{\strut{} 0.06}}%
      \put(1078,1216){\makebox(0,0)[r]{\strut{} 0.08}}%
      \put(1078,1472){\makebox(0,0)[r]{\strut{} 0.1}}%
      \put(1078,1728){\makebox(0,0)[r]{\strut{} 0.12}}%
      \put(1078,1984){\makebox(0,0)[r]{\strut{} 0.14}}%
      \put(1078,2239){\makebox(0,0)[r]{\strut{} 0.16}}%
      \put(1078,2495){\makebox(0,0)[r]{\strut{} 0.18}}%
      \put(1078,2751){\makebox(0,0)[r]{\strut{} 0.2}}%
      \put(1078,3007){\makebox(0,0)[r]{\strut{} 0.22}}%
      \put(1078,3263){\makebox(0,0)[r]{\strut{} 0.24}}%
      \put(1210,484){\makebox(0,0){\strut{} 0}}%
      \put(2009,484){\makebox(0,0){\strut{} 1}}%
      \put(2808,484){\makebox(0,0){\strut{} 2}}%
      \put(3607,484){\makebox(0,0){\strut{} 3}}%
      \put(4406,484){\makebox(0,0){\strut{} 4}}%
      \put(5205,484){\makebox(0,0){\strut{} 5}}%
      \put(6004,484){\makebox(0,0){\strut{} 6}}%
      \put(6803,484){\makebox(0,0){\strut{} 7}}%
      \put(176,1983){\rotatebox{-270}{\makebox(0,0){\strut{}$\mu$}}}%
      \put(4006,154){\makebox(0,0){\strut{}$m_{\rm BH}$}}%
    }%
    \gplgaddtomacro\gplfronttext{%
    }%
    \gplbacktext
    \put(0,0){\includegraphics{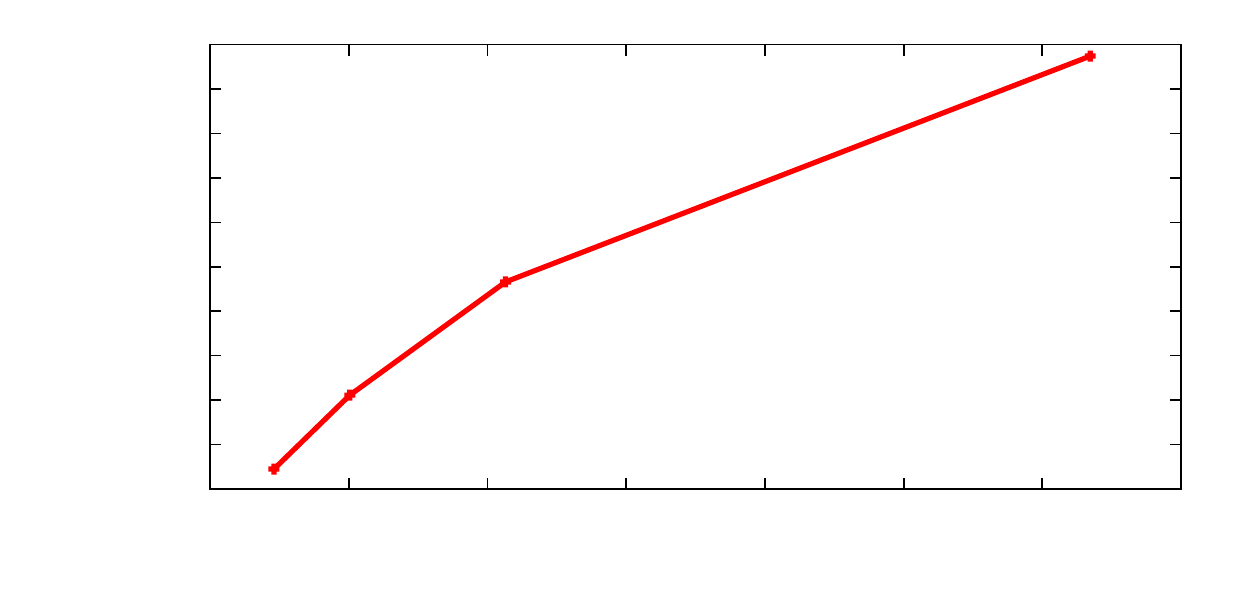}}%
    \gplfronttext
  \end{picture}%
\endgroup

%% file: figs/x.tex
\begingroup
  \makeatletter
  \providecommand\color[2][]{%
    \GenericError{(gnuplot) \space\space\space\@spaces}{%
      Package color not loaded in conjunction with
      terminal option `colourtext'%
    }{See the gnuplot documentation for explanation.%
    }{Either use 'blacktext' in gnuplot or load the package
      color.sty in LaTeX.}%
    \renewcommand\color[2][]{}%
  }%
  \providecommand\includegraphics[2][]{%
    \GenericError{(gnuplot) \space\space\space\@spaces}{%
      Package graphicx or graphics not loaded%
    }{See the gnuplot documentation for explanation.%
    }{The gnuplot epslatex terminal needs graphicx.sty or graphics.sty.}%
    \renewcommand\includegraphics[2][]{}%
  }%
  \providecommand\rotatebox[2]{#2}%
  \@ifundefined{ifGPcolor}{%
    \newif\ifGPcolor
    \GPcolortrue
  }{}%
  \@ifundefined{ifGPblacktext}{%
    \newif\ifGPblacktext
    \GPblacktexttrue
  }{}%
  \let\gplgaddtomacro\g@addto@macro
  \gdef\gplbacktext{}%
  \gdef\gplfronttext{}%
  \makeatother
  \ifGPblacktext
    \def\colorrgb#1{}%
    \def\colorgray#1{}%
  \else
    \ifGPcolor
      \def\colorrgb#1{\color[rgb]{#1}}%
      \def\colorgray#1{\color[gray]{#1}}%
      \expandafter\def\csname LTw\endcsname{\color{white}}%
      \expandafter\def\csname LTb\endcsname{\color{black}}%
      \expandafter\def\csname LTa\endcsname{\color{black}}%
      \expandafter\def\csname LT0\endcsname{\color[rgb]{1,0,0}}%
      \expandafter\def\csname LT1\endcsname{\color[rgb]{0,1,0}}%
      \expandafter\def\csname LT2\endcsname{\color[rgb]{0,0,1}}%
      \expandafter\def\csname LT3\endcsname{\color[rgb]{1,0,1}}%
      \expandafter\def\csname LT4\endcsname{\color[rgb]{0,1,1}}%
      \expandafter\def\csname LT5\endcsname{\color[rgb]{1,1,0}}%
      \expandafter\def\csname LT6\endcsname{\color[rgb]{0,0,0}}%
      \expandafter\def\csname LT7\endcsname{\color[rgb]{1,0.3,0}}%
      \expandafter\def\csname LT8\endcsname{\color[rgb]{0.5,0.5,0.5}}%
    \else
      \def\colorrgb#1{\color{black}}%
      \def\colorgray#1{\color[gray]{#1}}%
      \expandafter\def\csname LTw\endcsname{\color{white}}%
      \expandafter\def\csname LTb\endcsname{\color{black}}%
      \expandafter\def\csname LTa\endcsname{\color{black}}%
      \expandafter\def\csname LT0\endcsname{\color{black}}%
      \expandafter\def\csname LT1\endcsname{\color{black}}%
      \expandafter\def\csname LT2\endcsname{\color{black}}%
      \expandafter\def\csname LT3\endcsname{\color{black}}%
      \expandafter\def\csname LT4\endcsname{\color{black}}%
      \expandafter\def\csname LT5\endcsname{\color{black}}%
      \expandafter\def\csname LT6\endcsname{\color{black}}%
      \expandafter\def\csname LT7\endcsname{\color{black}}%
      \expandafter\def\csname LT8\endcsname{\color{black}}%
    \fi
  \fi
  \setlength{\unitlength}{0.0500bp}%
  \begin{picture}(7200.00,3528.00)%
    \gplgaddtomacro\gplbacktext{%
      \csname LTb\endcsname%
      \put(946,704){\makebox(0,0)[r]{\strut{}-0.8}}%
      \put(946,1024){\makebox(0,0)[r]{\strut{}-0.7}}%
      \put(946,1344){\makebox(0,0)[r]{\strut{}-0.6}}%
      \put(946,1664){\makebox(0,0)[r]{\strut{}-0.5}}%
      \put(946,1983){\makebox(0,0)[r]{\strut{}-0.4}}%
      \put(946,2303){\makebox(0,0)[r]{\strut{}-0.3}}%
      \put(946,2623){\makebox(0,0)[r]{\strut{}-0.2}}%
      \put(946,2943){\makebox(0,0)[r]{\strut{}-0.1}}%
      \put(946,3263){\makebox(0,0)[r]{\strut{} 0}}%
      \put(1078,484){\makebox(0,0){\strut{} 0.04}}%
      \put(1651,484){\makebox(0,0){\strut{} 0.06}}%
      \put(2223,484){\makebox(0,0){\strut{} 0.08}}%
      \put(2796,484){\makebox(0,0){\strut{} 0.1}}%
      \put(3368,484){\makebox(0,0){\strut{} 0.12}}%
      \put(3940,484){\makebox(0,0){\strut{} 0.14}}%
      \put(4513,484){\makebox(0,0){\strut{} 0.16}}%
      \put(5086,484){\makebox(0,0){\strut{} 0.18}}%
      \put(5658,484){\makebox(0,0){\strut{} 0.2}}%
      \put(6231,484){\makebox(0,0){\strut{} 0.22}}%
      \put(6803,484){\makebox(0,0){\strut{} 0.24}}%
      \put(176,1983){\rotatebox{-270}{\makebox(0,0){\strut{}$x$}}}%
      \put(3940,154){\makebox(0,0){\strut{}$\mu$}}%
    }%
    \gplgaddtomacro\gplfronttext{%
    }%
    \gplbacktext
    \put(0,0){\includegraphics{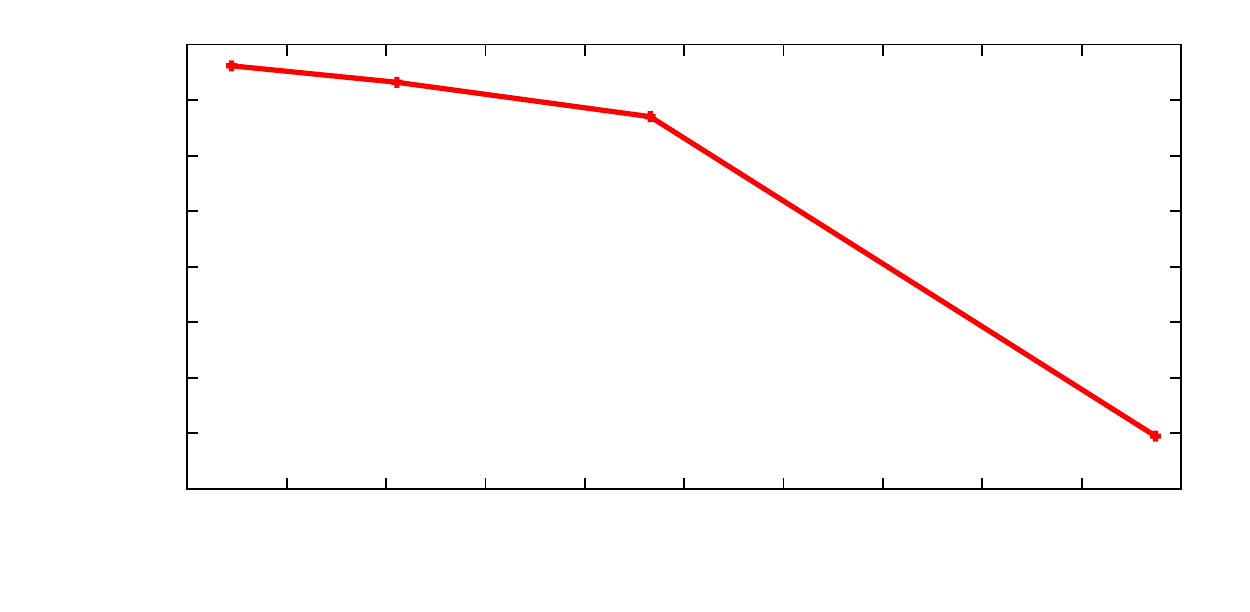}}%
    \gplfronttext
  \end{picture}%
\endgroup

%% file: figs/meff.tex
\begingroup
  \makeatletter
  \providecommand\color[2][]{%
    \GenericError{(gnuplot) \space\space\space\@spaces}{%
      Package color not loaded in conjunction with
      terminal option `colourtext'%
    }{See the gnuplot documentation for explanation.%
    }{Either use 'blacktext' in gnuplot or load the package
      color.sty in LaTeX.}%
    \renewcommand\color[2][]{}%
  }%
  \providecommand\includegraphics[2][]{%
    \GenericError{(gnuplot) \space\space\space\@spaces}{%
      Package graphicx or graphics not loaded%
    }{See the gnuplot documentation for explanation.%
    }{The gnuplot epslatex terminal needs graphicx.sty or graphics.sty.}%
    \renewcommand\includegraphics[2][]{}%
  }%
  \providecommand\rotatebox[2]{#2}%
  \@ifundefined{ifGPcolor}{%
    \newif\ifGPcolor
    \GPcolortrue
  }{}%
  \@ifundefined{ifGPblacktext}{%
    \newif\ifGPblacktext
    \GPblacktexttrue
  }{}%
  \let\gplgaddtomacro\g@addto@macro
  \gdef\gplbacktext{}%
  \gdef\gplfronttext{}%
  \makeatother
  \ifGPblacktext
    \def\colorrgb#1{}%
    \def\colorgray#1{}%
  \else
    \ifGPcolor
      \def\colorrgb#1{\color[rgb]{#1}}%
      \def\colorgray#1{\color[gray]{#1}}%
      \expandafter\def\csname LTw\endcsname{\color{white}}%
      \expandafter\def\csname LTb\endcsname{\color{black}}%
      \expandafter\def\csname LTa\endcsname{\color{black}}%
      \expandafter\def\csname LT0\endcsname{\color[rgb]{1,0,0}}%
      \expandafter\def\csname LT1\endcsname{\color[rgb]{0,1,0}}%
      \expandafter\def\csname LT2\endcsname{\color[rgb]{0,0,1}}%
      \expandafter\def\csname LT3\endcsname{\color[rgb]{1,0,1}}%
      \expandafter\def\csname LT4\endcsname{\color[rgb]{0,1,1}}%
      \expandafter\def\csname LT5\endcsname{\color[rgb]{1,1,0}}%
      \expandafter\def\csname LT6\endcsname{\color[rgb]{0,0,0}}%
      \expandafter\def\csname LT7\endcsname{\color[rgb]{1,0.3,0}}%
      \expandafter\def\csname LT8\endcsname{\color[rgb]{0.5,0.5,0.5}}%
    \else
      \def\colorrgb#1{\color{black}}%
      \def\colorgray#1{\color[gray]{#1}}%
      \expandafter\def\csname LTw\endcsname{\color{white}}%
      \expandafter\def\csname LTb\endcsname{\color{black}}%
      \expandafter\def\csname LTa\endcsname{\color{black}}%
      \expandafter\def\csname LT0\endcsname{\color{black}}%
      \expandafter\def\csname LT1\endcsname{\color{black}}%
      \expandafter\def\csname LT2\endcsname{\color{black}}%
      \expandafter\def\csname LT3\endcsname{\color{black}}%
      \expandafter\def\csname LT4\endcsname{\color{black}}%
      \expandafter\def\csname LT5\endcsname{\color{black}}%
      \expandafter\def\csname LT6\endcsname{\color{black}}%
      \expandafter\def\csname LT7\endcsname{\color{black}}%
      \expandafter\def\csname LT8\endcsname{\color{black}}%
    \fi
  \fi
  \setlength{\unitlength}{0.0500bp}%
  \begin{picture}(7200.00,12096.00)%
    \gplgaddtomacro\gplbacktext{%
      \csname LTb\endcsname%
      \put(858,9776){\makebox(0,0)[r]{\strut{} 0.43}}%
      \put(858,10069){\makebox(0,0)[r]{\strut{} 0.44}}%
      \put(858,10363){\makebox(0,0)[r]{\strut{} 0.45}}%
      \put(858,10656){\makebox(0,0)[r]{\strut{} 0.46}}%
      \put(858,10950){\makebox(0,0)[r]{\strut{} 0.47}}%
      \put(858,11243){\makebox(0,0)[r]{\strut{} 0.48}}%
      \put(858,11537){\makebox(0,0)[r]{\strut{} 0.49}}%
      \put(858,11830){\makebox(0,0)[r]{\strut{} 0.5}}%
      \put(990,9556){\makebox(0,0){\strut{} 0}}%
      \put(1959,9556){\makebox(0,0){\strut{} 5}}%
      \put(2928,9556){\makebox(0,0){\strut{} 10}}%
      \put(3897,9556){\makebox(0,0){\strut{} 15}}%
      \put(4865,9556){\makebox(0,0){\strut{} 20}}%
      \put(5834,9556){\makebox(0,0){\strut{} 25}}%
      \put(6803,9556){\makebox(0,0){\strut{} 30}}%
      \put(3896,9226){\makebox(0,0){\strut{}$\tau$}}%
    }%
    \gplgaddtomacro\gplfronttext{%
      \put(1945,11657){\makebox(0,0){\strut{}$m=0.5$}}%
      \csname LTb\endcsname%
      \put(1914,11437){\makebox(0,0)[r]{\strut{}$M_{\rm eff}(\tau)$}}%
      \csname LTb\endcsname%
      \put(1914,11217){\makebox(0,0)[r]{\strut{}$m_{\rm BH}$}}%
    }%
    \gplgaddtomacro\gplbacktext{%
      \csname LTb\endcsname%
      \put(858,6752){\makebox(0,0)[r]{\strut{} 0.95}}%
      \put(858,6957){\makebox(0,0)[r]{\strut{} 0.96}}%
      \put(858,7163){\makebox(0,0)[r]{\strut{} 0.97}}%
      \put(858,7368){\makebox(0,0)[r]{\strut{} 0.98}}%
      \put(858,7574){\makebox(0,0)[r]{\strut{} 0.99}}%
      \put(858,7779){\makebox(0,0)[r]{\strut{} 1}}%
      \put(858,7984){\makebox(0,0)[r]{\strut{} 1.01}}%
      \put(858,8190){\makebox(0,0)[r]{\strut{} 1.02}}%
      \put(858,8395){\makebox(0,0)[r]{\strut{} 1.03}}%
      \put(858,8601){\makebox(0,0)[r]{\strut{} 1.04}}%
      \put(858,8806){\makebox(0,0)[r]{\strut{} 1.05}}%
      \put(990,6532){\makebox(0,0){\strut{} 0}}%
      \put(1820,6532){\makebox(0,0){\strut{} 10}}%
      \put(2651,6532){\makebox(0,0){\strut{} 20}}%
      \put(3481,6532){\makebox(0,0){\strut{} 30}}%
      \put(4312,6532){\makebox(0,0){\strut{} 40}}%
      \put(5142,6532){\makebox(0,0){\strut{} 50}}%
      \put(5973,6532){\makebox(0,0){\strut{} 60}}%
      \put(6803,6532){\makebox(0,0){\strut{} 70}}%
      \put(3896,6202){\makebox(0,0){\strut{}$\tau$}}%
    }%
    \gplgaddtomacro\gplfronttext{%
      \put(1945,8633){\makebox(0,0){\strut{}$m=1$}}%
      \csname LTb\endcsname%
      \put(1914,8413){\makebox(0,0)[r]{\strut{}$M_{\rm eff}(\tau)$}}%
      \csname LTb\endcsname%
      \put(1914,8193){\makebox(0,0)[r]{\strut{}$m_{\rm BH}$}}%
    }%
    \gplgaddtomacro\gplbacktext{%
      \csname LTb\endcsname%
      \put(858,3728){\makebox(0,0)[r]{\strut{} 1.9}}%
      \put(858,4139){\makebox(0,0)[r]{\strut{} 1.95}}%
      \put(858,4550){\makebox(0,0)[r]{\strut{} 2}}%
      \put(858,4961){\makebox(0,0)[r]{\strut{} 2.05}}%
      \put(858,5372){\makebox(0,0)[r]{\strut{} 2.1}}%
      \put(858,5783){\makebox(0,0)[r]{\strut{} 2.15}}%
      \put(990,3508){\makebox(0,0){\strut{} 0}}%
      \put(2153,3508){\makebox(0,0){\strut{} 20}}%
      \put(3315,3508){\makebox(0,0){\strut{} 40}}%
      \put(4478,3508){\makebox(0,0){\strut{} 60}}%
      \put(5640,3508){\makebox(0,0){\strut{} 80}}%
      \put(6803,3508){\makebox(0,0){\strut{} 100}}%
      \put(3896,3178){\makebox(0,0){\strut{}$\tau$}}%
    }%
    \gplgaddtomacro\gplfronttext{%
      \put(5847,4341){\makebox(0,0){\strut{}$m=2$}}%
      \csname LTb\endcsname%
      \put(5816,4121){\makebox(0,0)[r]{\strut{}$M_{\rm eff}(\tau)$}}%
      \csname LTb\endcsname%
      \put(5816,3901){\makebox(0,0)[r]{\strut{}$m_{\rm BH}$}}%
    }%
    \gplgaddtomacro\gplbacktext{%
      \csname LTb\endcsname%
      \put(726,704){\makebox(0,0)[r]{\strut{} 4}}%
      \put(726,932){\makebox(0,0)[r]{\strut{} 4.5}}%
      \put(726,1161){\makebox(0,0)[r]{\strut{} 5}}%
      \put(726,1389){\makebox(0,0)[r]{\strut{} 5.5}}%
      \put(726,1617){\makebox(0,0)[r]{\strut{} 6}}%
      \put(726,1846){\makebox(0,0)[r]{\strut{} 6.5}}%
      \put(726,2074){\makebox(0,0)[r]{\strut{} 7}}%
      \put(726,2302){\makebox(0,0)[r]{\strut{} 7.5}}%
      \put(726,2531){\makebox(0,0)[r]{\strut{} 8}}%
      \put(726,2759){\makebox(0,0)[r]{\strut{} 8.5}}%
      \put(858,484){\makebox(0,0){\strut{} 0}}%
      \put(1849,484){\makebox(0,0){\strut{} 50}}%
      \put(2840,484){\makebox(0,0){\strut{} 100}}%
      \put(3831,484){\makebox(0,0){\strut{} 150}}%
      \put(4821,484){\makebox(0,0){\strut{} 200}}%
      \put(5812,484){\makebox(0,0){\strut{} 250}}%
      \put(6803,484){\makebox(0,0){\strut{} 300}}%
      \put(3830,154){\makebox(0,0){\strut{}$\tau$}}%
    }%
    \gplgaddtomacro\gplfronttext{%
      \put(5847,1317){\makebox(0,0){\strut{}$m=5$}}%
      \csname LTb\endcsname%
      \put(5816,1097){\makebox(0,0)[r]{\strut{}$M_{\rm eff}(\tau)$}}%
      \csname LTb\endcsname%
      \put(5816,877){\makebox(0,0)[r]{\strut{}$m_{\rm BH}$}}%
    }%
    \gplbacktext
    \put(0,0){\includegraphics{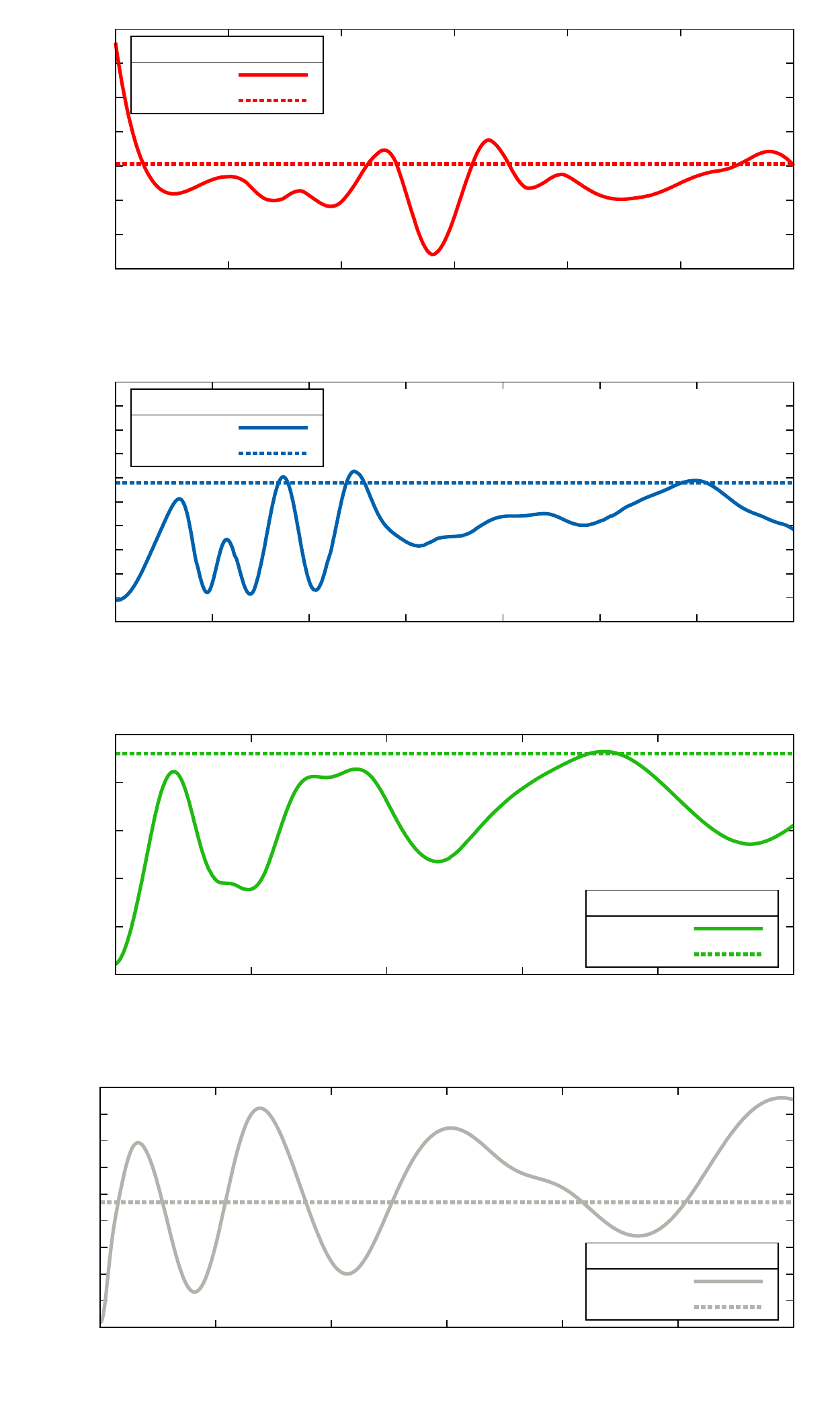}}%
    \gplfronttext
  \end{picture}%
\endgroup

%% file: figs/peffcmf.tex
\begingroup
  \makeatletter
  \providecommand\color[2][]{%
    \GenericError{(gnuplot) \space\space\space\@spaces}{%
      Package color not loaded in conjunction with
      terminal option `colourtext'%
    }{See the gnuplot documentation for explanation.%
    }{Either use 'blacktext' in gnuplot or load the package
      color.sty in LaTeX.}%
    \renewcommand\color[2][]{}%
  }%
  \providecommand\includegraphics[2][]{%
    \GenericError{(gnuplot) \space\space\space\@spaces}{%
      Package graphicx or graphics not loaded%
    }{See the gnuplot documentation for explanation.%
    }{The gnuplot epslatex terminal needs graphicx.sty or graphics.sty.}%
    \renewcommand\includegraphics[2][]{}%
  }%
  \providecommand\rotatebox[2]{#2}%
  \@ifundefined{ifGPcolor}{%
    \newif\ifGPcolor
    \GPcolortrue
  }{}%
  \@ifundefined{ifGPblacktext}{%
    \newif\ifGPblacktext
    \GPblacktexttrue
  }{}%
  \let\gplgaddtomacro\g@addto@macro
  \gdef\gplbacktext{}%
  \gdef\gplfronttext{}%
  \makeatother
  \ifGPblacktext
    \def\colorrgb#1{}%
    \def\colorgray#1{}%
  \else
    \ifGPcolor
      \def\colorrgb#1{\color[rgb]{#1}}%
      \def\colorgray#1{\color[gray]{#1}}%
      \expandafter\def\csname LTw\endcsname{\color{white}}%
      \expandafter\def\csname LTb\endcsname{\color{black}}%
      \expandafter\def\csname LTa\endcsname{\color{black}}%
      \expandafter\def\csname LT0\endcsname{\color[rgb]{1,0,0}}%
      \expandafter\def\csname LT1\endcsname{\color[rgb]{0,1,0}}%
      \expandafter\def\csname LT2\endcsname{\color[rgb]{0,0,1}}%
      \expandafter\def\csname LT3\endcsname{\color[rgb]{1,0,1}}%
      \expandafter\def\csname LT4\endcsname{\color[rgb]{0,1,1}}%
      \expandafter\def\csname LT5\endcsname{\color[rgb]{1,1,0}}%
      \expandafter\def\csname LT6\endcsname{\color[rgb]{0,0,0}}%
      \expandafter\def\csname LT7\endcsname{\color[rgb]{1,0.3,0}}%
      \expandafter\def\csname LT8\endcsname{\color[rgb]{0.5,0.5,0.5}}%
    \else
      \def\colorrgb#1{\color{black}}%
      \def\colorgray#1{\color[gray]{#1}}%
      \expandafter\def\csname LTw\endcsname{\color{white}}%
      \expandafter\def\csname LTb\endcsname{\color{black}}%
      \expandafter\def\csname LTa\endcsname{\color{black}}%
      \expandafter\def\csname LT0\endcsname{\color{black}}%
      \expandafter\def\csname LT1\endcsname{\color{black}}%
      \expandafter\def\csname LT2\endcsname{\color{black}}%
      \expandafter\def\csname LT3\endcsname{\color{black}}%
      \expandafter\def\csname LT4\endcsname{\color{black}}%
      \expandafter\def\csname LT5\endcsname{\color{black}}%
      \expandafter\def\csname LT6\endcsname{\color{black}}%
      \expandafter\def\csname LT7\endcsname{\color{black}}%
      \expandafter\def\csname LT8\endcsname{\color{black}}%
    \fi
  \fi
  \setlength{\unitlength}{0.0500bp}%
  \begin{picture}(7200.00,7056.00)%
    \gplgaddtomacro\gplbacktext{%
      \csname LTb\endcsname%
      \put(682,3415){\makebox(0,0)[r]{\strut{}-2}}%
      \put(682,3822){\makebox(0,0)[r]{\strut{}-1}}%
      \put(682,4229){\makebox(0,0)[r]{\strut{} 0}}%
      \put(682,4636){\makebox(0,0)[r]{\strut{} 1}}%
      \put(682,5043){\makebox(0,0)[r]{\strut{} 2}}%
      \put(682,5450){\makebox(0,0)[r]{\strut{} 3}}%
      \put(176,4510){\rotatebox{-270}{\makebox(0,0){\strut{}$p_{\rm eff} \; (\cdot 10^5)$}}}%
    }%
    \gplgaddtomacro\gplfronttext{%
      \csname LTb\endcsname%
      \put(5816,5585){\makebox(0,0)[r]{\strut{}L1}}%
      \csname LTb\endcsname%
      \put(5816,5365){\makebox(0,0)[r]{\strut{}L2}}%
      \csname LTb\endcsname%
      \put(5816,5145){\makebox(0,0)[r]{\strut{}L3}}%
    }%
    \gplgaddtomacro\gplbacktext{%
      \csname LTb\endcsname%
      \put(682,914){\makebox(0,0)[r]{\strut{}-8}}%
      \put(682,1319){\makebox(0,0)[r]{\strut{}-6}}%
      \put(682,1725){\makebox(0,0)[r]{\strut{}-4}}%
      \put(682,2130){\makebox(0,0)[r]{\strut{}-2}}%
      \put(682,2536){\makebox(0,0)[r]{\strut{} 0}}%
      \put(682,2941){\makebox(0,0)[r]{\strut{} 2}}%
      \put(814,484){\makebox(0,0){\strut{} 0}}%
      \put(1670,484){\makebox(0,0){\strut{} 10}}%
      \put(2525,484){\makebox(0,0){\strut{} 20}}%
      \put(3381,484){\makebox(0,0){\strut{} 30}}%
      \put(4236,484){\makebox(0,0){\strut{} 40}}%
      \put(5092,484){\makebox(0,0){\strut{} 50}}%
      \put(5947,484){\makebox(0,0){\strut{} 60}}%
      \put(6803,484){\makebox(0,0){\strut{} 70}}%
      \put(176,1983){\rotatebox{-270}{\makebox(0,0){\strut{}$p_{\rm eff} \; (\cdot 10^5)$}}}%
      \put(3808,154){\makebox(0,0){\strut{}$\tau$}}%
    }%
    \gplgaddtomacro\gplfronttext{%
      \csname LTb\endcsname%
      \put(5816,1317){\makebox(0,0)[r]{\strut{}L1}}%
      \csname LTb\endcsname%
      \put(5816,1097){\makebox(0,0)[r]{\strut{}L5}}%
      \csname LTb\endcsname%
      \put(5816,877){\makebox(0,0)[r]{\strut{}L6}}%
    }%
    \gplbacktext
    \put(0,0){\includegraphics{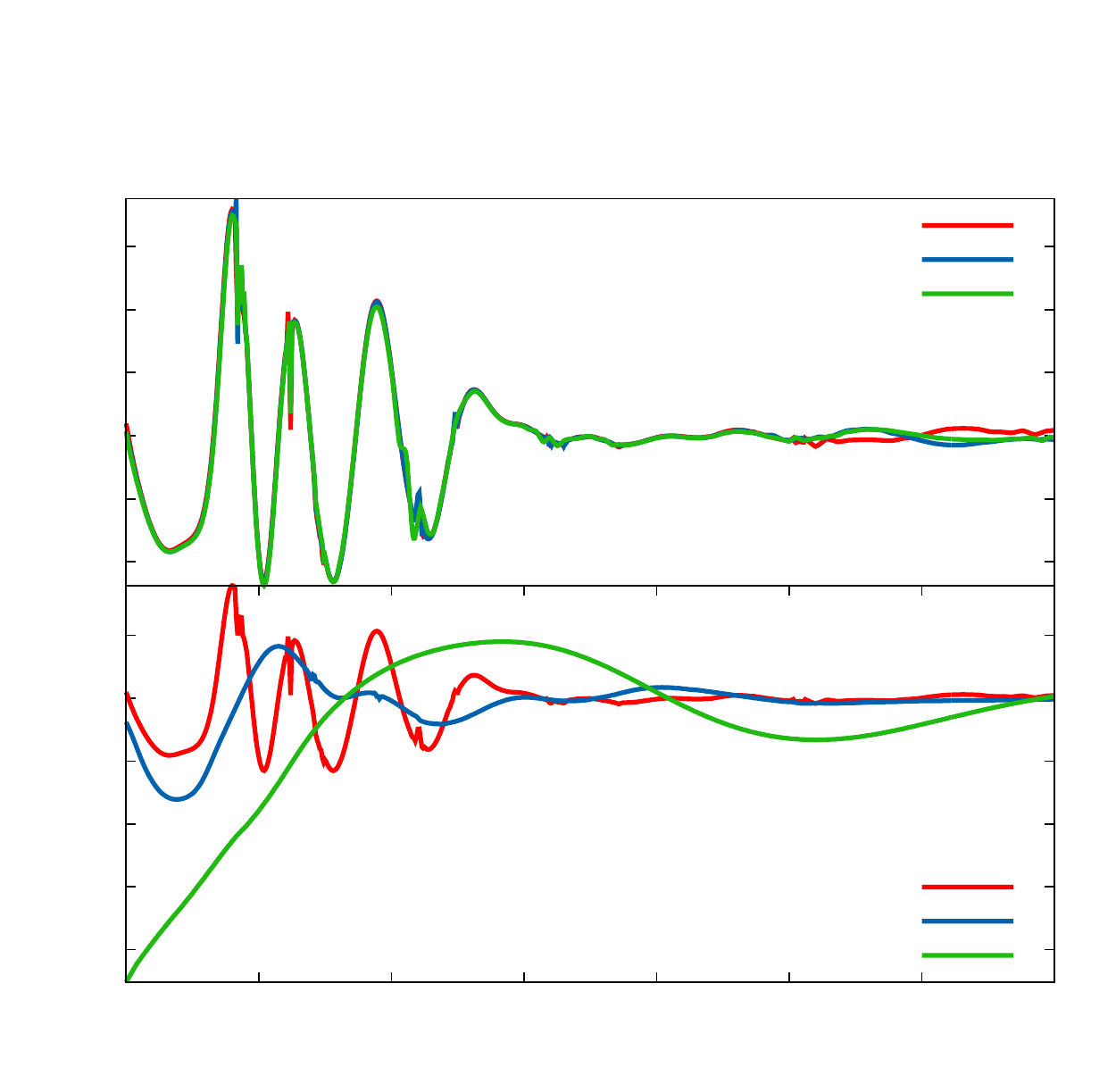}}%
    \gplfronttext
  \end{picture}%
\endgroup

%% file: figs/cfact.tex
\begingroup
  \makeatletter
  \providecommand\color[2][]{%
    \GenericError{(gnuplot) \space\space\space\@spaces}{%
      Package color not loaded in conjunction with
      terminal option `colourtext'%
    }{See the gnuplot documentation for explanation.%
    }{Either use 'blacktext' in gnuplot or load the package
      color.sty in LaTeX.}%
    \renewcommand\color[2][]{}%
  }%
  \providecommand\includegraphics[2][]{%
    \GenericError{(gnuplot) \space\space\space\@spaces}{%
      Package graphicx or graphics not loaded%
    }{See the gnuplot documentation for explanation.%
    }{The gnuplot epslatex terminal needs graphicx.sty or graphics.sty.}%
    \renewcommand\includegraphics[2][]{}%
  }%
  \providecommand\rotatebox[2]{#2}%
  \@ifundefined{ifGPcolor}{%
    \newif\ifGPcolor
    \GPcolortrue
  }{}%
  \@ifundefined{ifGPblacktext}{%
    \newif\ifGPblacktext
    \GPblacktexttrue
  }{}%
  \let\gplgaddtomacro\g@addto@macro
  \gdef\gplbacktext{}%
  \gdef\gplfronttext{}%
  \makeatother
  \ifGPblacktext
    \def\colorrgb#1{}%
    \def\colorgray#1{}%
  \else
    \ifGPcolor
      \def\colorrgb#1{\color[rgb]{#1}}%
      \def\colorgray#1{\color[gray]{#1}}%
      \expandafter\def\csname LTw\endcsname{\color{white}}%
      \expandafter\def\csname LTb\endcsname{\color{black}}%
      \expandafter\def\csname LTa\endcsname{\color{black}}%
      \expandafter\def\csname LT0\endcsname{\color[rgb]{1,0,0}}%
      \expandafter\def\csname LT1\endcsname{\color[rgb]{0,1,0}}%
      \expandafter\def\csname LT2\endcsname{\color[rgb]{0,0,1}}%
      \expandafter\def\csname LT3\endcsname{\color[rgb]{1,0,1}}%
      \expandafter\def\csname LT4\endcsname{\color[rgb]{0,1,1}}%
      \expandafter\def\csname LT5\endcsname{\color[rgb]{1,1,0}}%
      \expandafter\def\csname LT6\endcsname{\color[rgb]{0,0,0}}%
      \expandafter\def\csname LT7\endcsname{\color[rgb]{1,0.3,0}}%
      \expandafter\def\csname LT8\endcsname{\color[rgb]{0.5,0.5,0.5}}%
    \else
      \def\colorrgb#1{\color{black}}%
      \def\colorgray#1{\color[gray]{#1}}%
      \expandafter\def\csname LTw\endcsname{\color{white}}%
      \expandafter\def\csname LTb\endcsname{\color{black}}%
      \expandafter\def\csname LTa\endcsname{\color{black}}%
      \expandafter\def\csname LT0\endcsname{\color{black}}%
      \expandafter\def\csname LT1\endcsname{\color{black}}%
      \expandafter\def\csname LT2\endcsname{\color{black}}%
      \expandafter\def\csname LT3\endcsname{\color{black}}%
      \expandafter\def\csname LT4\endcsname{\color{black}}%
      \expandafter\def\csname LT5\endcsname{\color{black}}%
      \expandafter\def\csname LT6\endcsname{\color{black}}%
      \expandafter\def\csname LT7\endcsname{\color{black}}%
      \expandafter\def\csname LT8\endcsname{\color{black}}%
    \fi
  \fi
  \setlength{\unitlength}{0.0500bp}%
  \begin{picture}(8640.00,3528.00)%
    \gplgaddtomacro\gplbacktext{%
      \csname LTb\endcsname%
      \put(946,704){\makebox(0,0)[r]{\strut{} 0}}%
      \put(946,960){\makebox(0,0)[r]{\strut{} 0.5}}%
      \put(946,1216){\makebox(0,0)[r]{\strut{} 1}}%
      \put(946,1472){\makebox(0,0)[r]{\strut{} 1.5}}%
      \put(946,1728){\makebox(0,0)[r]{\strut{} 2}}%
      \put(946,1984){\makebox(0,0)[r]{\strut{} 2.5}}%
      \put(946,2239){\makebox(0,0)[r]{\strut{} 3}}%
      \put(946,2495){\makebox(0,0)[r]{\strut{} 3.5}}%
      \put(946,2751){\makebox(0,0)[r]{\strut{} 4}}%
      \put(946,3007){\makebox(0,0)[r]{\strut{} 4.5}}%
      \put(946,3263){\makebox(0,0)[r]{\strut{} 5}}%
      \put(1078,484){\makebox(0,0){\strut{} 0}}%
      \put(2523,484){\makebox(0,0){\strut{} 50}}%
      \put(3969,484){\makebox(0,0){\strut{} 100}}%
      \put(5414,484){\makebox(0,0){\strut{} 150}}%
      \put(6859,484){\makebox(0,0){\strut{} 200}}%
      \put(176,1983){\rotatebox{-270}{\makebox(0,0){\strut{}$c$}}}%
      \put(3968,154){\makebox(0,0){\strut{}$\tau$}}%
    }%
    \gplgaddtomacro\gplfronttext{%
      \csname LTb\endcsname%
      \put(7651,2313){\makebox(0,0)[r]{\strut{}1st}}%
      \csname LTb\endcsname%
      \put(7651,2093){\makebox(0,0)[r]{\strut{}2nd}}%
      \csname LTb\endcsname%
      \put(7651,1873){\makebox(0,0)[r]{\strut{}4th}}%
      \csname LTb\endcsname%
      \put(7651,1653){\makebox(0,0)[r]{\strut{}6th}}%
    }%
    \gplbacktext
    \put(0,0){\includegraphics{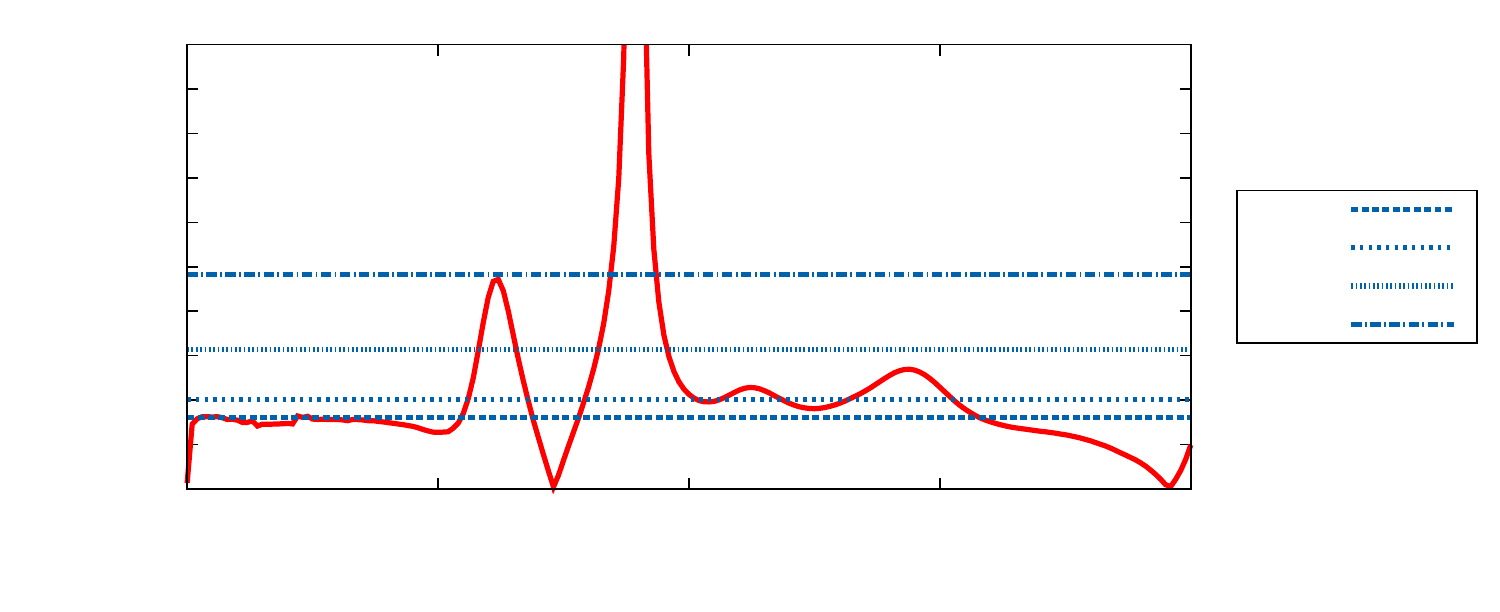}}%
    \gplfronttext
  \end{picture}%
\endgroup

%% file: figs/cont.tex
\begingroup
  \makeatletter
  \providecommand\color[2][]{%
    \GenericError{(gnuplot) \space\space\space\@spaces}{%
      Package color not loaded in conjunction with
      terminal option `colourtext'%
    }{See the gnuplot documentation for explanation.%
    }{Either use 'blacktext' in gnuplot or load the package
      color.sty in LaTeX.}%
    \renewcommand\color[2][]{}%
  }%
  \providecommand\includegraphics[2][]{%
    \GenericError{(gnuplot) \space\space\space\@spaces}{%
      Package graphicx or graphics not loaded%
    }{See the gnuplot documentation for explanation.%
    }{The gnuplot epslatex terminal needs graphicx.sty or graphics.sty.}%
    \renewcommand\includegraphics[2][]{}%
  }%
  \providecommand\rotatebox[2]{#2}%
  \@ifundefined{ifGPcolor}{%
    \newif\ifGPcolor
    \GPcolortrue
  }{}%
  \@ifundefined{ifGPblacktext}{%
    \newif\ifGPblacktext
    \GPblacktexttrue
  }{}%
  \let\gplgaddtomacro\g@addto@macro
  \gdef\gplbacktext{}%
  \gdef\gplfronttext{}%
  \makeatother
  \ifGPblacktext
    \def\colorrgb#1{}%
    \def\colorgray#1{}%
  \else
    \ifGPcolor
      \def\colorrgb#1{\color[rgb]{#1}}%
      \def\colorgray#1{\color[gray]{#1}}%
      \expandafter\def\csname LTw\endcsname{\color{white}}%
      \expandafter\def\csname LTb\endcsname{\color{black}}%
      \expandafter\def\csname LTa\endcsname{\color{black}}%
      \expandafter\def\csname LT0\endcsname{\color[rgb]{1,0,0}}%
      \expandafter\def\csname LT1\endcsname{\color[rgb]{0,1,0}}%
      \expandafter\def\csname LT2\endcsname{\color[rgb]{0,0,1}}%
      \expandafter\def\csname LT3\endcsname{\color[rgb]{1,0,1}}%
      \expandafter\def\csname LT4\endcsname{\color[rgb]{0,1,1}}%
      \expandafter\def\csname LT5\endcsname{\color[rgb]{1,1,0}}%
      \expandafter\def\csname LT6\endcsname{\color[rgb]{0,0,0}}%
      \expandafter\def\csname LT7\endcsname{\color[rgb]{1,0.3,0}}%
      \expandafter\def\csname LT8\endcsname{\color[rgb]{0.5,0.5,0.5}}%
    \else
      \def\colorrgb#1{\color{black}}%
      \def\colorgray#1{\color[gray]{#1}}%
      \expandafter\def\csname LTw\endcsname{\color{white}}%
      \expandafter\def\csname LTb\endcsname{\color{black}}%
      \expandafter\def\csname LTa\endcsname{\color{black}}%
      \expandafter\def\csname LT0\endcsname{\color{black}}%
      \expandafter\def\csname LT1\endcsname{\color{black}}%
      \expandafter\def\csname LT2\endcsname{\color{black}}%
      \expandafter\def\csname LT3\endcsname{\color{black}}%
      \expandafter\def\csname LT4\endcsname{\color{black}}%
      \expandafter\def\csname LT5\endcsname{\color{black}}%
      \expandafter\def\csname LT6\endcsname{\color{black}}%
      \expandafter\def\csname LT7\endcsname{\color{black}}%
      \expandafter\def\csname LT8\endcsname{\color{black}}%
    \fi
  \fi
  \setlength{\unitlength}{0.0500bp}%
  \begin{picture}(7200.00,3528.00)%
    \gplgaddtomacro\gplbacktext{%
      \csname LTb\endcsname%
      \put(726,704){\makebox(0,0)[r]{\strut{} 0}}%
      \put(726,960){\makebox(0,0)[r]{\strut{} 10}}%
      \put(726,1216){\makebox(0,0)[r]{\strut{} 20}}%
      \put(726,1472){\makebox(0,0)[r]{\strut{} 30}}%
      \put(726,1728){\makebox(0,0)[r]{\strut{} 40}}%
      \put(726,1983){\makebox(0,0)[r]{\strut{} 50}}%
      \put(726,2239){\makebox(0,0)[r]{\strut{} 60}}%
      \put(726,2495){\makebox(0,0)[r]{\strut{} 70}}%
      \put(726,2751){\makebox(0,0)[r]{\strut{} 80}}%
      \put(726,3007){\makebox(0,0)[r]{\strut{} 90}}%
      \put(726,3263){\makebox(0,0)[r]{\strut{} 100}}%
      \put(858,484){\makebox(0,0){\strut{} 0}}%
      \put(2344,484){\makebox(0,0){\strut{} 50}}%
      \put(3831,484){\makebox(0,0){\strut{} 100}}%
      \put(5317,484){\makebox(0,0){\strut{} 150}}%
      \put(6803,484){\makebox(0,0){\strut{} 200}}%
      \put(3830,154){\makebox(0,0){\strut{}$\tau$}}%
    }%
    \gplgaddtomacro\gplfronttext{%
      \csname LTb\endcsname%
      \put(2970,3090){\makebox(0,0)[r]{\strut{}$a_{\rm FLRW}(\tau)$}}%
      \csname LTb\endcsname%
      \put(2970,2870){\makebox(0,0)[r]{\strut{}$D_{\rm edge}^{(\rm c)}(\tau)$}}%
    }%
    \gplbacktext
    \put(0,0){\includegraphics{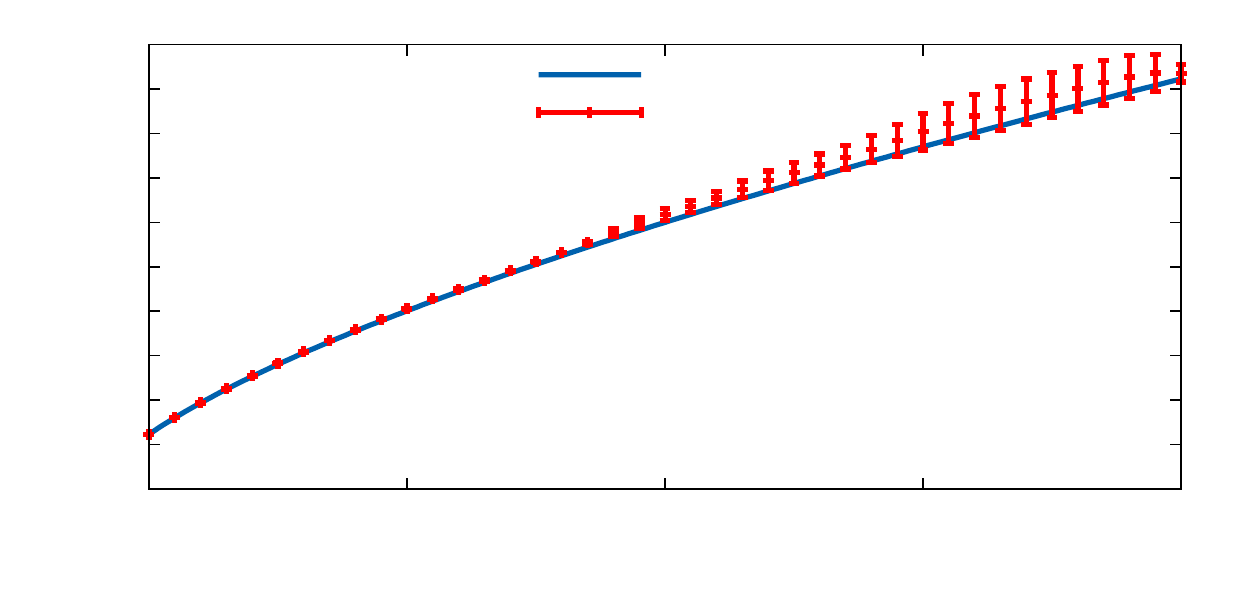}}%
    \gplfronttext
  \end{picture}%
\endgroup

%% file: figs/dvsFLRW.tex
\begingroup
  \makeatletter
  \providecommand\color[2][]{%
    \GenericError{(gnuplot) \space\space\space\@spaces}{%
      Package color not loaded in conjunction with
      terminal option `colourtext'%
    }{See the gnuplot documentation for explanation.%
    }{Either use 'blacktext' in gnuplot or load the package
      color.sty in LaTeX.}%
    \renewcommand\color[2][]{}%
  }%
  \providecommand\includegraphics[2][]{%
    \GenericError{(gnuplot) \space\space\space\@spaces}{%
      Package graphicx or graphics not loaded%
    }{See the gnuplot documentation for explanation.%
    }{The gnuplot epslatex terminal needs graphicx.sty or graphics.sty.}%
    \renewcommand\includegraphics[2][]{}%
  }%
  \providecommand\rotatebox[2]{#2}%
  \@ifundefined{ifGPcolor}{%
    \newif\ifGPcolor
    \GPcolortrue
  }{}%
  \@ifundefined{ifGPblacktext}{%
    \newif\ifGPblacktext
    \GPblacktexttrue
  }{}%
  \let\gplgaddtomacro\g@addto@macro
  \gdef\gplbacktext{}%
  \gdef\gplfronttext{}%
  \makeatother
  \ifGPblacktext
    \def\colorrgb#1{}%
    \def\colorgray#1{}%
  \else
    \ifGPcolor
      \def\colorrgb#1{\color[rgb]{#1}}%
      \def\colorgray#1{\color[gray]{#1}}%
      \expandafter\def\csname LTw\endcsname{\color{white}}%
      \expandafter\def\csname LTb\endcsname{\color{black}}%
      \expandafter\def\csname LTa\endcsname{\color{black}}%
      \expandafter\def\csname LT0\endcsname{\color[rgb]{1,0,0}}%
      \expandafter\def\csname LT1\endcsname{\color[rgb]{0,1,0}}%
      \expandafter\def\csname LT2\endcsname{\color[rgb]{0,0,1}}%
      \expandafter\def\csname LT3\endcsname{\color[rgb]{1,0,1}}%
      \expandafter\def\csname LT4\endcsname{\color[rgb]{0,1,1}}%
      \expandafter\def\csname LT5\endcsname{\color[rgb]{1,1,0}}%
      \expandafter\def\csname LT6\endcsname{\color[rgb]{0,0,0}}%
      \expandafter\def\csname LT7\endcsname{\color[rgb]{1,0.3,0}}%
      \expandafter\def\csname LT8\endcsname{\color[rgb]{0.5,0.5,0.5}}%
    \else
      \def\colorrgb#1{\color{black}}%
      \def\colorgray#1{\color[gray]{#1}}%
      \expandafter\def\csname LTw\endcsname{\color{white}}%
      \expandafter\def\csname LTb\endcsname{\color{black}}%
      \expandafter\def\csname LTa\endcsname{\color{black}}%
      \expandafter\def\csname LT0\endcsname{\color{black}}%
      \expandafter\def\csname LT1\endcsname{\color{black}}%
      \expandafter\def\csname LT2\endcsname{\color{black}}%
      \expandafter\def\csname LT3\endcsname{\color{black}}%
      \expandafter\def\csname LT4\endcsname{\color{black}}%
      \expandafter\def\csname LT5\endcsname{\color{black}}%
      \expandafter\def\csname LT6\endcsname{\color{black}}%
      \expandafter\def\csname LT7\endcsname{\color{black}}%
      \expandafter\def\csname LT8\endcsname{\color{black}}%
    \fi
  \fi
  \setlength{\unitlength}{0.0500bp}%
  \begin{picture}(7200.00,3528.00)%
    \gplgaddtomacro\gplbacktext{%
      \csname LTb\endcsname%
      \put(1210,704){\makebox(0,0)[r]{\strut{}-0.004}}%
      \put(1210,960){\makebox(0,0)[r]{\strut{}-0.002}}%
      \put(1210,1216){\makebox(0,0)[r]{\strut{} 0}}%
      \put(1210,1472){\makebox(0,0)[r]{\strut{} 0.002}}%
      \put(1210,1728){\makebox(0,0)[r]{\strut{} 0.004}}%
      \put(1210,1984){\makebox(0,0)[r]{\strut{} 0.006}}%
      \put(1210,2239){\makebox(0,0)[r]{\strut{} 0.008}}%
      \put(1210,2495){\makebox(0,0)[r]{\strut{} 0.01}}%
      \put(1210,2751){\makebox(0,0)[r]{\strut{} 0.012}}%
      \put(1210,3007){\makebox(0,0)[r]{\strut{} 0.014}}%
      \put(1210,3263){\makebox(0,0)[r]{\strut{} 0.016}}%
      \put(1342,484){\makebox(0,0){\strut{} 0}}%
      \put(2122,484){\makebox(0,0){\strut{} 10}}%
      \put(2902,484){\makebox(0,0){\strut{} 20}}%
      \put(3682,484){\makebox(0,0){\strut{} 30}}%
      \put(4463,484){\makebox(0,0){\strut{} 40}}%
      \put(5243,484){\makebox(0,0){\strut{} 50}}%
      \put(6023,484){\makebox(0,0){\strut{} 60}}%
      \put(6803,484){\makebox(0,0){\strut{} 70}}%
      \put(176,1983){\rotatebox{-270}{\makebox(0,0){\strut{}$|D_{\rm edge}(\tau)-a_{\rm FLRW}(\tau)|/D_{\rm edge}(\tau)$}}}%
      \put(4072,154){\makebox(0,0){\strut{}$\tau$}}%
    }%
    \gplgaddtomacro\gplfronttext{%
      \csname LTb\endcsname%
      \put(5816,1537){\makebox(0,0)[r]{\strut{}L1}}%
      \csname LTb\endcsname%
      \put(5816,1317){\makebox(0,0)[r]{\strut{}L2}}%
      \csname LTb\endcsname%
      \put(5816,1097){\makebox(0,0)[r]{\strut{}L3}}%
      \csname LTb\endcsname%
      \put(5816,877){\makebox(0,0)[r]{\strut{}Continuum}}%
    }%
    \gplbacktext
    \put(0,0){\includegraphics{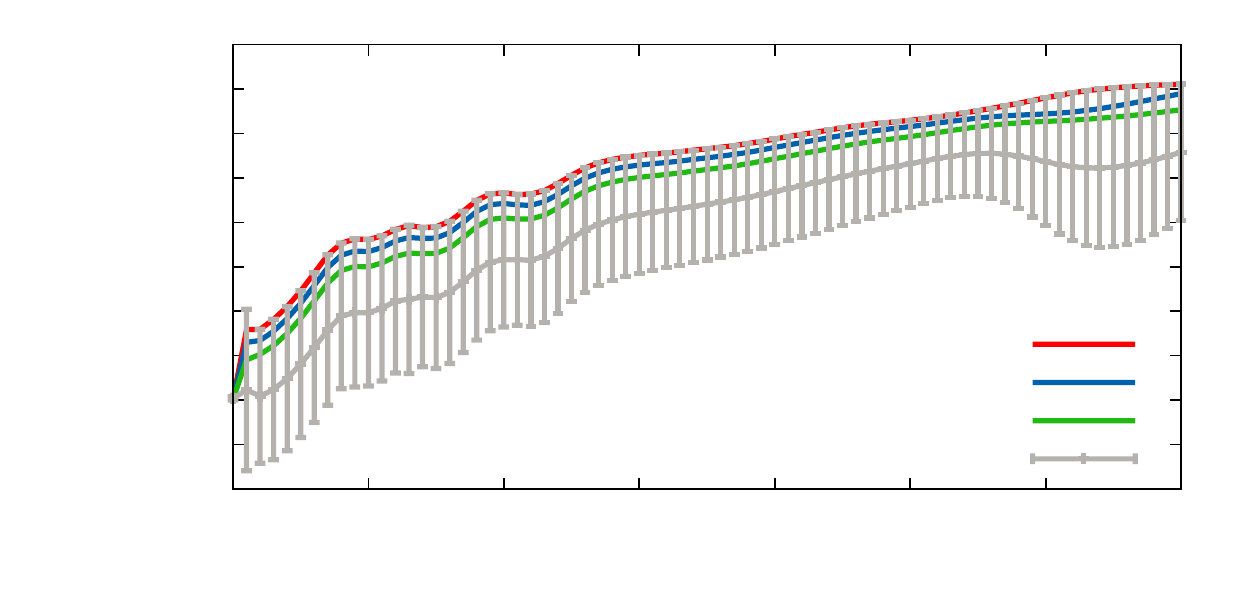}}%
    \gplfronttext
  \end{picture}%
\endgroup

%% file: figs/Mcont.tex
\begingroup
  \makeatletter
  \providecommand\color[2][]{%
    \GenericError{(gnuplot) \space\space\space\@spaces}{%
      Package color not loaded in conjunction with
      terminal option `colourtext'%
    }{See the gnuplot documentation for explanation.%
    }{Either use 'blacktext' in gnuplot or load the package
      color.sty in LaTeX.}%
    \renewcommand\color[2][]{}%
  }%
  \providecommand\includegraphics[2][]{%
    \GenericError{(gnuplot) \space\space\space\@spaces}{%
      Package graphicx or graphics not loaded%
    }{See the gnuplot documentation for explanation.%
    }{The gnuplot epslatex terminal needs graphicx.sty or graphics.sty.}%
    \renewcommand\includegraphics[2][]{}%
  }%
  \providecommand\rotatebox[2]{#2}%
  \@ifundefined{ifGPcolor}{%
    \newif\ifGPcolor
    \GPcolortrue
  }{}%
  \@ifundefined{ifGPblacktext}{%
    \newif\ifGPblacktext
    \GPblacktexttrue
  }{}%
  \let\gplgaddtomacro\g@addto@macro
  \gdef\gplbacktext{}%
  \gdef\gplfronttext{}%
  \makeatother
  \ifGPblacktext
    \def\colorrgb#1{}%
    \def\colorgray#1{}%
  \else
    \ifGPcolor
      \def\colorrgb#1{\color[rgb]{#1}}%
      \def\colorgray#1{\color[gray]{#1}}%
      \expandafter\def\csname LTw\endcsname{\color{white}}%
      \expandafter\def\csname LTb\endcsname{\color{black}}%
      \expandafter\def\csname LTa\endcsname{\color{black}}%
      \expandafter\def\csname LT0\endcsname{\color[rgb]{1,0,0}}%
      \expandafter\def\csname LT1\endcsname{\color[rgb]{0,1,0}}%
      \expandafter\def\csname LT2\endcsname{\color[rgb]{0,0,1}}%
      \expandafter\def\csname LT3\endcsname{\color[rgb]{1,0,1}}%
      \expandafter\def\csname LT4\endcsname{\color[rgb]{0,1,1}}%
      \expandafter\def\csname LT5\endcsname{\color[rgb]{1,1,0}}%
      \expandafter\def\csname LT6\endcsname{\color[rgb]{0,0,0}}%
      \expandafter\def\csname LT7\endcsname{\color[rgb]{1,0.3,0}}%
      \expandafter\def\csname LT8\endcsname{\color[rgb]{0.5,0.5,0.5}}%
    \else
      \def\colorrgb#1{\color{black}}%
      \def\colorgray#1{\color[gray]{#1}}%
      \expandafter\def\csname LTw\endcsname{\color{white}}%
      \expandafter\def\csname LTb\endcsname{\color{black}}%
      \expandafter\def\csname LTa\endcsname{\color{black}}%
      \expandafter\def\csname LT0\endcsname{\color{black}}%
      \expandafter\def\csname LT1\endcsname{\color{black}}%
      \expandafter\def\csname LT2\endcsname{\color{black}}%
      \expandafter\def\csname LT3\endcsname{\color{black}}%
      \expandafter\def\csname LT4\endcsname{\color{black}}%
      \expandafter\def\csname LT5\endcsname{\color{black}}%
      \expandafter\def\csname LT6\endcsname{\color{black}}%
      \expandafter\def\csname LT7\endcsname{\color{black}}%
      \expandafter\def\csname LT8\endcsname{\color{black}}%
    \fi
  \fi
  \setlength{\unitlength}{0.0500bp}%
  \begin{picture}(7200.00,3528.00)%
    \gplgaddtomacro\gplbacktext{%
      \csname LTb\endcsname%
      \put(1078,704){\makebox(0,0)[r]{\strut{} 0.85}}%
      \put(1078,1216){\makebox(0,0)[r]{\strut{} 0.9}}%
      \put(1078,1728){\makebox(0,0)[r]{\strut{} 0.95}}%
      \put(1078,2239){\makebox(0,0)[r]{\strut{} 1}}%
      \put(1078,2751){\makebox(0,0)[r]{\strut{} 1.05}}%
      \put(1078,3263){\makebox(0,0)[r]{\strut{} 1.1}}%
      \put(1210,484){\makebox(0,0){\strut{} 0}}%
      \put(2009,484){\makebox(0,0){\strut{} 10}}%
      \put(2808,484){\makebox(0,0){\strut{} 20}}%
      \put(3607,484){\makebox(0,0){\strut{} 30}}%
      \put(4406,484){\makebox(0,0){\strut{} 40}}%
      \put(5205,484){\makebox(0,0){\strut{} 50}}%
      \put(6004,484){\makebox(0,0){\strut{} 60}}%
      \put(6803,484){\makebox(0,0){\strut{} 70}}%
      \put(176,1983){\rotatebox{-270}{\makebox(0,0){\strut{}$M_{\rm eff}$}}}%
      \put(4006,154){\makebox(0,0){\strut{}$\tau$}}%
    }%
    \gplgaddtomacro\gplfronttext{%
      \csname LTb\endcsname%
      \put(2530,3090){\makebox(0,0)[r]{\strut{}L1}}%
      \csname LTb\endcsname%
      \put(2530,2870){\makebox(0,0)[r]{\strut{}L2}}%
      \csname LTb\endcsname%
      \put(2530,2650){\makebox(0,0)[r]{\strut{}L3}}%
      \csname LTb\endcsname%
      \put(2530,2430){\makebox(0,0)[r]{\strut{}Continuum}}%
    }%
    \gplbacktext
    \put(0,0){\includegraphics{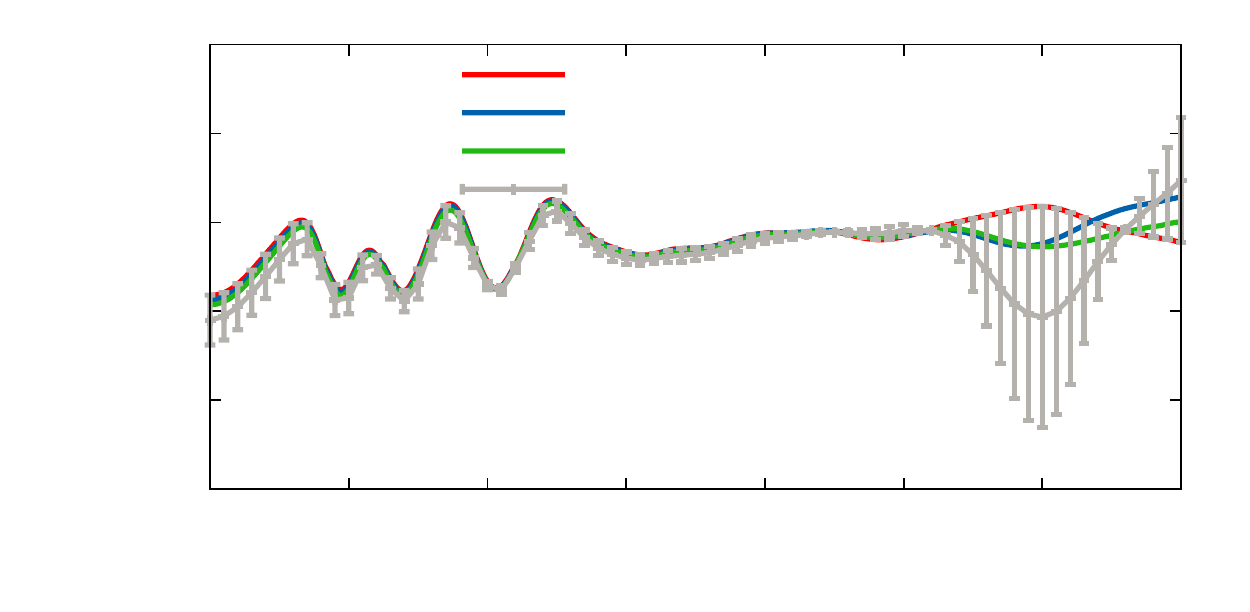}}%
    \gplfronttext
  \end{picture}%
\endgroup

%% file: figs/pcont.tex
\begingroup
  \makeatletter
  \providecommand\color[2][]{%
    \GenericError{(gnuplot) \space\space\space\@spaces}{%
      Package color not loaded in conjunction with
      terminal option `colourtext'%
    }{See the gnuplot documentation for explanation.%
    }{Either use 'blacktext' in gnuplot or load the package
      color.sty in LaTeX.}%
    \renewcommand\color[2][]{}%
  }%
  \providecommand\includegraphics[2][]{%
    \GenericError{(gnuplot) \space\space\space\@spaces}{%
      Package graphicx or graphics not loaded%
    }{See the gnuplot documentation for explanation.%
    }{The gnuplot epslatex terminal needs graphicx.sty or graphics.sty.}%
    \renewcommand\includegraphics[2][]{}%
  }%
  \providecommand\rotatebox[2]{#2}%
  \@ifundefined{ifGPcolor}{%
    \newif\ifGPcolor
    \GPcolortrue
  }{}%
  \@ifundefined{ifGPblacktext}{%
    \newif\ifGPblacktext
    \GPblacktexttrue
  }{}%
  \let\gplgaddtomacro\g@addto@macro
  \gdef\gplbacktext{}%
  \gdef\gplfronttext{}%
  \makeatother
  \ifGPblacktext
    \def\colorrgb#1{}%
    \def\colorgray#1{}%
  \else
    \ifGPcolor
      \def\colorrgb#1{\color[rgb]{#1}}%
      \def\colorgray#1{\color[gray]{#1}}%
      \expandafter\def\csname LTw\endcsname{\color{white}}%
      \expandafter\def\csname LTb\endcsname{\color{black}}%
      \expandafter\def\csname LTa\endcsname{\color{black}}%
      \expandafter\def\csname LT0\endcsname{\color[rgb]{1,0,0}}%
      \expandafter\def\csname LT1\endcsname{\color[rgb]{0,1,0}}%
      \expandafter\def\csname LT2\endcsname{\color[rgb]{0,0,1}}%
      \expandafter\def\csname LT3\endcsname{\color[rgb]{1,0,1}}%
      \expandafter\def\csname LT4\endcsname{\color[rgb]{0,1,1}}%
      \expandafter\def\csname LT5\endcsname{\color[rgb]{1,1,0}}%
      \expandafter\def\csname LT6\endcsname{\color[rgb]{0,0,0}}%
      \expandafter\def\csname LT7\endcsname{\color[rgb]{1,0.3,0}}%
      \expandafter\def\csname LT8\endcsname{\color[rgb]{0.5,0.5,0.5}}%
    \else
      \def\colorrgb#1{\color{black}}%
      \def\colorgray#1{\color[gray]{#1}}%
      \expandafter\def\csname LTw\endcsname{\color{white}}%
      \expandafter\def\csname LTb\endcsname{\color{black}}%
      \expandafter\def\csname LTa\endcsname{\color{black}}%
      \expandafter\def\csname LT0\endcsname{\color{black}}%
      \expandafter\def\csname LT1\endcsname{\color{black}}%
      \expandafter\def\csname LT2\endcsname{\color{black}}%
      \expandafter\def\csname LT3\endcsname{\color{black}}%
      \expandafter\def\csname LT4\endcsname{\color{black}}%
      \expandafter\def\csname LT5\endcsname{\color{black}}%
      \expandafter\def\csname LT6\endcsname{\color{black}}%
      \expandafter\def\csname LT7\endcsname{\color{black}}%
      \expandafter\def\csname LT8\endcsname{\color{black}}%
    \fi
  \fi
  \setlength{\unitlength}{0.0500bp}%
  \begin{picture}(7200.00,3528.00)%
    \gplgaddtomacro\gplbacktext{%
      \csname LTb\endcsname%
      \put(1210,704){\makebox(0,0)[r]{\strut{}-3e-05}}%
      \put(1210,1070){\makebox(0,0)[r]{\strut{}-2e-05}}%
      \put(1210,1435){\makebox(0,0)[r]{\strut{}-1e-05}}%
      \put(1210,1801){\makebox(0,0)[r]{\strut{} 0}}%
      \put(1210,2166){\makebox(0,0)[r]{\strut{} 1e-05}}%
      \put(1210,2532){\makebox(0,0)[r]{\strut{} 2e-05}}%
      \put(1210,2897){\makebox(0,0)[r]{\strut{} 3e-05}}%
      \put(1210,3263){\makebox(0,0)[r]{\strut{} 4e-05}}%
      \put(1342,484){\makebox(0,0){\strut{} 0}}%
      \put(2122,484){\makebox(0,0){\strut{} 10}}%
      \put(2902,484){\makebox(0,0){\strut{} 20}}%
      \put(3682,484){\makebox(0,0){\strut{} 30}}%
      \put(4463,484){\makebox(0,0){\strut{} 40}}%
      \put(5243,484){\makebox(0,0){\strut{} 50}}%
      \put(6023,484){\makebox(0,0){\strut{} 60}}%
      \put(6803,484){\makebox(0,0){\strut{} 70}}%
      \put(176,1983){\rotatebox{-270}{\makebox(0,0){\strut{}$p_{\rm eff}$}}}%
      \put(4072,154){\makebox(0,0){\strut{}$\tau$}}%
    }%
    \gplgaddtomacro\gplfronttext{%
      \csname LTb\endcsname%
      \put(5816,3090){\makebox(0,0)[r]{\strut{}L1}}%
      \csname LTb\endcsname%
      \put(5816,2870){\makebox(0,0)[r]{\strut{}L2}}%
      \csname LTb\endcsname%
      \put(5816,2650){\makebox(0,0)[r]{\strut{}L3}}%
      \csname LTb\endcsname%
      \put(5816,2430){\makebox(0,0)[r]{\strut{}Continuum}}%
    }%
    \gplbacktext
    \put(0,0){\includegraphics{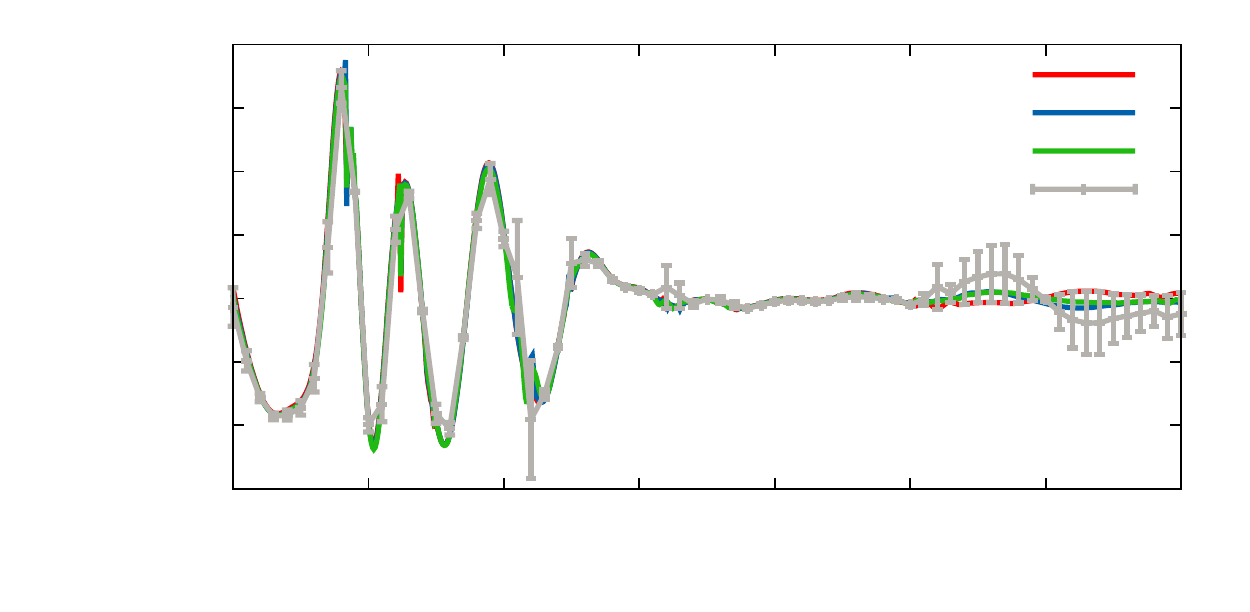}}%
    \gplfronttext
  \end{picture}%
\endgroup